\pdfoutput=1
\documentclass[11pt,twoside,a4paper,cmspaper,final]{cms-tdr}
\def\svnVersion{ecc7345-D}\def\svnDate{2023/10/06}\def\cmsCernNoTag{CERN-EP-2021-221}\def\cmsCernDate{\today}\def\cmsMessage{Published in the Journal of High Energy Physics as \href{http://dx.doi.org/10.1007/JHEP02(2022)142}{\doi{10.1007/JHEP02(2022)142}.}}
\begin{document}\cmsNoteHeader{SMP-20-011}

\newlength\cmsTabSkip\setlength{\cmsTabSkip}{1ex}
\renewcommand{\arraystretch}{1.3}
\providecommand{\cmsTable}[1]{\resizebox{\textwidth}{!}{#1}}

\newcommand{\chisq}{\ensuremath{\chi^2}\xspace}
\newcommand{\Ep}{\ensuremath{E_\Pp}\xspace}
\newcommand{\absy}{\ensuremath{\abs{y}}\xspace}
\newcommand{\sqrts}{\ensuremath{\sqrt{s}}}

\newcommand{\alpSZ}{\ensuremath{\alpS(m_\PZ)}\xspace}
\newcommand{\ptHLT}{\ensuremath{\pt^\mathrm{HLT}}\xspace}
\newcommand{\ptPF}{\ensuremath{\pt^\mathrm{PF}}\xspace}
\newcommand{\Ndof}{\ensuremath{N_\mathrm{dof}}\xspace}
\newcommand{\Ndp}{\ensuremath{N_\mathrm{dp}}\xspace}
\newcommand{\ptgen}{\ensuremath{\pt^\text{gen}}\xspace}
\newcommand{\ptrec}{\ensuremath{\pt^\text{rec}}\xspace}
\newcommand{\absygen}{\ensuremath{\abs{y}^\text{gen}}\xspace}
\newcommand{\absyrec}{\ensuremath{\abs{y}^\text{rec}}\xspace}
\newcommand{\muf}{\ensuremath{\mu_\mathrm{f}}\xspace}
\newcommand{\mur}{\ensuremath{\mu_\mathrm{r}}\xspace}
\newcommand{\sigNLO}{\ensuremath{\sigma^{\mathrm{NLO}}}\xspace}
\newcommand{\sigNLL}{\ensuremath{\sigma^{\mathrm{NLL}}}\xspace}
\newcommand{\sigNNLO}{\ensuremath{\sigma^{\mathrm{NNLO}}}\xspace}
\newcommand{\mt}{\ensuremath{m_\PQt}}
\newcommand{\mb}{\ensuremath{m_\PQb}}
\newcommand{\mc}{\ensuremath{m_\PQc}}
\newcommand{\mtpole}{\ensuremath{\mt^{\text{pole}}}\xspace}
\newcommand{\Ubar}{\overline{\mathrm{U}}\xspace}
\newcommand{\Dbar}{\overline{\mathrm{D}}\xspace}
\newcommand{\ptmax}{\ensuremath{\pt^\mathrm{max}}\xspace}
\newcommand{\Qmin}{\ensuremath{Q^2_\mathrm{min}}\xspace}
\newcommand{\fastNLO}{\ensuremath{\textsc{fastNLO}}\xspace}
\newcommand{\xFitter}{\textsc{xFitter}\xspace}
\newcommand{\QCDNUM}{\textsc{qcdnum}\xspace}
\newcommand{\CIJET}{\textsc{cijet}\xspace}
\newcommand{\TUnfold}{\textsc{TUnfold}\xspace}
\newcommand{\pp}{\ensuremath{\Pp\Pp}\xspace}
\providecommand{\suppMaterial}{Appendix~\ref{app:suppMat}}

\cmsNoteHeader{SMP-20-011}
\title{Measurement and QCD analysis of double-differential inclusive jet cross sections in proton-proton collisions at \texorpdfstring{$\sqrt{s} = 13\TeV$}{sqrt(s) = 13 TeV}}

\author{The CMS Collaboration}

\date{\today}

\abstract{A measurement of the inclusive jet production in proton-proton collisions at the LHC at $\sqrt{s}=13\TeV$ is presented. 
The double-differential cross sections are measured as a function of the jet transverse momentum $\pt$ and the absolute jet rapidity $\abs{y}$. The anti-$\kt$ clustering algorithm is used with distance parameter of 0.4 (0.7) in a phase space region with jet \pt from 97\GeV up to 3.1\TeV and $\abs{y}<2.0$. Data collected with the CMS detector are used, corresponding to an integrated luminosity of 36.3\fbinv (33.5\fbinv). The measurement is used in a comprehensive QCD analysis at next-to-next-to-leading order, which results in significant improvement in the accuracy of the parton distributions in the proton. Simultaneously, the value of the strong coupling constant at the \PZ boson mass is extracted as $\alpSZ= 0.1170 \pm 0.0019$. For the first time, these data are used in a standard model effective field theory analysis at next-to-leading order, where parton distributions and the QCD parameters are extracted simultaneously with imposed constraints on the Wilson coefficient $c_1$ of 4-quark contact interactions.\\[1cm]
\textbf{Note added: in the Addendum to this paper \DOI{10.1007/JHEP12(2022)035}, available as Appendix B in this document, an improved value of $\alpSZ = 0.1166 \pm 0.0017$ has been extracted. This result supersedes the number in the above abstract of the original publication \DOI{10.1007/JHEP02(2022)142}.}  
}

\hypersetup{%
pdfauthor={CMS Collaboration},%
pdftitle={Measurement and QCD analysis of double-differential inclusive jet cross sections in proton-proton collisions at sqrt(s) = 13 TeV},%
pdfsubject={CMS},%
pdfkeywords={CMS, QCD, jet, proton, collisions, PDF, NLO, NNLO, NLL}}

\maketitle

\section{Introduction} \label{sec:intro}

Quantum chromodynamics (QCD) is the theory describing strong interactions among partons~(quarks and gluons), the fundamental constituents of hadrons.
In high-energy proton-proton (\pp) collisions, partons from both colliding protons interact, producing energetic collimated sprays of hadrons~(jets) in the final state. 
Inclusive jet production in $\Pp+\Pp \to \text{jet} + \mathrm{X}$, consisting of events with at least one jet, is a key process to test QCD predictions at the highest achievable energy scales.
At the CERN LHC, inclusive jet production in $\Pp\Pp$~collisions has been extensively measured by both the ATLAS and CMS Collaborations at several centre-of-mass energies $\sqrt{s}$.
The present status of the measurements with corresponding integrated luminosities is summarised in Table~\ref{tab:prevIncJets}; the earlier 7\TeV measurements~\cite{Aad:2010ad,Aad:2011fc,CMS:2011ab} with lower integrated luminosities are omitted.
In Ref.~\cite{Khachatryan:2016wdh}, the first data collected by the CMS Collaboration at 13\TeV were analysed.

\begin{table}[ht]
    \centering
    \topcaption[Past measurements]{Recent measurements of inclusive jet production, performed by the ATLAS and CMS Collaborations at different $\sqrt{s}$, with the corresponding integrated luminosities.  }
    \begin{tabular}{ccc}
        $\sqrt{s}$         & ATLAS       &   CMS \\
        \hline
        2.76\TeV          & 0.0002\fbinv~\cite{Aad:2013lpa}   &  0.0054\fbinv~\cite{Khachatryan:2015luy} \\
        7\TeV             & 4.5\fbinv~\cite{Aad:2014vwa}      &  5.0\fbinv~\cite{Chatrchyan:2012bja,Chatrchyan:2014gia} \\
        8\TeV             & 20\fbinv~\cite{Aaboud:2017dvo}    &  20\fbinv~\cite{Khachatryan:2016mlc}\\
        13\TeV            & 3.2\fbinv~\cite{Aaboud:2017wsi}   &  0.071\fbinv~\cite{Khachatryan:2016wdh}
    \end{tabular}
    \label{tab:prevIncJets}
\end{table}
These measurements were compared with fixed-order predictions in perturbative QCD (pQCD) at next-to-leading order (NLO).
Predictions at next-to-leading-logarithmic order (NLL)~\cite{Liu:2018ktv} and at next-to-next-to-leading order (NNLO) are available~\cite{Currie:2016bfm,Currie:2018xkj} and describe the LHC data~\cite{Aaboud:2017wsi} well using the transverse momentum \pt of an individual jet as the renormalisation and factorisation scales.

Inclusive jet production at high momenta probes the proton structure in the kinematic range of high fraction $x$ of the proton momentum carried by the parton. In particular, it is directly sensitive to the gluon distribution in the proton at high $x$. Measurements of the inclusive jet cross sections provide additional constraints on the parton distribution functions (PDFs) and the value of the strong coupling constant \alpS, as demonstrated \eg in previous CMS publications~\cite{Khachatryan:2014waa,Khachatryan:2016mlc}.

In this paper, the data collected by the CMS experiment in 2016, corresponding to an integrated luminosity of up to 36.3\fbinv, are analysed.
The measurement of the double-differential inclusive jet cross sections is presented as a function of the jet~\pt and jet rapidity $\abs{y}$.
The jets are clustered with the anti-\kt jet algorithm~\cite{Cacciari:2008gp}, as implemented in the \FASTJET package~\cite{Cacciari:2011ma}.
Two jet distance parameters, $R=0.4$ and 0.7, are used:
\begin{itemize}
    \item  $R=0.4$ is a default in most recent analyses both in ATLAS and CMS at 13\TeV.
    \item  $R=0.7$ is chosen for most of QCD~analyses using jet measurements because the effects of out-of-cone radiation are smaller. In particular, this value is used in the analogous analysis~\cite{Khachatryan:2016mlc} with CMS data at 8\TeV.
\end{itemize}
The size of the nonperturbative (NP) corrections as a function of $R$ is discussed in Ref.~\cite{Dasgupta:2007wa}.
Experimentally, the impact of the jet radius on the inclusive jet cross sections is studied by the CMS Collaboration in Ref.~\cite{Sirunyan:2020uoj}.

The impact of the present measurements on the proton PDFs is illustrated in a QCD analysis, where the measured double-differential cross sections of inclusive jet production are used together with data from deep inelastic scattering (DIS) at HERA~\cite{Abramowicz:2015mha}.
In addition, the CMS measurements~\cite{Sirunyan:2019zvx} of normalised triple-differential top quark-antiquark (\ttbar) production cross sections are used, which provide additional sensitivities to the gluon distribution, \alpS, and the top~quark mass \mt.

Furthermore, the effect of beyond the standard model (BSM) particle exchanges between the quarks is studied, using the model of contact interactions (CI)~\cite{Gao:2012qpa,Gao:2013kp}, added to the standard model (SM) process via effective couplings. 
The earlier searches for CI by the CMS Collaboration were performed using inclusive jet~\cite{Chatrchyan:2013muj} and dijet~\cite{Chatrchyan:1427761,CMS:2014eaj,CMS:2017caz} production.
In those analyses, fixed values for the Wilson coefficients were assumed and the limits on the scale of the new interactions were set based on a comparison of data to the SM+CI prediction. However, the PDFs used in the SM prediction are derived assuming the validity of the SM at high jet \pt, where the effects of new physics are expected to be most pronounced.
Consequently, the BSM effects might be absorbed into PDFs and the interpretation of high-\pt jet data may be biased.
In the present analysis, the CI Wilson coefficient $c_1$ is a free parameter in the effective field theory (EFT)-improved SM (SMEFT) fit and is extracted simultaneously with the PDFs. The scenarios investigated correspond to purely left-handed, vector- and axial vector-like CI.

The paper is organised as follows.
In Section~\ref{sec:detector}, a brief description of the detector is given.
In Section~\ref{sec:analysis}, the measurement of the double-differential cross sections is detailed.
Theoretical predictions are explained in Section~\ref{sec:theory}.
Experimental and theoretical cross sections are compared in Section~\ref{sec:comparison}. Finally, the QCD interpretation is given in Section~\ref{sec:QCD}. The paper is summarised in Section~\ref{sec:summary}.

\section{The CMS detector} \label{sec:detector}

The central feature of the CMS apparatus is a superconducting solenoid of 6\unit{m} internal diameter, providing a magnetic field of 3.8\unit{T}. Within the solenoid volume are a silicon pixel and strip tracker, a lead tungstate crystal electromagnetic calorimeter (ECAL), and a brass and scintillator hadron calorimeter~(HCAL), each composed of a barrel and two endcap sections. Forward calorimeters extend the pseudorapidity $\eta$ coverage provided by the barrel and endcap detectors. Muons are detected in gas-ionisation chambers embedded in the steel flux-return yoke outside the solenoid. 

The ECAL consists of 75\,848~lead tungstate crystals, which provide coverage in $\abs{\eta} < 1.48 $ in the barrel region and $1.48 < \abs{\eta} < 3.00$ in two endcap regions. Preshower detectors consisting of two planes of silicon sensors interleaved with a total of $3 X_0$ of lead are located in front of each ECAL endcap detector.

In the region $\abs{\eta} < 1.74$, the HCAL cells have widths of 0.087 in $\eta$ and 0.087 in azimuth~($\phi$). In the $\eta$-$\phi$ plane, and for $\abs{\eta} < 1.48$, the HCAL cells map on to $5{\times}5$ arrays of ECAL crystals to form calorimeter towers projecting radially outwards from close to the nominal interaction point. For $\abs{\eta} > 1.74$, the coverage of the towers increases progressively to a maximum of 0.174 in $\Delta \eta$ and $\Delta \phi$. Within each tower, the energy deposits in ECAL and HCAL cells are summed to define the calorimeter tower energies.

The reconstructed vertex with the largest value of summed physics-object $\pt^2$ is taken to be the primary \pp interaction vertex.

The particle-flow (PF) algorithm~\cite{CMS-PRF-14-001} reconstructs and identifies each individual particle in an event, with an optimised combination of information from the various elements of the CMS detector.
The energy of charged hadrons is determined from a combination of their momentum measured in the tracker and the matching ECAL and HCAL energy deposits, corrected for zero-suppression effects and for the response function of the calorimeters to hadronic showers. Finally, the energy of neutral hadrons is obtained from the corresponding corrected ECAL and HCAL energies. 
Jets are reconstructed offline from PF objects using the anti-\kt algorithm~\cite{Cacciari:2008gp, Cacciari:2011ma} with $R$ of 0.4 and 0.7.

Jet momentum is the vector sum of all PF candidate momenta in the jet, and is determined from simulation to be, on average, within 5--10\% of the true momentum over the entire \pt spectrum and detector acceptance.
Additional \pp interactions within the same or nearby bunch crossings (pileup) can contribute additional tracks and calorimetric energy depositions that increase the detector-level jet momentum. To mitigate this effect, tracks identified as originating from pileup vertices are discarded and an offset correction is applied to correct for remaining contributions. Jet energy corrections  are derived from simulation studies so the average measured response of jets becomes identical to that of particle-level jets. In situ measurements of the momentum balance in dijet, photon+jet, \PZ+jet, and multijet events are used to determine any residual differences between the jet energy scale (JES) in data and in simulation, and appropriate corrections are derived~\cite{Khachatryan:2016kdb}. Additional selection criteria are applied to each jet to remove jets potentially affected by instrumental effects or reconstruction failures~\cite{CMS-PAS-JME-16-003}.

The missing transverse momentum vector \ptvecmiss is computed as the negative vector \pt sum of all the PF candidates in an event, and its magnitude is denoted as \ptmiss~\cite{Sirunyan:2019kia}. The \ptvecmiss is modified to account for corrections to the energy scale of the reconstructed objects in the event. 
Anomalous high-\ptmiss events can be due to a variety of reconstruction failures, detector malfunctions, or noncollisions backgrounds. Such events are rejected by event filters that are designed to identify more than 85--90\% of the spurious high-\ptmiss events with a mistagging rate less than 0.1\%.

Events of interest are selected online using a two-tiered trigger system~\cite{Khachatryan:2016bia}. The first level, composed of custom hardware processors, uses information from the calorimeters and muon detectors to select events at a rate of up to 100\unit{kHz} within a latency of less than 4\mus~\cite{Sirunyan_2020}. The second level, known as the high-level trigger (HLT), consists of a farm of processors running a version of the full event reconstruction software optimised for fast processing, and reduces the event rate to around 1\unit{kHz} before data storage.

During the 2016 data taking, a gradual shift in the timing of the inputs of the ECAL first-level trigger in the region $\abs{\eta} > 2.0$, referred to as prefiring, caused a specific trigger inefficiency. For events containing a jet with \pt larger than $\approx$100\GeV in the region $2.5 < \abs{\eta} < 3.0$, the efficiency loss is $\approx$10--20\%, depending on \pt, $\eta$, and data taking period. Correction factors were computed from data and applied to the acceptance evaluated by simulation. 

A more detailed description of the CMS detector, together with a definition of the coordinate system used and the relevant kinematic variables, can be found in Ref.~\cite{Chatrchyan:2008zzk}. 

\section{Data analysis} \label{sec:dataReduction} \label{sec:analysis}

The inclusive jet double-differential cross section, as a function of the individual jet \pt and $y$ is defined as follows:
\begin{equation}
    \frac{\rd^2 \sigma}{\rd \pt \, \rd y} = \frac{1}{\mathcal{L}} \frac{N_\text{jets}^\text{eff}}{\Delta \pt \, \Delta y},
\end{equation}
where $\mathcal{L}$ corresponds to the integrated luminosity, $\Delta\pt$ ($\Delta y$) to the bin width of the jet \pt ($y$), and $N_\text{jets}^\text{eff}$ to the effective number of jets per bin estimated at the particle level, \ie after corrections for detector effects.
The binning scheme coincides with the one chosen in former publications~\cite{Chatrchyan:2012bja,Chatrchyan:2014gia,Khachatryan:2016mlc,Khachatryan:2016wdh}.

The details of the data and simulation are described in the following: the selection of the events and jets (Section~\ref{sec:selection}), the triggers (Section~\ref{sec:triggers}), the calibrations (Section~\ref{sec:calibration}), corrections for efficiencies, misidentification, and migrations due to the limited resolution of the detector (Section~\ref{sec:unfolding}), and the experimental uncertainties (Section~\ref{sec:uncertainties}).

\subsection{Event selection} \label{sec:selection}

The data samples recorded in 2016 correspond to an integrated luminosity of 36.3\fbinv (33.5\fbinv) for events with jets clustered with $R=0.4$ (0.7)~\cite{CMS-LUM-17-003}.
Triggers for the larger jet distance parameter were activated after 2.8\fbinv of data had been taken, which explains the difference in the integrated luminosities.

The detector response in simulations is modelled with~\GEANTfour~\cite{GEANT4}.
The simulations in Table~\ref{tab:MC} include a simulation of the pileup produced with \PYTHIA~8+CUETP8M1, which correspond to the pileup conditions in the data. 

The primary vertex (PV) must satisfy $\abs{z_\mathrm{PV}} < 24\cm$ and $\rho_\mathrm{PV} < 2\cm$, where $z_\mathrm{PV}$ ($\rho_\mathrm{PV}$) corresponds to the longitudinal (radial) distance from the nominal interaction point.
Event filters mentioned in Section~\ref{sec:detector} are applied to reduce the noise from the detector.

In addition, the jets must satisfy quality criteria based on the jet constituents to remove the effect of detector noise, and must be reconstructed within $\abs{y} < 2.5$, corresponding to the tracker acceptance.
Jets reconstructed in regions of the detector corresponding to defective zones in the calorimeters are excluded from the measurement and recovered later in the unfolding procedure; the effect is of the order of a percent and is uniform as a function of \pt.

\begin{table}[ht]
    \centering
    
    \topcaption[Simulation samples]{Description of the simulations used in the analysis.}
    \cmsTable{
    \begin{tabular}{llll} 
        generator & PDF & matrix element & tune \\
        \hline
        \PYTHIA~8~(230)~\cite{Sjostrand:2014zea}  & NNPDF~2.3~\cite{Ball:2012cx} & LO $2\to 2$ & CUETP8M1~\cite{Khachatryan:2015pea} \\
        \MGvATNLO~(2.4.3)~\cite{Alwall:2011uj,Alwall:2014hca}  & NNPDF~2.3~\cite{Ball:2012cx} & LO $2\to 2,3,4$ & CUETP8M1~\cite{Khachatryan:2015pea} \\
        \HERWIGpp~(2.7.1)~\cite{Bahr:2008pv}  & CTEQ6L1~\cite{Pumplin:2002vw}& LO $2\to 2$ & CUETHppS1~\cite{Khachatryan:2015pea} \\
    \end{tabular}
    }
    \label{tab:MC}
\end{table}

\subsection{Triggers} \label{sec:triggers}

The prescaled single-jet triggers are used, requiring at least one jet in the event with jet \ptHLT larger than a certain threshold.
All triggers are prescaled, except the one with the highest threshold, and correspond to different effective integrated luminosities.
Two different series of triggers are used for each value of the $R$ parameter, shown in Tables~\ref{table:trigger_emulation_turnons_AK4}-\ref{table:trigger_emulation_turnons_AK7}.
The jets are weighted event by event in contrast to previous measurements~\cite{Chatrchyan:2012bja,Chatrchyan:2014gia,Khachatryan:2016mlc,Khachatryan:2016wdh} of inclusive jet production by the CMS Collaboration, where the whole contribution of each trigger is normalised with respect to its effective luminosity. This makes a difference, especially in terms of smoothness of the spectrum even after unfolding, since the trigger rate is nonlinear as a function of the instantaneous luminosity and the JES corrections generally vary with time and with pileup conditions.

An event is considered for the measurement only if the leading jet reconstructed with the offline PF algorithm is matched to an HLT jet.
The data contain events selected with a combination of triggers in different, exclusive intervals of the leading jet \pt.
The edges of each interval are determined in such a way that the corresponding trigger has an efficiency above 99.5\% in all \pt and $\abs{y}$ bins.
The efficiency of each trigger is determined from the data set recorded by the next single-jet trigger (with lower but closest threshold), except for the most inclusive trigger. To determine the region of efficiency of the most inclusive trigger and to cross-check the efficiencies obtained for the other triggers, a tag-and-probe method is applied to dijet topologies, which counts the leading and subleading offline PF jets that can be matched to HLT jets.
The measured spectrum is corrected for the residual trigger inefficiency as a function of \pt and $\abs{y}$.
Finally, to control possible steps in the distribution caused by passing the trigger thresholds, the spectrum is fitted in $\abs{y}$ regions using a truncated Taylor expansion with Chebyshev polynomials of the first kind as basis~\cite{connor2021step}; the $\chi^2/\Ndof$ (where dof is degree of freedom) is compatible or close to one within statistical uncertainties. \label{sec:Chebyshev}

The additional trigger inefficiency due to the prefiring effect, mentioned in Section~\ref{sec:detector}, is corrected event by event in the data before the unfolding procedure, using maps of prefiring probability in $2.0 < \abs{\eta} < 3.0$ as a function of \pt and $\eta$.
The total event weight is obtained as the product of the nonprefiring probability of all jets.
The resulting effect is typically 2\% for the spectrum at $\abs{y}<2.0$.

\begin{table}[ht]

    \centering

    \topcaption{The HLT ranges and effective integrated luminosities used in the jet cross section measurement for $R=0.4$.
        The first (second) row shows the \pt threshold for the HLT (offline PF) reconstruction; the third row corresponds to the effective luminosity of each trigger $\mathcal{L}$.
    }

    \cmsTable{
    \begin{tabular}{l c c c c c c c c c}
        \ptHLT (\GeVns{}) &  40 &  60  &  80 & 140 & 200 & 260 & 320 & 400 & 450 \\
        \hline
        \ptPF (\GeVns{}) & 74--97 & 97--133 & 133--196 & 196--272 & 272--362 & 362--430 & 430--548 & 548--592 & $>$592 \\
        $\mathcal{L}$ (\pbinv) & 0.267 & 0.726 & 2.76 & 24.2 & 103 & 594 & 1770 & 5190 & 36300 \\
    \end{tabular}
    }

    \label{table:trigger_emulation_turnons_AK4}
\end{table}

\begin{table}[ht]

    \centering

    \topcaption{The HLT ranges and effective integrated luminosities used in the jet cross section measurement for $R=0.7$.
        The first (second) row shows the \pt threshold for the HLT (offline PF) reconstruction; the third row corresponds to the effective luminosity of each trigger $\mathcal{L}$.
    }

    \cmsTable{
    \begin{tabular}{l c c c c c c c c c}
        \ptHLT (\GeVns{}) &  40 &  60  &  80 & 140 & 200 & 260 & 320 & 400 & 450 \\
        \hline
        \ptPF (\GeVns{}) & 74--97 & 97--114 & 114--196 & 196--272 & 272--330 & 330--395 & 395--507 & 507--592 & $>$592 \\
        $\mathcal{L}$ (\pbinv) & 0.0497 & 0.328 & 1.00 & 10.1 & 85.8 & 518 & 1526 & 4590 & 33500 \\
    \end{tabular}
    }
    \label{table:trigger_emulation_turnons_AK7}
\end{table}

\subsection{Calibration} \label{sec:calibration}

The JES corrections are applied according to the standard CMS procedure~\cite{Khachatryan:2016kdb}.
An additional smoothing procedure is applied to the JES corrections (originally parameterised with linear splines in bins of \pt) to ensure and preserve the smoothness of the cross sections using the same fit method described in Section~\ref{sec:Chebyshev}.

The detector-level \pt (\ptrec) is rescaled such that the jet energy resolution (JER) in the simulated samples matches the JER in the data; this procedure is also known as JER smearing.
A matching between the particle- and detector-level jets is performed for each event.
The particle-level jets are ordered by decreasing \pt.
Each particle-level jet is matched to the highest-\pt detector-level jet present in a cone with $\Delta R = 0.2$ (0.35) for jets clustered with $R=0.4$ (0.7); a particle-level jet may be matched to only one detector-level jet.
The response is fitted from matched jets using a double Crystal-Ball function~\cite{oreglia1980,gaiser1983charmonium} to account for the presence of non-Gaussian tails. The width of the resolution is extracted and fitted as a function of \ptrec, in bins of $\eta$ and the variable $\rho$ defined in Ref.~\cite{Cacciari:2007fd}. A modified NSC~function is used, where the second term in Eq.~(8.10) of Ref.~\cite{Khachatryan:2016kdb} is extended by $\pt^d$ with $d$ being an additional fit parameter.
The \ptrec is then rescaled as a function of the response with a scale factor; if no matching can be performed in the Gaussian core of the double Crystal-Ball function, the response is estimated with a Gaussian of width obtained from the modified NSC~function.

The simulation of the pileup also includes the signal, but with a lower event count.
To avoid double-counting and ensure a good statistical precision over the whole phase space, events with a hard scatter from the pileup simulation and events with anomalously large weights are discarded from the simulated samples.
To ensure correct normalisation, the remaining events are reweighted to restore the originally generated spectrum. 
Finally, the pileup profile used in the simulation is corrected to match that in data.

\subsection{Correction for detector effects} \label{sec:unfolding}

The measured detector-level distribution is unfolded to the particle level using corrections derived from the simulated events.

Sequentially, corrections for the background, migrations and inefficiencies are applied.
To estimate the different effects, the same matching algorithm as described in the context of the JER smearing is used.
Successfully matched jets, both in the Gaussian core and in the tails of the response, are used to estimate the response matrix, whereas unmatched jets at the particle (detector) level are used to estimate the inefficiencies (background). The background contribution is at 1--2\%-level at low \pt and is negligible at medium at high \pt, while the maximum inefficiency reaches 2--5\% at low and high \pt.

Various types of backgrounds and inefficiencies are considered: the migrations in/out of the phase space, and the unmatched jets that correspond to either pileup or objects wrongly identified as jets by the reconstruction algorithm.

\begin{figure}

    \centering

    \includegraphics[width=\textwidth]{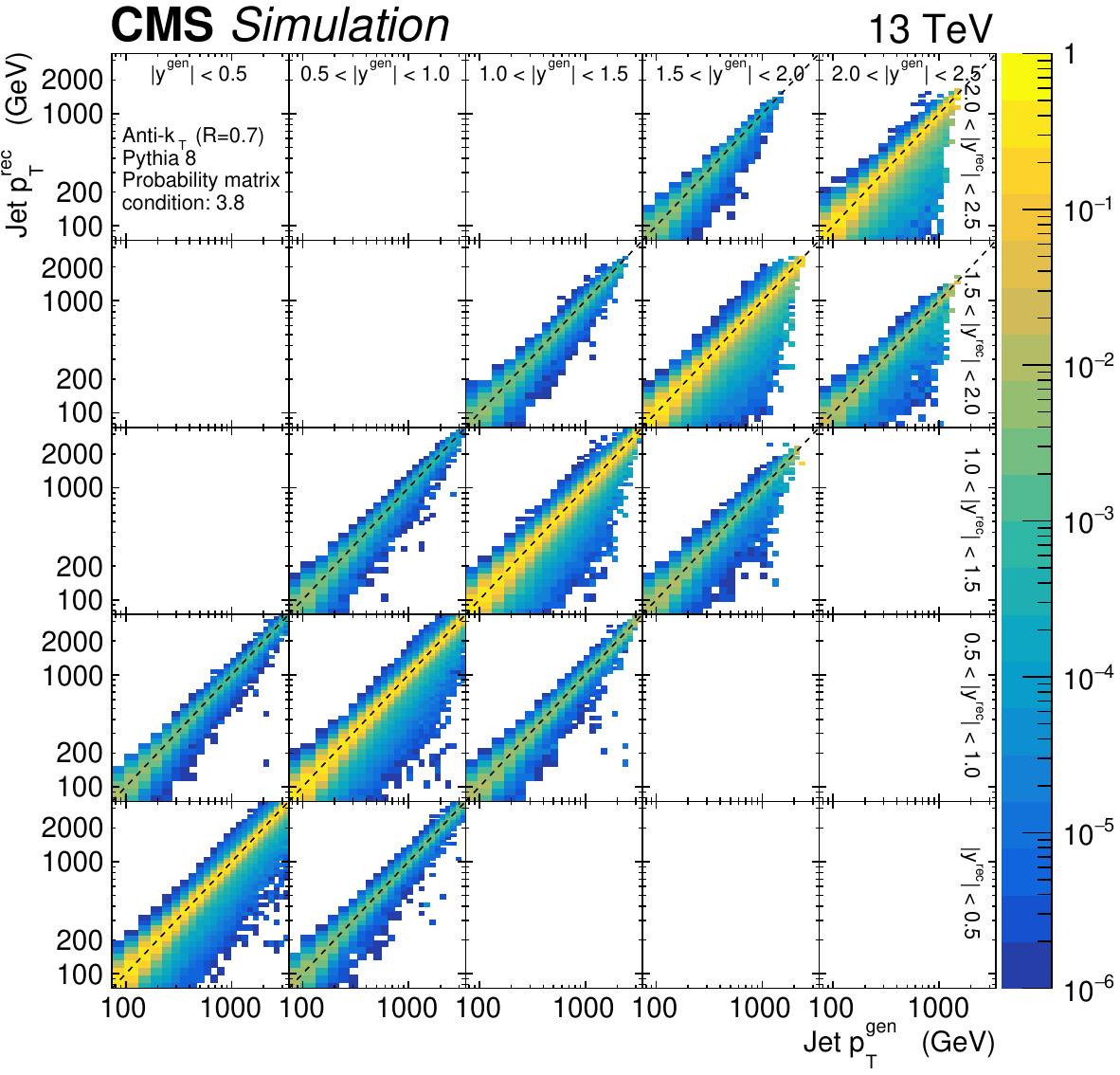}

    \caption[Probability matrix]{The probability matrix, estimated with a simulated sample based on \PYTHIA~8, for jets clustered using the anti-\kt algorithm with $R = 0.7$.
        The horizontal (vertical) axis corresponds to jets at the particle~(detector) level.
        The global $5\times5$ structure corresponds to the bins of rapidity $y$ of the jets, indicated by the labels in the uppermost row and rightmost column; the horizontal and vertical~axes of each cell correspond to the transverse momentum \pt of the jets.
        The colour range covers a range from $10^{-6}$ to 1 and the rows are normalised to unity, indicating the probability for a particle-level jet generated with values of \ptgen and \absygen to be reconstructed at the detector level with values of \ptrec and \absyrec.
        Migrations outside of the phase space are not included; migrations across rapidity bins only occur among adjacent rapidity bins.
        The dashed lines indicate the diagonal bins in each rapidity cell.
    }
    
    \label{fig:PM}

\end{figure}

The probability matrix (PM), shown in Fig.~\ref{fig:PM} for jets clustered with $R=0.7$, is obtained by normalising the response matrix row by row.
It contains the probability for a given particle-level jet to be reconstructed as a given detector-level jet.
Assuming a PM $\mathbf{A}$ and the sum of the backgrounds $\mathbf{b}$, both obtained from the simulation, and given a measured detector-level distribution $\mathbf{y}$, the particle-level distribution $\mathbf{x}$ is determined by minimising the following objective function:
\begin{equation} \label{chi2Unf}
    \chi^2 = \min_{\mathbf{x}} \left[(\mathbf{A} \mathbf{x} - \mathbf{y} - \mathbf{b})^\intercal\, \mathbf{V}^{-1}\, (\mathbf{A} \mathbf{x} - \mathbf{y} - \mathbf{b}) \right],
\end{equation}
where $\mathbf{V}$ is the covariance matrix of the detector-level data describing the statistical uncertainties associated with the data (including correlations), as well as with the backgrounds.
The detector-level distribution has twice the number of bins as in the particle-level distribution.
The matrix condition number of the PM, \ie the ratio of the highest and lowest eigenvalues, is close to 4, assuring that the PM is not ill-conditioned; therefore no additional regularisation is applied.
The whole procedure is performed with the \TUnfold package~\cite{Schmitt_2012}, version 17.9.

Finally, the residual inefficiencies are obtained from the rate of particle-level jets that cannot be matched to any detector-level jets passing the event selection, and corrected bin-by-bin on the particle-level distribution $\mathbf{x}$.

\subsection{Uncertainties} \label{sec:uncertainties}

\begin{figure}

    \includegraphics[width=\textwidth]{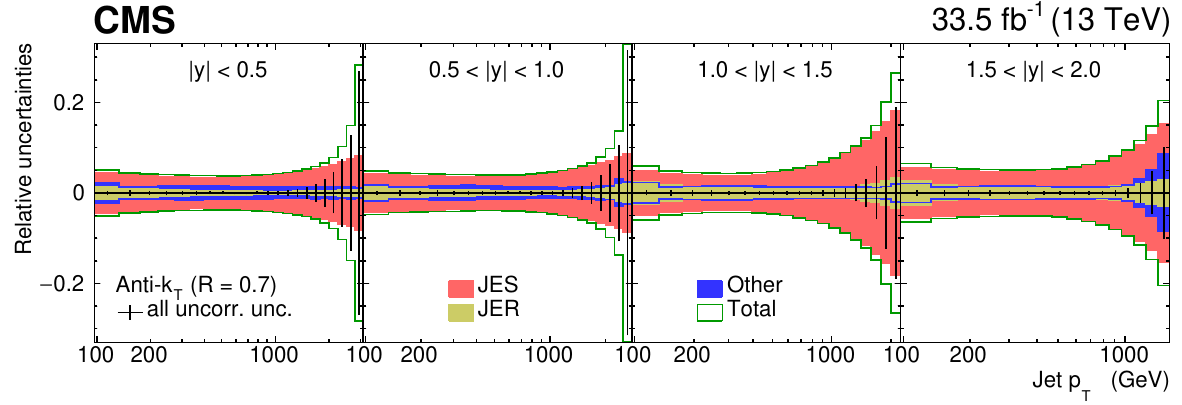}

    \caption[Experimental uncertainties]{Relative uncertainties in the double-differential cross section, as functions of jet transverse momentum ($x$~axis) and rapidity (cells), for jets clustered using the anti-\kt algorithm with $R = 0.7$.
        The systematic uncertainties are shown in different, noncumulative colour bands: the red bands correspond to JES uncertainties, the yellow bands to the JER uncertainties, and the blue bands to all other sources, including the integrated luminosity uncertainty, the model uncertainty, uncertainties in the migrations in and out of the phase space, and uncertainties in various inefficiencies and backgrounds.
        The vertical error bars include the statistical uncertainties from the data and from the \PYTHIA~8 simulated sample used for the unfolding, as well as the binwise systematic uncertainties, all summed in quadrature.
        The total uncertainty, shown in green, includes all systematic and statistical uncertainties summed in quadrature.
    }
    
    \label{fig:uncertainties}
\end{figure}

The measurement is affected by systematic and statistical uncertainties.
The contributions of the various uncertainties are shown in Fig.~\ref{fig:uncertainties}, where the coloured band indicates the bin-to-bin fully correlated uncertainties and the vertical error bars indicate the uncorrelated uncertainties.

The bin-to-bin fully correlated uncertainties are determined as follows:
\begin{itemize}
    \item   Variations of the JES corrections and of the prefiring correction are applied to the data at the detector level and are mapped to the particle level by repeating the unfolding procedure.
        The JES uncertainties are the dominant uncertainties in this measurement.
    \item   Systematic effects related to the JER and to the pileup profile correction are varied in the simulated sample and propagated to the particle level by repeating the unfolding procedure.
    \item   The normalisation of the estimates of the inefficiencies and backgrounds, obtained from the Monte Carlo (MC) simulation, are varied separately within a conservative estimate of 5\%, covering a potential model dependence in migrations in the phase space and an impact from the matching algorithm in the unfolding procedure.
    \item   The model dependence in the unfolding for \pt and $\abs{y}$ is assessed by a model uncertainty derived as the difference between the nominal cross section obtained with the original \PYTHIA~8 simulation and a modified version in which the inclusive jet spectrum is corrected to match the data. 
        The modified version is obtained by applying a smooth correction as a function of the \pt, $y$, and jet multiplicity.
        The effect is strongest for $\abs{y} > 1.5$, where \PYTHIA~8 does not describe the data well. Alternatively to \PYTHIA~8, \MGvATNLO simulation is used and agrees well with the \PYTHIA~8 results.
    \item   A fully correlated 1.2\% uncertainty in the integrated luminosity calibration is applied to the nominal variation of the unfolded spectrum~\cite{CMS-LUM-17-003}.
\end{itemize}
Bin-to-bin fluctuations in the systematic variations are removed by applying a smoothing procedure based on Chebyshev polynomials following the method described in Section~\ref{sec:Chebyshev}. 

\begin{figure}

    \centering

    \includegraphics[width=\textwidth]{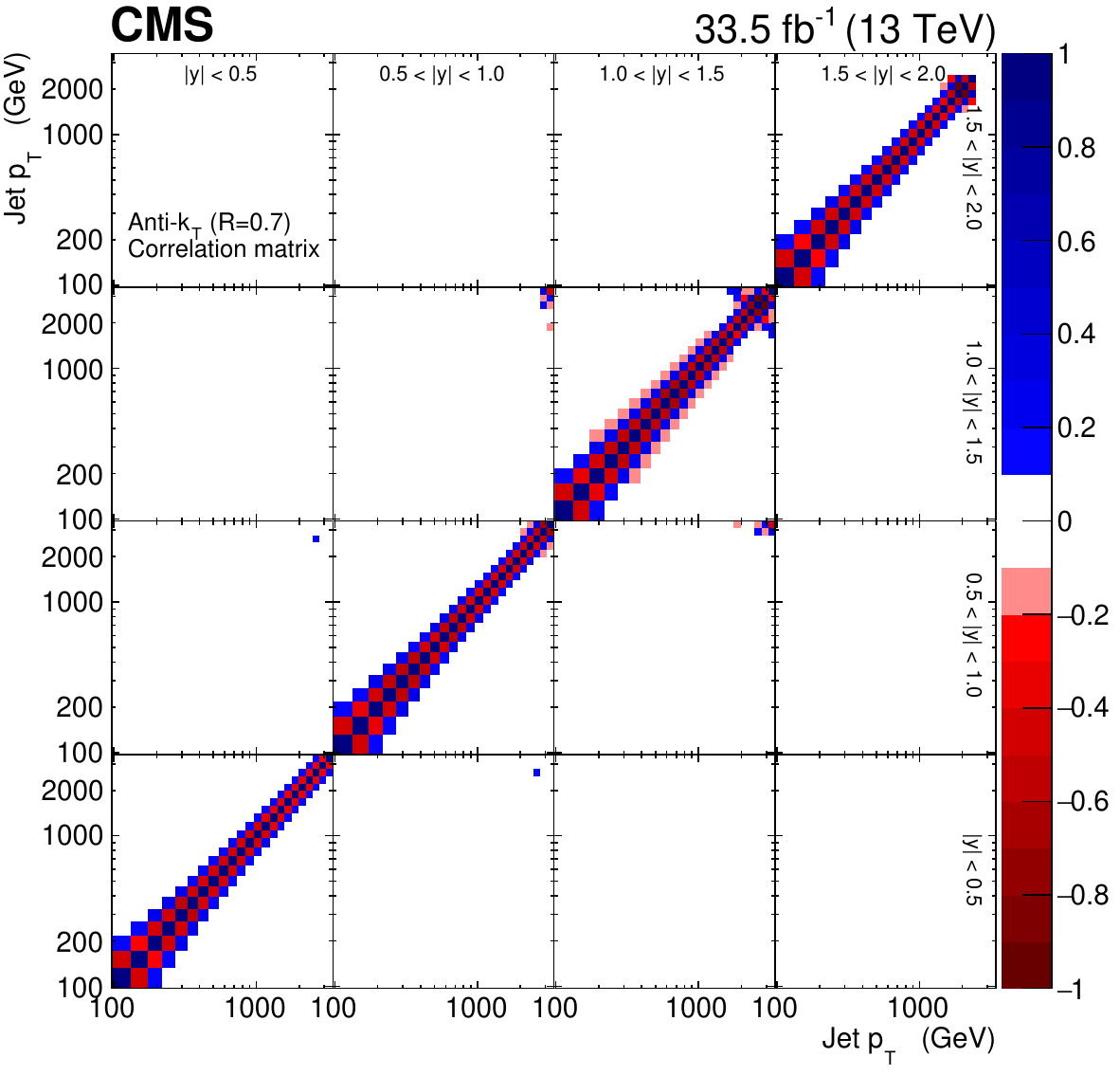}

    \caption[Correlation matrix]{The correlation matrix at the particle level, for jets clustered using the anti-\kt algorithm with $R = 0.7$.
        It contains contributions from the data and from the \PYTHIA~8 sample used to perform the unfolding.
        The global $4\times4$ structure corresponds to the bins of rapidity $y$ of the jets, indicated by the labels in the uppermost row and rightmost column; the horizontal and vertical axes of each cell correspond to the transverse momentum \pt of the jets.
        The colour range covers a range from $-1$ to 1 and indicates correlations in blue shades and anti-correlations in red shades, except for values between $-0.1$ and 0.1.
        Correlations across rapidity bins reach significant values mostly at the edges of the \pt range.
    }

    \label{fig:cov}
    
\end{figure}

Uncorrelated and partly correlated uncertainties among \pt and $y$ bins arise from various origins:
\begin{itemize}
    \item   The inclusive jet measurement is based on multiple jets recorded in each event.
        This is the dominant contribution to the uncorrelated uncertainties shown in Fig.~\ref{fig:uncertainties}.
    \item   An additional uncorrelated systematic uncertainty of 0.2\% is added before the unfolding to account for differences in alternative methods of determining the trigger efficiency.
    \item   Statistical fluctuations in the simulated distributions.
\end{itemize}
After unfolding, the different sources are no longer distinguished from one another; in addition, the unfolding procedure introduces anti-correlations among directly neighbouring bins.
The correlation matrix after the unfolding is shown in Fig.~\ref{fig:cov}.

\section{Theoretical predictions} \label{sec:theory}

In the following, the fixed-order pQCD predictions, the electroweak (EW) corrections, and the NP corrections are described.

\subsection{Fixed-order predictions} \label{sec:theory_fixOrder}

Fixed-order pQCD predictions for the inclusive jet production are available at NLO and NNLO accuracy, obtained using the NLOJet++~\cite{Nagy:2001fj,Nagy:2003tz} and NNLOJET~(rev5918)~\cite{Currie:2016bfm,Currie:2018xkj,Gehrmann:2018szu} programs, with NLO calculations implemented in \fastNLO~\cite{Britzger:2012bs}. The calculations are performed for five active massless quark flavours.
The renormalisation (\mur) and factorisation (\muf) scales are set to the individual jet \pt. Alternative prediction using \mur and \muf set to the scalar sum of the parton \pt (\HT parton) is used for comparison. In Ref.~\cite{Currie:2018xkj} the individual jet \pt was found to be a better choice for the scale than the transverse momentum of the leading-jet \ptmax. Furthermore in Refs.~\cite{Cacciari:2019qjx,Dasgupta:2016bnd,Bellm:2019yyh} it was discussed that NNLO calculations with jet distance parameter of the anti-\kt clustering algorithm $R=0.7$ are more stable than those with $R=0.4$.

To estimate possible uncertainty due to missing higher-order contributions, the scales are varied independently by a factor of 2 up and down, avoiding cases with $\muf/\mur = 4^{\pm1}$. The largest deviation of the cross section from the result obtained with the central scale choice is used as an estimate of the scale uncertainty. In general, the scale uncertainties for $R=0.7$ are larger than for $R=0.4$.

In the QCD predictions at NLO and NNLO, the proton structure is described by several alternative PDF sets:
CT14~\cite{Dulat:2015mca},
NNPDF~3.1~\cite{Ball:2017nwa},
MMHT2014~\cite{Harland-Lang:2014zoa},
ABMP~16~\cite{Alekhin:2017kpj}, and
HERAPDF~2.0~\cite{Abramowicz:2015mha}, obtained at NLO or NNLO, respectively, and each using their default value of \alpSZ. 

The NLO QCD prediction is improved to NLO+NLL accuracy with a simultaneous jet radius and threshold resummation $k$-factor for each bin $i$:
\begin{equation} \label{resummation_K-factor}
k^{\mathrm{NLO+NLL}}_i = \frac{ \sigNLO_i -\sigNLO_{\text{sing.},i} + \sigNLL_i}{\sigNLO_i},
\end{equation}
where the singular NLO terms $\sigNLO_{\text{sing.}}$ and the resummed contributions \sigNLL are obtained using the \textsc{NLL-Jet} calculation, provided by the authors of Ref.~\cite{Liu:2018ktv}.
Following their approach, the \sigNLO for the resummation factor is computed using the modified Ellis Kunszt Soper (MEKS) code version 1.0~\cite{Gao:2012he}, and choosing the renormalisation and factorisation scales $\mur = \muf = \ptmax$.

\subsection{Electroweak corrections} \label{sec:ewCorrs} \label{sec:EW}

\begin{figure}
    \centering
    \includegraphics[width=0.49\textwidth]{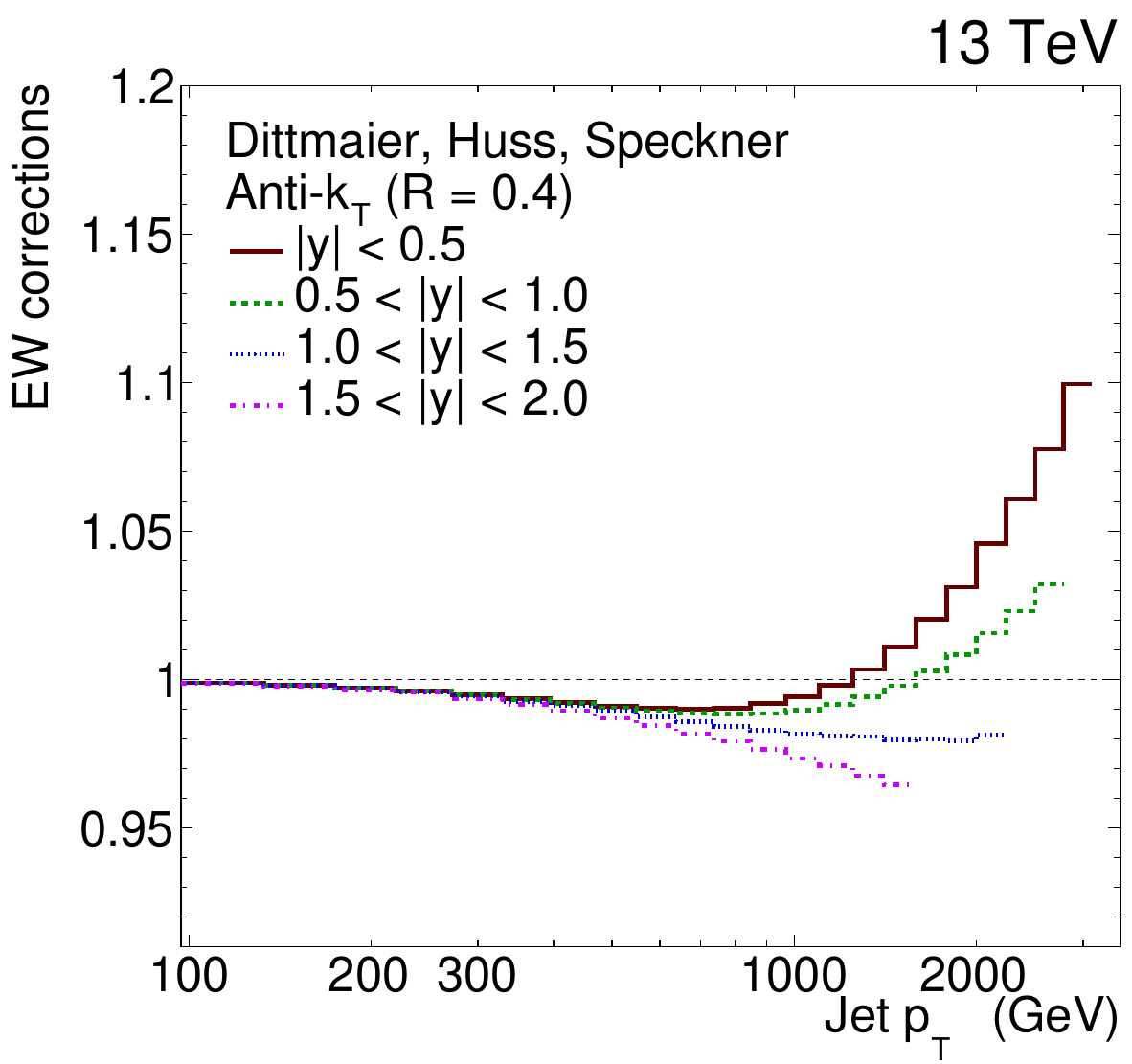}
    \includegraphics[width=0.49\textwidth]{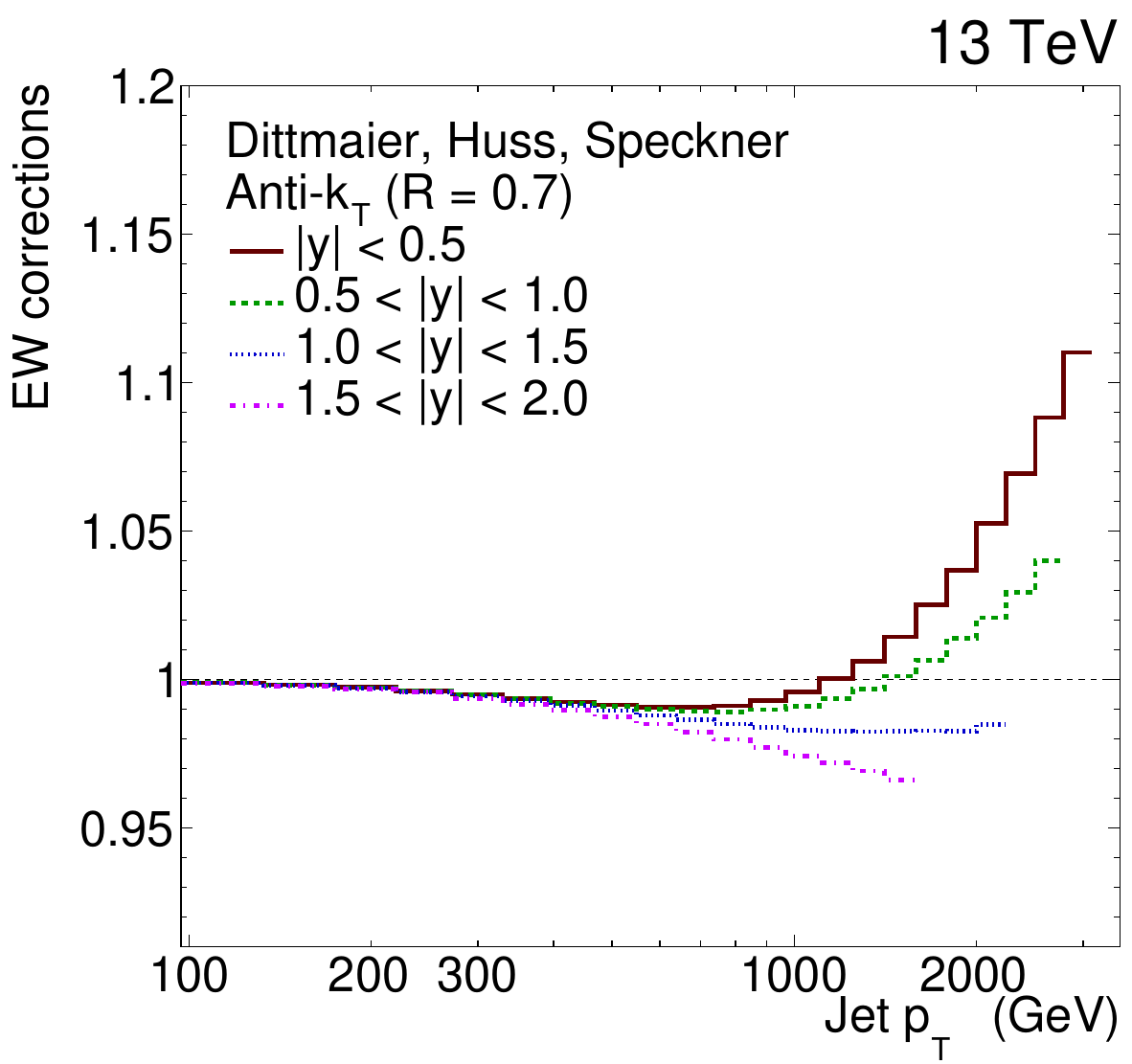}

    \caption[EW corrections]{The EW corrections for inclusive jet cross sections, as reported in Ref.~\cite{Dittmaier:2012kx}. 
        The values for jets clustered using the anti-\kt algorithm with $R=0.4$ (0.7) are shown on the left (right); each curve corresponds to a rapidity bin.
    }
    \label{fig:ewCorrs}
\end{figure}

The EW effects, which arise from the virtual exchange of the massive \PW and \PZ gauge bosons are calculated to NLO accuracy~\cite{Dittmaier:2012kx} and are applied to the fixed-order QCD predictions. In the high-\pt region, these EW effects grow to 11\%, as illustrated in Fig.~\ref{fig:ewCorrs}. No uncertainty associated with these corrections is available yet. 

The contribution of real production of EW boson production in association with jets is estimated at NLO using \MCFM~\cite{Campbell:1999ah,Campbell:2015qma,Campbell:2011bn} program to be at most at percent level which is negligible for the present analysis.

\subsection{Nonperturbative corrections} \label{sec:NP}

The NP corrections are defined for each bin $i$ as
\begin{equation} \label{eq:NPs}
    \mathrm{NP}_i = \frac{ \sigma_i^\mathrm{MC}(\text{PS \& MPI \& HAD}) }{ \sigma_i^\mathrm{MC} (\mathrm{PS}) },
\end{equation}
where PS stands for parton shower, HAD for hadronisation, and MPI for multiparton interaction.
The NP factors correct for the hadronisation and the MPI effects that are not included in the fixed-order pQCD predictions.
At low \pt, the NP corrections are dominated by MPI, which increases the radiation in the jet cone by a constant offset.
This is especially important for $R=0.7$.
On the other hand, hadronisation plays a role at smaller $R$.
The effects of perturbative radiation are partially considered in the higher-order predictions; for this reason the PS simulation is included in both the numerator and the denominator. It is stronger for smaller $R$, where out-of-cone radiation plays a larger role, which NLL corrections can account for.

To define final NP~corrections, \PYTHIA~8~CP1 tune~\cite{Sirunyan:2019dfx} and \HERWIGpp~EE5C tune~\cite{Khachatryan:2015pea} are fitted with a smooth function $a_0 + a_1/\pt^{a_2}$. The correction is obtained from the resulting envelope with the central value taken in the middle of the envelope and the uncertainties from its edges.

\begin{figure}
    \centering
    \includegraphics[width=0.49\textwidth]{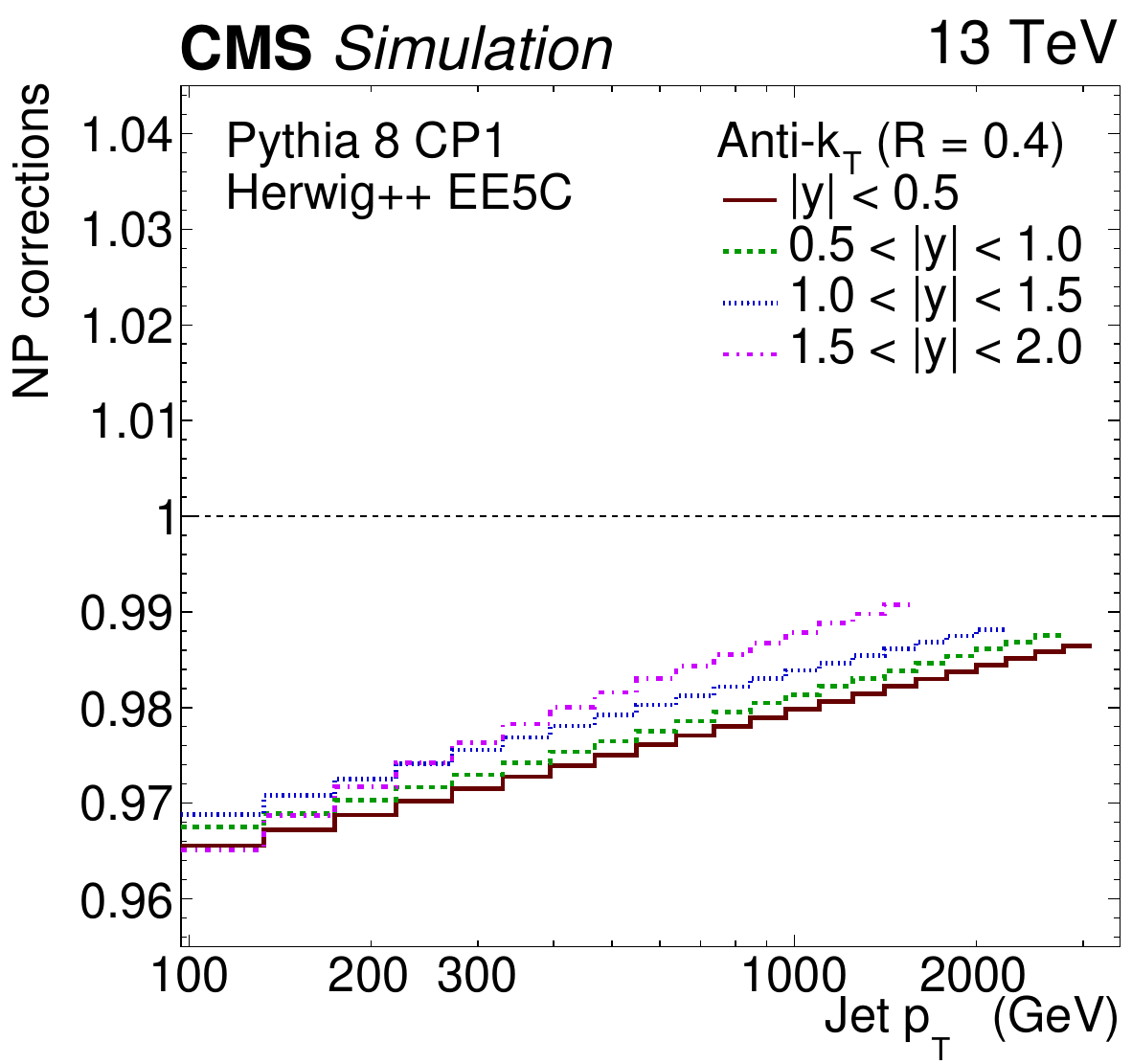}
    \includegraphics[width=0.49\textwidth]{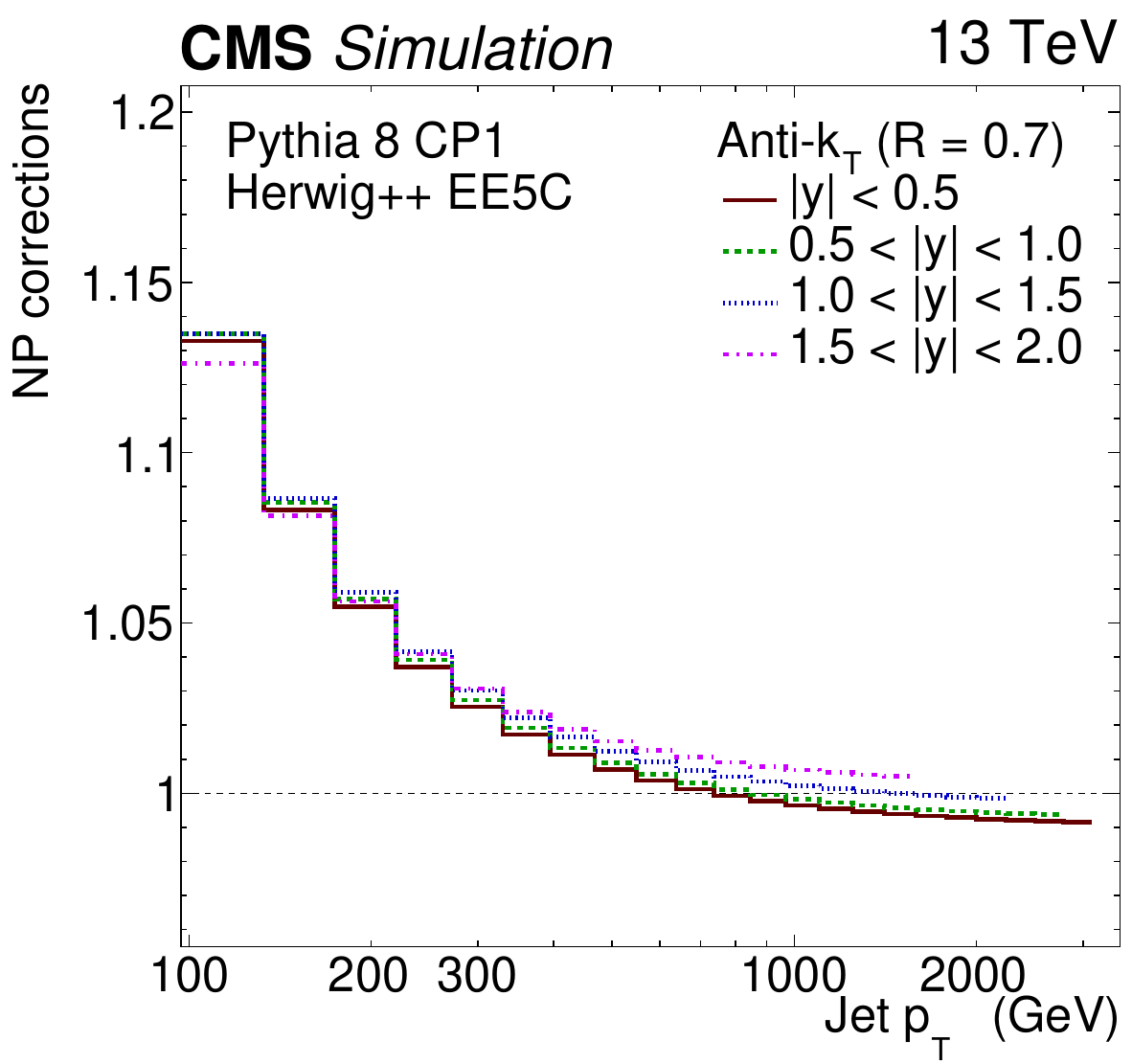}

    \caption[NP corrections]{The values for NP corrections for inclusive jet cross sections. 
        The values for jets with $R=0.4$ (0.7) are shown on the left~(right); each curve corresponds to a rapidity bin.
        The values correspond to the average of the corrections obtained with \PYTHIA~8 and with \HERWIGpp.
    }
    \label{fig:npCorrs}
\end{figure}

The NP corrections defined in Eq.~\eqref{eq:NPs} are shown in Fig.~\ref{fig:npCorrs}.
The corrections are larger for $R=0.7$ than for $R=0.4$, since a larger cone size includes more effects from the underlying event.

\section{Results} \label{sec:theoryVSdata} \label{sec:comparison}

\begin{figure}
    \centering
    \includegraphics[width=0.65\textwidth]{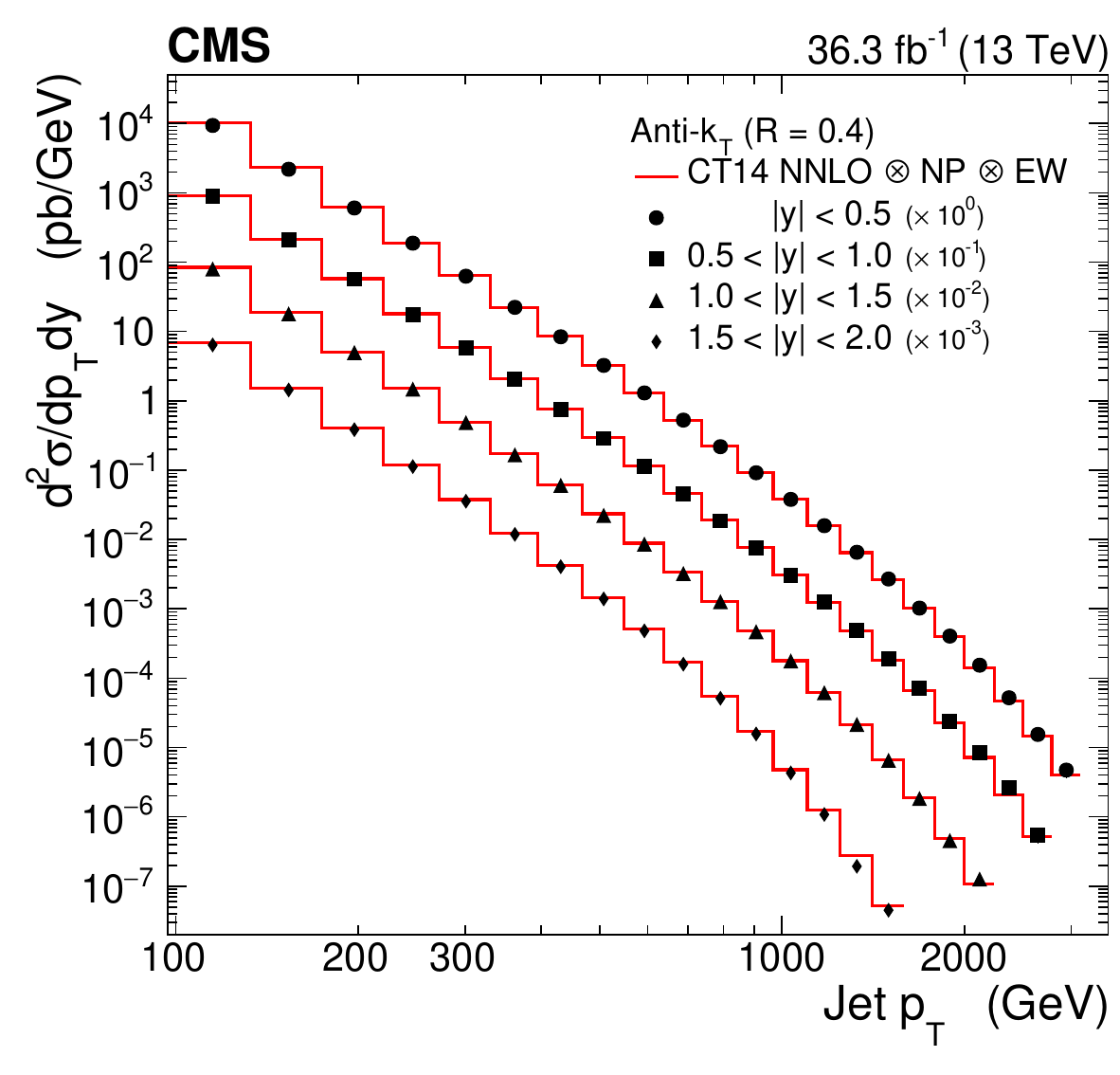}

    \includegraphics[width=0.65\textwidth]{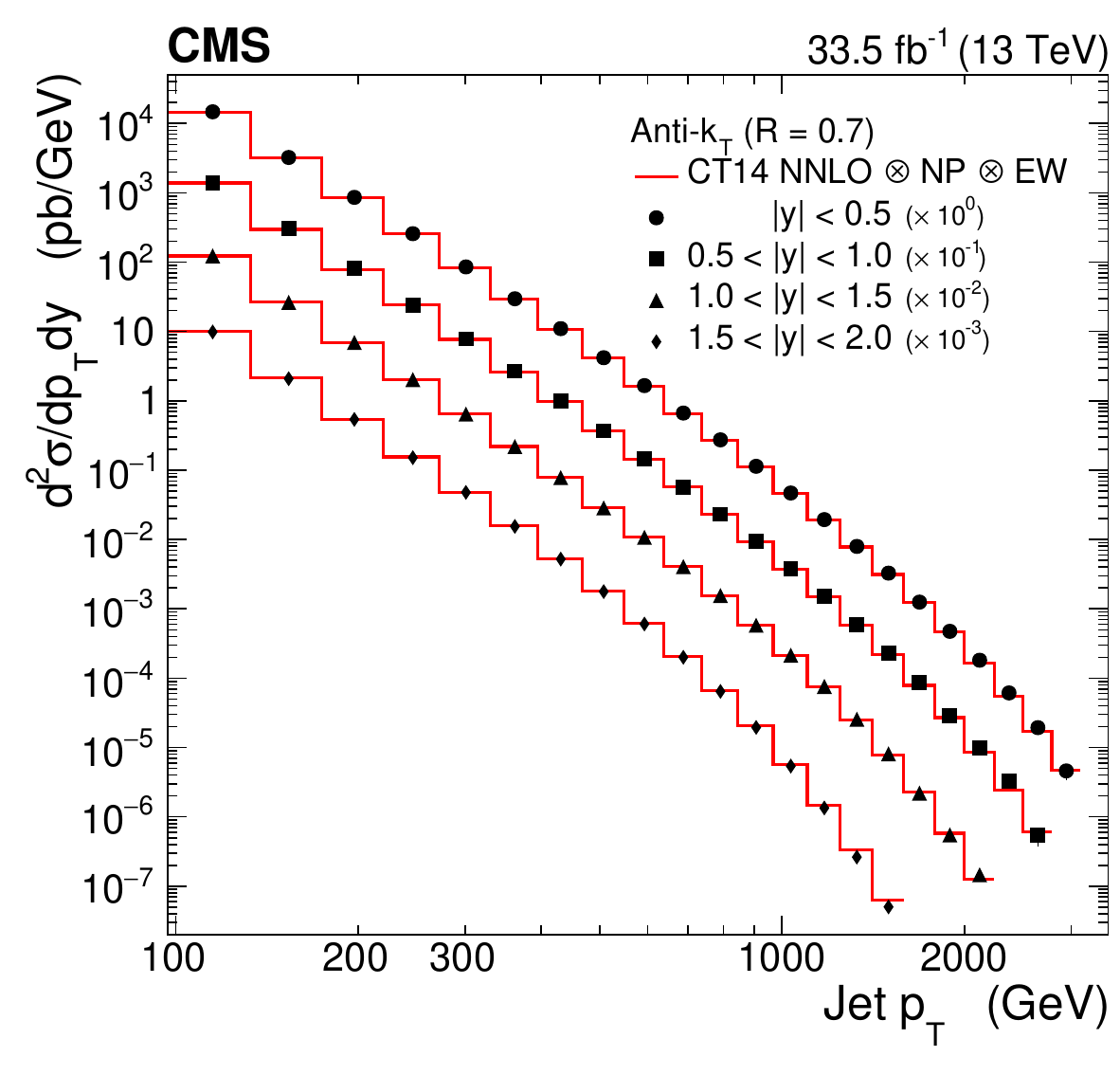}

    \caption[Absolute double-differential cross sections]{The inclusive jet production cross sections as a function of the jet transverse momentum \pt measured in intervals of the absolute rapidity $\abs{y}$. The cross section obtained for jets clustered using the anti-\kt algorithm with $R=0.4$ (0.7) is shown on the upper (lower) plot.
        The results in different $\abs{y}$ intervals are scaled by a constant factor for presentation purpose.
        The data in different $\abs{y}$ intervals are shown by markers of different style.
        The statistical uncertainties are too small to be visible; the systematic uncertainties are not shown.
        The measurements are compared with fixed-order NNLO QCD predictions (solid line) using CT14nnlo PDF and corrected for EW and NP effects.  
    }
    \label{fig:diffCrossSec}
\end{figure}

In Fig.~\ref{fig:diffCrossSec}, the inclusive jet cross sections are presented as functions of the jet \pt and $\abs{y}$ for $R=0.4$ and 0.7.
The cross sections are shown for four absolute rapidity intervals: $\abs{y} < 0.5$, $0.5 <\abs{y} < 1.0$, $1.0 <\abs{y} < 1.5$, and $1.5 <\abs{y} < 2.0$ with jet $\pt > 97\GeV$.
The data are compared with fixed-order NNLO QCD predictions using CT14 PDF, corrected for NP and EW effects. The data cover a wide range of the jet \pt from 97\GeV up to 3.1\TeV.

\begin{figure}
    \centering

    \includegraphics[width=\textwidth]{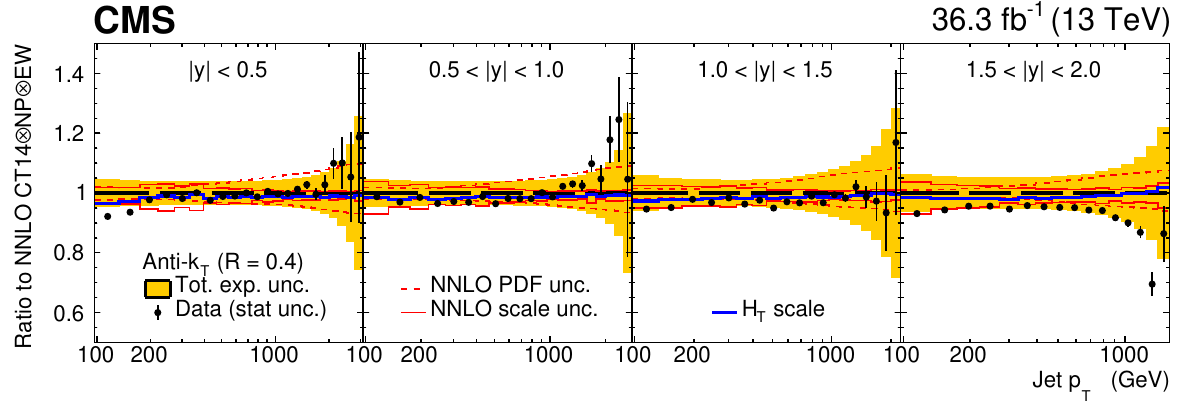}

    \includegraphics[width=\textwidth]{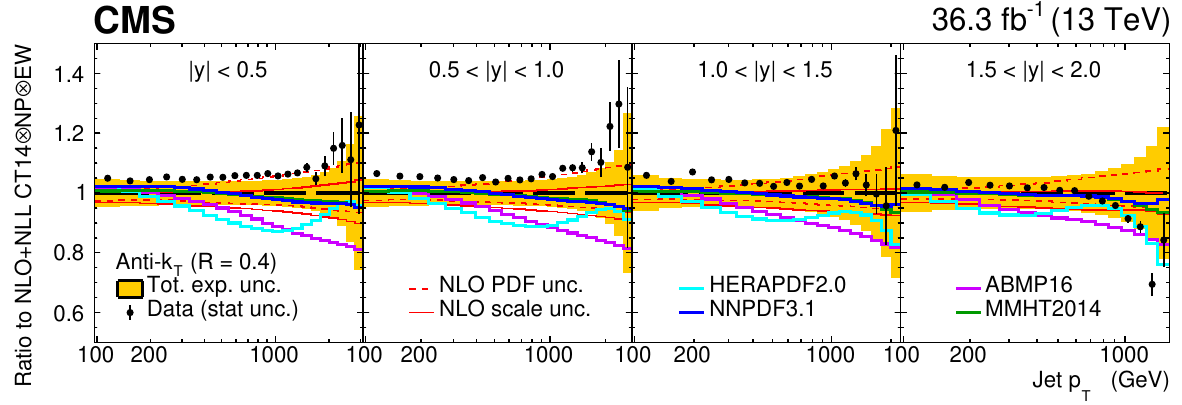}

    \caption[Comparison of result to theory ($R=0.4$)]{The double-differential cross section of inclusive jet production, as a function of \pt and $\abs{y}$, for jets clustered using the anti-\kt algorithm with $R=0.4$, presented as ratios to the QCD predictions. The data points are shown by filled circles, with statistic uncertainties shown by vertical error bars, while the total experimental uncertainty is centred at one and is presented by the orange band.
In the upper panel, the data are divided by the NNLO prediction, corrected for NP and EW effects, using CT14nnlo PDF and with renormalisation and factorisation scales jet \pt and, alternatively \HT (blue solid line). 
In the lower panel, the data are shown as ratio to NLO+NLL prediction, calculated with CT14nlo PDF, and corrected for NP and EW effects. The scale (PDF) uncertainties are shown by red solid (dashed) lines. NLO+NLL predictions obtained with alternative PDF sets are displayed in different colours as a ratio to the central prediction using CT14nlo.
    }

    \label{fig:16ak4}
\end{figure}

\begin{figure}
    \centering

    \includegraphics[width=\textwidth]{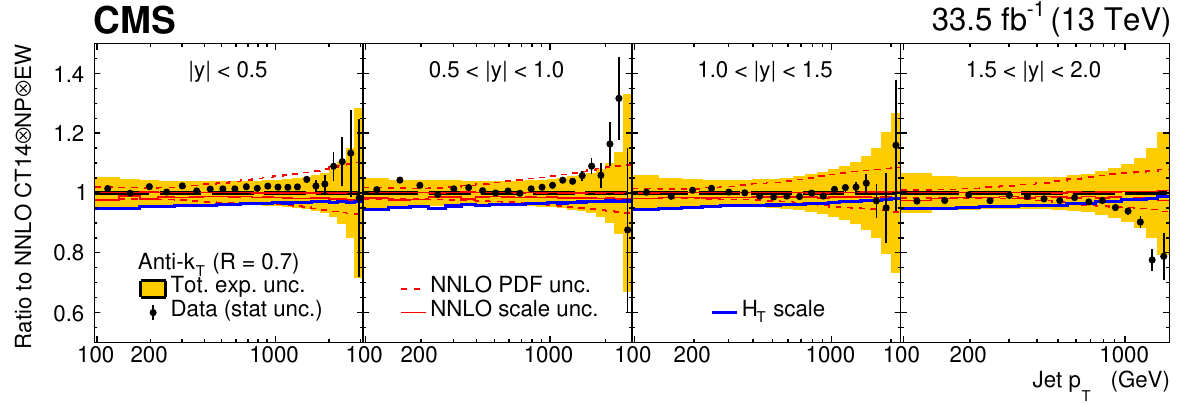}

    \includegraphics[width=\textwidth]{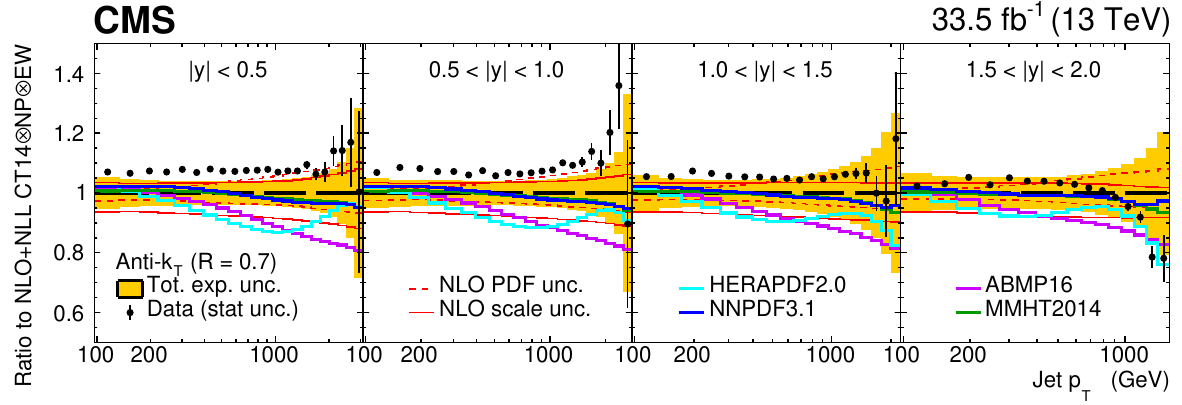}

    \caption[Comparison of result to theory ($R=0.7$)]{The double-differential cross section of inclusive jet production, as a function of \pt and 
    $\abs{y}$, for jets clustered using the anti-\kt algorithm with $R=0.7$, presented as ratios to the QCD predictions. The notations are identical to those of Fig.~\ref{fig:16ak4}.
    }

    \label{fig:16ak7}
\end{figure}

In Fig.~\ref{fig:16ak4} (Fig.~\ref{fig:16ak7}), the ratios of the measured cross sections to the NNLO QCD predictions using different scale choices and to the NLO+NLL predictions with various PDFs sets are shown for jets with $R=0.4$ (0.7).
In general at high \pt, smaller (higher) cross sections are predicted than experimentally measured for the central (forward) rapidities. The theoretical uncertainties are larger at high \pt and are dominated by the PDF uncertainties. The scale uncertainties are significantly smaller at NNLO as compared to NLO+NLL. The prediction using jet \pt as renormalisation and factorisation scale results in a harder \pt spectrum than in case of scale set to \HT parton.
The NLO+NLL calculations predict harder \pt spectrum than the NNLO calculations. All predictions describe the data well within the experimental and theory uncertainties.

The CT14, NNPDF~3.1, and MMHT 2014 PDF sets include CMS and ATLAS measurements of inclusive jet cross sections at 7\TeV, whereas ABMP~16 does not include LHC jet measurements and HERAPDF~2.0 is based only on the HERA DIS data. Predictions obtained with these PDFs are similar at low \pt, although significant differences at high \pt are observed. These differences arise from differences in the gluon distribution at high $x$ in the available PDFs and point to high sensitivity of the present measurement to the proton PDFs. 

Additional comparisons to theoretical predictions are available in~\suppMaterial.

\clearpage

\section{The QCD analysis}
\label{sec:QCD}

The sensitivity of the presented measurement of the inclusive jet production to the proton PDFs and to the value of the strong coupling constant at the mass of the \PZ boson \alpSZ is investigated in a comprehensive QCD analysis. The jet cross section for $R=0.7$ is used because of reduced out-of-cone radiation and a better description of the measurement by the pQCD predictions. The QCD analysis is performed either assuming only the standard model or applying effective corrections to the QCD calculation of the inclusive jet production to include 4-quark contact interactions.    

\subsection{Data sets used}
\label{QCD_analysis_data}

In this QCD analysis, the double-differential inclusive jet production cross section with $R=0.7$ is used together 
with the charged- and neutral-current DIS cross sections of HERA~\cite{Abramowicz:2015mha}. In addition, the normalised triple-differential \ttbar cross section~\cite{Sirunyan:2019zvx} from CMS is used. As demonstrated in Ref.~\cite{Sirunyan:2019zvx}, the top quark pole mass \mtpole, the gluon distribution, and the value of \alpSZ are closely correlated in the triple-differential \ttbar production cross section, thus providing additional sensitivity to the gluon distribution and to \alpSZ.
Combining this data, the proton PDFs and the values of \alpSZ and of \mtpole are extracted simultaneously. Although the inclusive jet production has no sensitivity to the value of~\mtpole, the strong constraints it has on the gluon distribution and on the value of \alpSZ are reflected in the improved value and uncertainty in \mtpole, since both \ttbar and inclusive jet production cross sections are used in the fit. 

For the QCD analysis, the open-source framework \xFitter~\cite{Alekhin:2014irh, Bertone:2017tig, xFitter:web} version 2.2.1, extended to SMEFT prediction, is used. The DGLAP~\cite{Gribov:1972ri, ALTARELLI1977298, CURCI198027, FURMANSKI1980437, Moch:2004pa, Vogt:2004mw} evolution is implemented using \QCDNUM~\cite{Botje:2010ay} version 17-01/14. The analysis is performed at NLO or NNLO, depending on the physics case, as described in the following. The correlations of the experimental statistics and systematic uncertainties in each individual data set are included. The HERA DIS measurements and the CMS data are treated as uncorrelated. In the CMS \ttbar and jet measurements, the common systematic sources associated with the JES uncertainties are taken as 100\% correlated.

\subsection{Theoretical calculations used in QCD analysis}
\label{QCD_analysis_theory}
 
The SM theoretical predictions for the inclusive jet production cross section at NLO and NNLO are obtained as described in Section~\ref{sec:theory_fixOrder} and are corrected for NP and EW effects. 
The NNLO calculation is approximated by $k$-factors, obtained as a ratio of fixed-order NNLO to NLO calculations using CT14nnlo PDF in each bin in \pt and $\abs{y}$. These are applied to the NLOJet++ prediction interfaced to \xFitter using fast-grid techniques of \fastNLO~\cite{Britzger:2012bs}. In a similar way, the NLO prediction is improved to NLO+NLL as explained in Section~\ref{sec:theory}. The QCD prediction for the normalised triple-differential cross section of the \ttbar production is available only at NLO and is described in detail in Ref.~\cite{Sirunyan:2019zvx}. 

The renormalisation and factorisation scales are set to the four-momentum transfer $Q$ for the DIS data and to the individual jet \pt for inclusive jet cross section measurements. Following Ref.~\cite{Sirunyan:2019zvx}, in the case of \ttbar production, the scales are set to $\mur = \muf = \frac{1}{2}\sum_i \mT{}_i$. The sum over $i$ covers the final-state partons \PQt, \PAQt, and at most three light partons in a $\ttbar + 2$ jets scenario. The transverse mass $\mT{}_i \equiv \sqrt{\smash[b]{{m_i^2 + \pt{}_i^2}}}$ is computed using the mass $m_i$ and transverse momentum $\pt{}_i$ of the partons~\cite{Sirunyan:2019zvx}. 

The QCD analysis at NLO is extended into a SMEFT study by adding dimension-6 operators for colour-charged fermions to the SM Lagrangian $\mathcal{L}_\mathrm{SM}$~\cite{Gao:2012qpa, Gao:2013kp}, so that
\begin{equation} \label{SMEFT_Lagrangian}
\mathcal{L}_\mathrm{SMEFT} = \mathcal{L}_\mathrm{SM} + \frac{2\pi}{\Lambda^2} \hspace*{-3mm} \sum_{~~n\in\{1,3,5\}} \hspace*{-4mm} c_n O_n.
\end{equation}
Here, the $c_n$ are Wilson coefficients and $\Lambda$ is the scale of new physics. For the 4-quark CI, the nonrenormalisable operators $O_n$ are
\begin{align}
O_1 &= \delta_{ij}\delta_{kl}
       \left(\sum_{c=1}^3 \overline{q}_{Lci} \gamma_\mu q_{Lcj}
             \sum_{d=1}^3 \overline{q}_{Ldk} \gamma^\mu q_{Ldl} \right),
\label{SMEFT_ops_start}\\
O_3 &= \delta_{ij}\delta_{kl}
       \left(\sum_{c=1}^3 \overline{q}_{Lci} \gamma_\mu q_{Lcj}
             \sum_{d=1}^3 \overline{q}_{Rdk} \gamma^\mu q_{Rdl} \right),\\
O_5 &= \delta_{ij}\delta_{kl}
       \left(\sum_{c=1}^3 \overline{q}_{Rci} \gamma_\mu q_{Rcj}
             \sum_{d=1}^3 \overline{q}_{Rdk} \gamma^\mu q_{Rdl} \right),
\label{SMEFT_ops_end}
\end{align}
where the sums in $c$ and $d$ run over generations, whereas $i,j,k,l$ are colour indices. The $L$ and $R$ subscripts denote the handedness of the quarks.
The operators in Eqs.~\eqref{SMEFT_ops_start}--\eqref{SMEFT_ops_end} correspond to having integrated out a colour-singlet BSM exchange between two quark lines. The colour-singlet exchanges are dominant in quark compositeness~\cite{Eichten:1983hw} or \PZpr models~\cite{Langacker:2008yv}. The operators commonly denoted $O_2$, $O_4$ and $O_6$ in literature involve a product of Gell-Mann matrices in place of $\delta_{ij}\delta_{kl}$ and cause colour-octet; they are not considered here. 

The CI studied in the SMEFT fits is either purely left-handed, vector-like, or axial vector-like. The only free Wilson coefficient is $c_1$, which multiplies the operator $O_1$ in Eq.~\eqref{SMEFT_Lagrangian}. The coefficients $c_3$ and $c_5$ are determined from $c_1$ in accordance with how the quark-line handedness may change in the interaction. In the left-handed singlet model there are CI only between two left-handed lines, and hence $c_3=c_5=0$. Vector-like and axial vector-like exchanges allow interactions also between right-handed quarks, giving $c_5=c_1$ in both cases. For interactions between quark lines of different handedness, the vector-like exchange implies $c_3=2c_1$, whereas the axial vector-like model has $c_3=-2c_1$. Further details of the theoretical model are given in Ref.~\cite{Gao:2012qpa}. 

In the QCD analysis, the SMEFT prediction for the double-differential cross section of the inclusive jet production reads
\begin{equation}
\sigma^\mathrm{SMEFT} = \sigNLO_\fastNLO \, k^\mathrm{NLO+NLL} \, \mathrm{EW} \, \mathrm{NP} +\mathrm{CI},
\end{equation}
where $k^\mathrm{NLO+NLL}$ is given in Eq.~\eqref{resummation_K-factor} and EW and NP are explained in Sections~\ref{sec:EW}-\ref{sec:NP}; the CI term is computed at NLO using the \CIJET software~\cite{Gao:2013kp} interfaced to \xFitter.

\subsection{The general QCD analysis strategy and PDF uncertainties}
\label{QCD_analysis_strategy}

The procedure for determining the PDFs follows the approach of HERAPDF~\cite{H1:2009pze,Abramowicz:2015mha}. The contributions of charm and beauty quarks are treated in the Thorne--Roberts~\cite{Thorne:1997ga, Thorne:2006qt, Thorne:2012az} variable-flavour number scheme at NLO. The values of heavy quark masses are set to $\mc = 1.47\GeV$ and $\mb = 4.5\GeV$. The \mtpole and \alpSZ are free parameters in the PDF fits. The DIS data are restricted to high $Q^2$ by setting $\Qmin = 7.5\GeV^2$. 

The parameterised PDFs are the gluon distribution $x\Pg(x)$, the valence quark distributions $x\PQu_v(x)$, and $x\PQd_v(x)$, as well as $x\Ubar(x)$ for the up- and $x\Dbar(x)$ for the down-type antiquark distributions. At the starting scale of QCD evolution $Q_0^2 = 1.9\GeV^2$, the general form of the parameterisation for a PDF $\mathrm{f}$ is
\begin{equation}
x \Pf(x) = A_{\Pf} x^{B_\Pf}(1-x)^{C_\Pf} (1 + D_\Pf x + E_\Pf x^2),
\label{PDFgeneralForm}
\end{equation}
with the normalisation parameters $A_{\PQu_v}$, $A_{\PQd_v}$, and $A_\Pg$ determined from QCD sum rules. 
The small-$x$ behaviour of the PDFs is driven by the $B$ parameters, whereas the $C$ parameters are responsible for the shape of the distribution as $x \to 1$. 

The relations $x\Ubar(x) = x\PAQu(x)$ and 
$x\Dbar(x) = x\PAQd(x) + x\PAQs(x)$ are assumed, with 
$x\PAQu(x)$, $x\PAQd(x)$, and $x\PAQs(x)$ 
being the distributions for the up, down, and strange antiquarks, respectively. 
The sea quark distribution is defined as 
$x\Sigma(x) = 2 \cdot x\PAQu(x) + x\PAQd(x) + x\PAQs(x)$. 
Further constraints 
$B_{\Ubar} = B_{\Dbar}$ 
and $A_{\Ubar} = A_{\Dbar}(1 - f_\PQs)$ are imposed, so that 
the $x\PAQu$ and $x\PAQd$ distributions have the same normalisation as $x \to 0$. Here
$f_\PQs  = \PAQs/(\PAQd + \PAQs)$ 
is the strangeness fraction, fixed to $f_\PQs = 0.4$ as in the HERAPDF2.0 analysis~\cite{Abramowicz:2015mha}.

The $D_\mathrm{f}$ and $E_\mathrm{f}$ parameters probe the sensitivity of the results to the specific selected functional form. 
In general, the parameterisation is obtained by first setting all $D$ and $E$ parameters to zero and then including them in the fit, one at a time. The improvement in the $\chi^2$ of a fit is monitored and the procedure is stopped when no further improvement is observed. Differences in the data sets or theoretical predictions lead to differences in the resulting parameterisation.

In the full QCD fit, the uncertainties in the individual PDFs and in the extracted non-PDF parameters are estimated similarly to the approach of HERAPDF~\cite{H1:2009pze,Abramowicz:2015mha}, which accounts for the fit, model, and parameterisation uncertainties as follows. 

\textbf{Fit uncertainties} originate from the uncertainties in the used measurements and are obtained by using the Hessian method~\cite{Pumplin:2001ct} implying the tolerance criterion $\Delta\chi^2=1$, which corresponds to the 68\% confidence level (\CL). Alternatively, the fit uncertainties are estimated by using the MC method~\cite{Giele:1998gw,Giele:2001mr}, where MC replicas are created by randomly fluctuating the cross section values in the data within their statistical and systematic uncertainties. For each fluctuation, the fit is performed and the central values for the fitted parameters and their uncertainties are estimated using the mean and the root mean square values over the replicas. 

\textbf{Parameterisation uncertainty} is estimated by extending the functional forms of all PDFs with additional parameters $D$ and $E$, which are added, independently, one at a time. The resulting uncertainty is constructed as an envelope, built from the maximal differences between the PDFs (or non-PDF parameters) resulting from all the parameterisation variations and the results of the central fit.

\textbf{Model uncertainties} arise from the variations in the values assumed for the heavy quark masses \mb and \mc with 
$4.25 \leq \mb \leq 4.75\GeV$,
$1.41 \leq \mc \leq 1.53\GeV$, 
and the value of \Qmin imposed on the HERA data, which is varied in the interval 
$5.0 \leq \Qmin \leq 10.0\GeV^2$. 
The strangeness fraction is varied within 
$0.32 \leq f_\PQs \leq 0.48$, 
and the starting scale within 
$1.7 \leq Q^2_0 \leq 2.1\GeV^2$. 
In addition, the theoretical uncertainty in the QCD predictions due to missing higher order corrections (scale uncertainty) is considered as a part of the model uncertainty. The renormalisation and factorisation scales are varied in the theoretical predictions by a factor of 2 up and down independently, avoiding cases with $\muf/\mur = 4^{\pm1}$ and the fit is repeated for every variation. Maximum deviation from the central result is included as the scale uncertainty. The individual contributions of all model variations are added in quadrature into a single model uncertainty.  

The total PDF uncertainty is obtained by adding in quadrature the fit and the model uncertainties, while the parameterisation uncertainties are added linearly.

The QCD analysis is performed in few steps. First, the impact of the CMS data on global PDF set CT14~\cite{Dulat:2015mca} is investigated by using the profiling technique~\cite{Paukkunen:2014zia, Schmidt:2018hvu, HERAFitterdevelopersTeam:2015cre, AbdulKhalek:2018rok}, as implemented in \xFitter. The available implementation does not allow for a simultaneous profiling of the PDF and non-PDF parameters such as \alpSZ, \mtpole or $c_1$. Therefore, those parameters are profiled individually. Next, the HERA DIS and CMS jet data are used in a full QCD fit in SM at NNLO, where the PDFs and the value of \alpSZ are determined simultaneously. Further, the full QCD fit in SMEFT at NLO is performed using the HERA data and the CMS measurements of inclusive jet and \ttbar production at $\sqrt{s} = 13\TeV$, where the PDFs, \alpSZ, \mtpole, and $c_1$ are obtained at the same time. The individual steps of the QCD interpretation are described in detail in the following. 

\subsection{Results of profiling analysis}

The impact of new data on the available PDFs is assessed in a profiling analysis~\cite{Paukkunen:2014zia, Schmidt:2018hvu, HERAFitterdevelopersTeam:2015cre, AbdulKhalek:2018rok}. Here, the PDF profiling is performed at NLO or at NNLO, using the CT14 PDF sets~\cite{Dulat:2015mca} derived at  NLO or NNLO, respectively. These PDF sets do not include the CMS \ttbar measurements. The theoretical prediction for the triple-differential \ttbar cross section corresponding to the CMS measurement~\cite{Sirunyan:2019zvx} is available only at NLO. 

\begin{figure}[htbp!]
\centering
\includegraphics[width=0.459\textwidth]{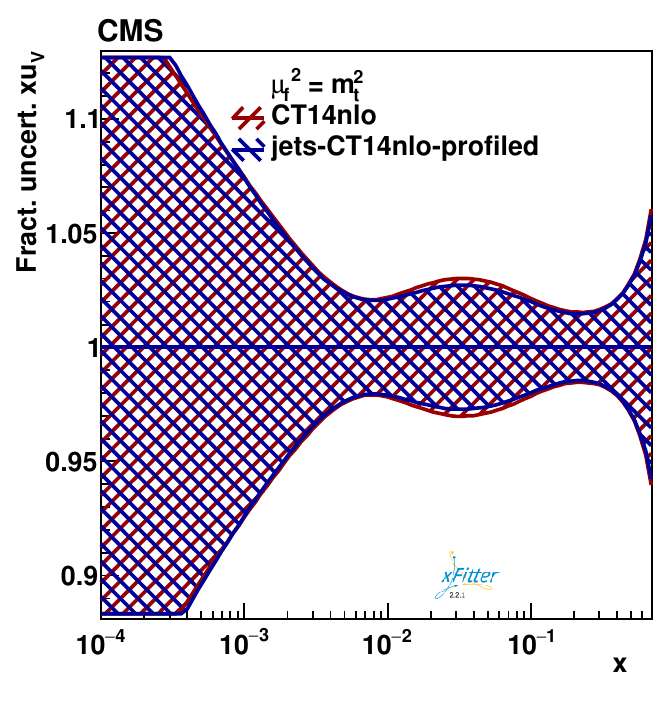}
\includegraphics[width=0.459\textwidth]{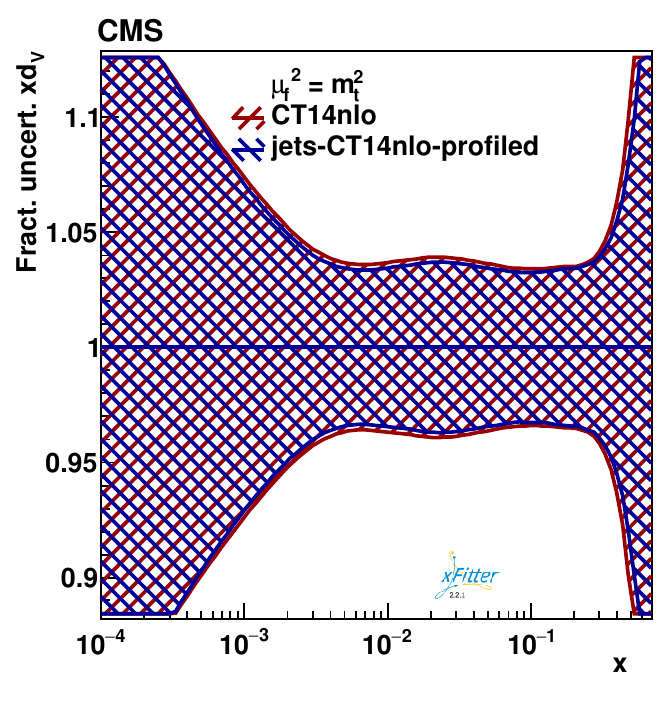}
\includegraphics[width=0.459\textwidth]{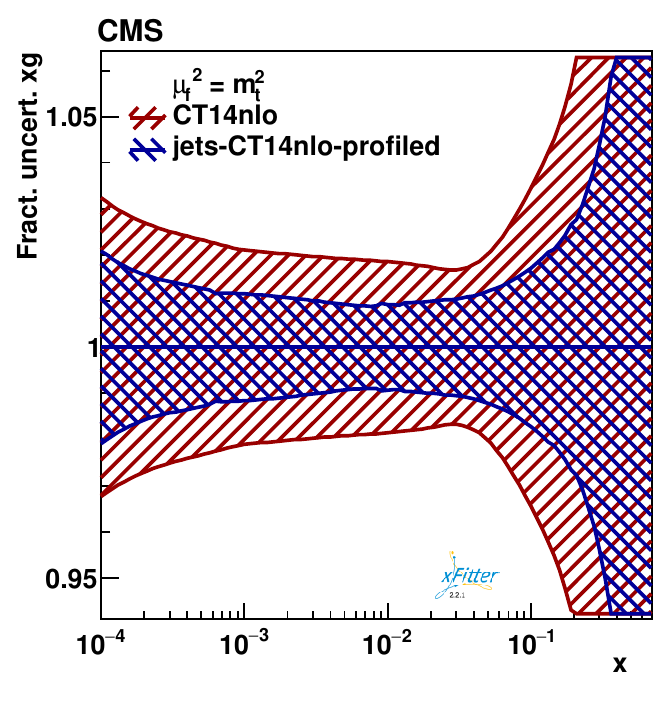}
\includegraphics[width=0.459\textwidth]{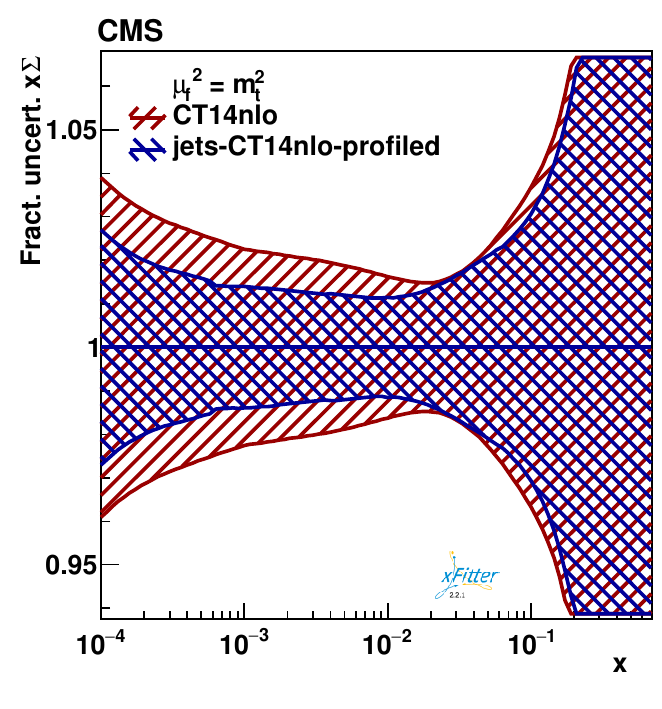}
\caption{Fractional uncertainties in the \PQu-valence (upper left), \PQd-valence (upper right), gluon (lower left), and the sea quark (lower right) distributions, shown as a function of $x$ for the scale $\muf=\mt$. 
The profiling is performed using CT14nlo PDF at NLO, by using CMS inclusive jet cross section at $\sqrt{s}=13\TeV$, implying the theoretical prediction for these data at NLO+NLL. The original uncertainty is shown in red, while the profiled result is shown in blue.}
\label{profiling_jets_NLO}
\end{figure}

\begin{figure}[htbp!]
\centering
\includegraphics[width=0.459\textwidth]{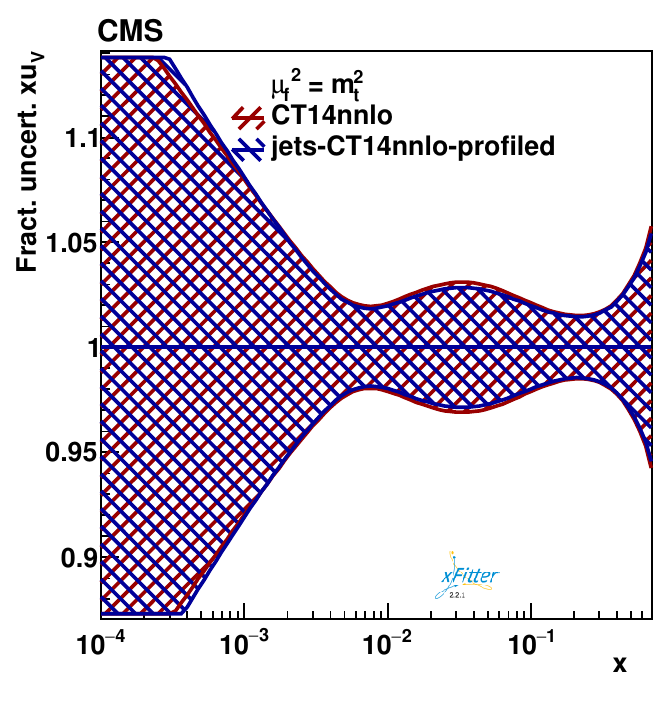}
\includegraphics[width=0.459\textwidth]{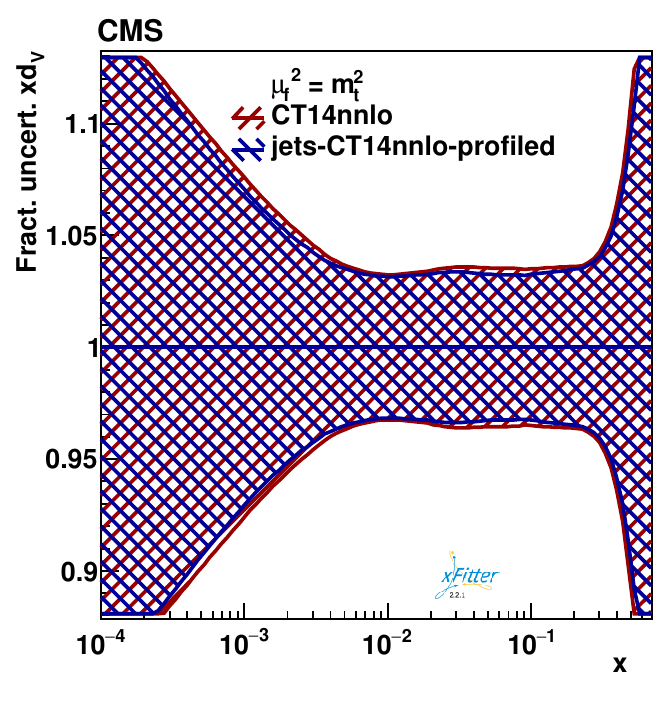}
\includegraphics[width=0.459\textwidth]{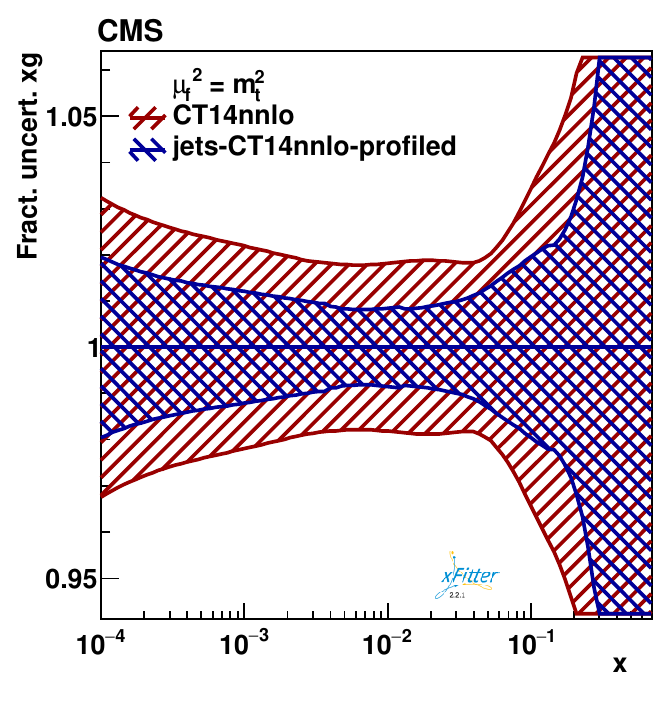}
\includegraphics[width=0.459\textwidth]{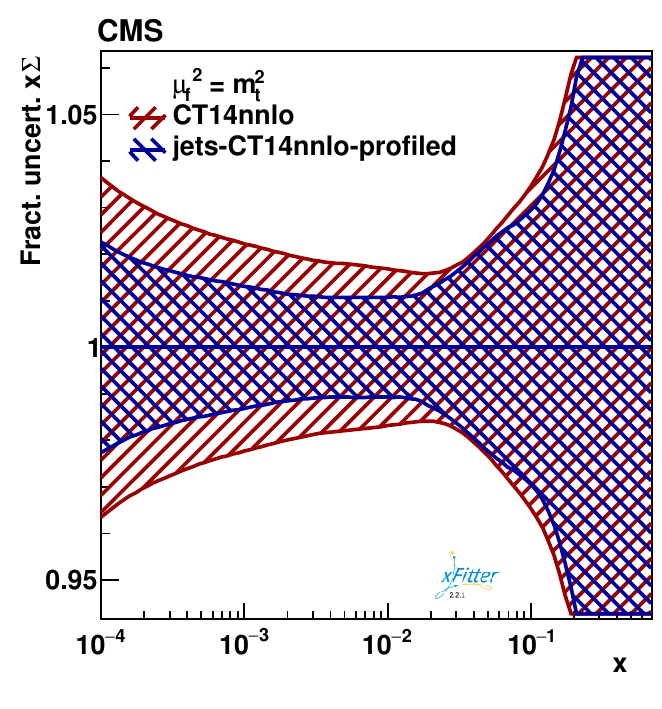}
\caption{Fractional uncertainties in the \PQu-valence (upper left), \PQd-valence (upper right), gluon (lower left), and the sea quark (lower right) distributions, shown as functions of $x$ for the scale $\muf=\mt$. 
The profiling is performed using CT14nnlo PDF at NNLO, by using the CMS inclusive jet cross section at $\sqrt{s}=13\TeV$, implying the theoretical prediction for these data at NNLO.  The original uncertainty is shown in red, while the profiled result is shown in blue.}
\label{profiling_jets_NNLO}
\end{figure}

In the PDF profiling, the strong coupling is fixed to the central value of the CT14 PDF set, $\alpSZ=0.118$. Once the \ttbar cross sections are used, the \mtpole is set to 170.5\GeV, corresponding to the result of Ref.~\cite{Sirunyan:2019zvx}. The results of the PDF profiling using the present inclusive jet cross section are shown in Fig.~\ref{profiling_jets_NLO} (Fig.~\ref{profiling_jets_NNLO}) at NLO (NNLO). According to the sensitivity of the data, the uncertainties in the PDFs are significantly improved by using the CMS jet measurement in the full $x$ range for the gluon and at medium $x$ for the sea quark distributions, whereas the valence distributions remain unchanged.

\begin{figure}[bh]
\centering
\includegraphics[width=0.487\textwidth]{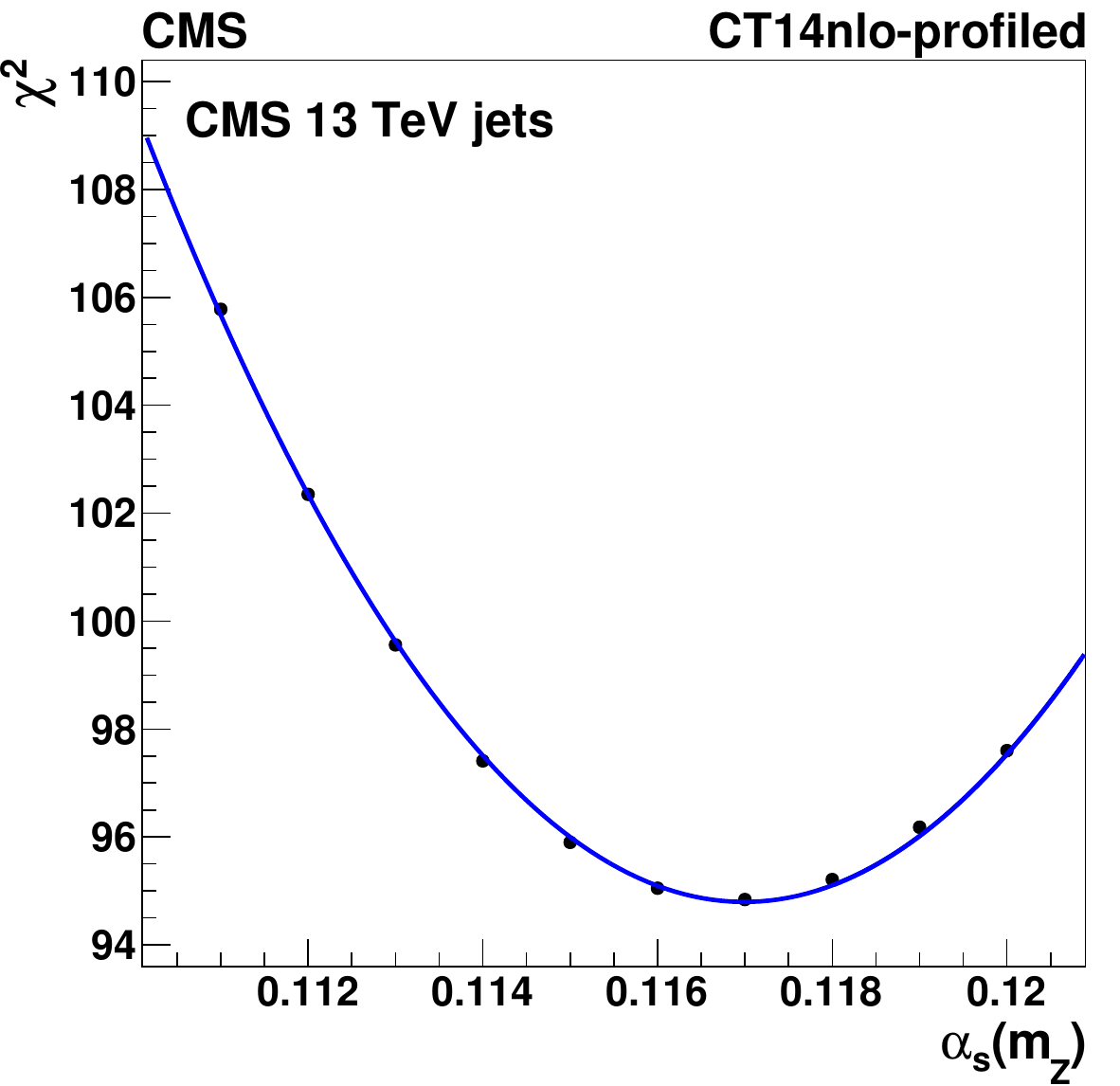}
\includegraphics[width=0.487\textwidth]{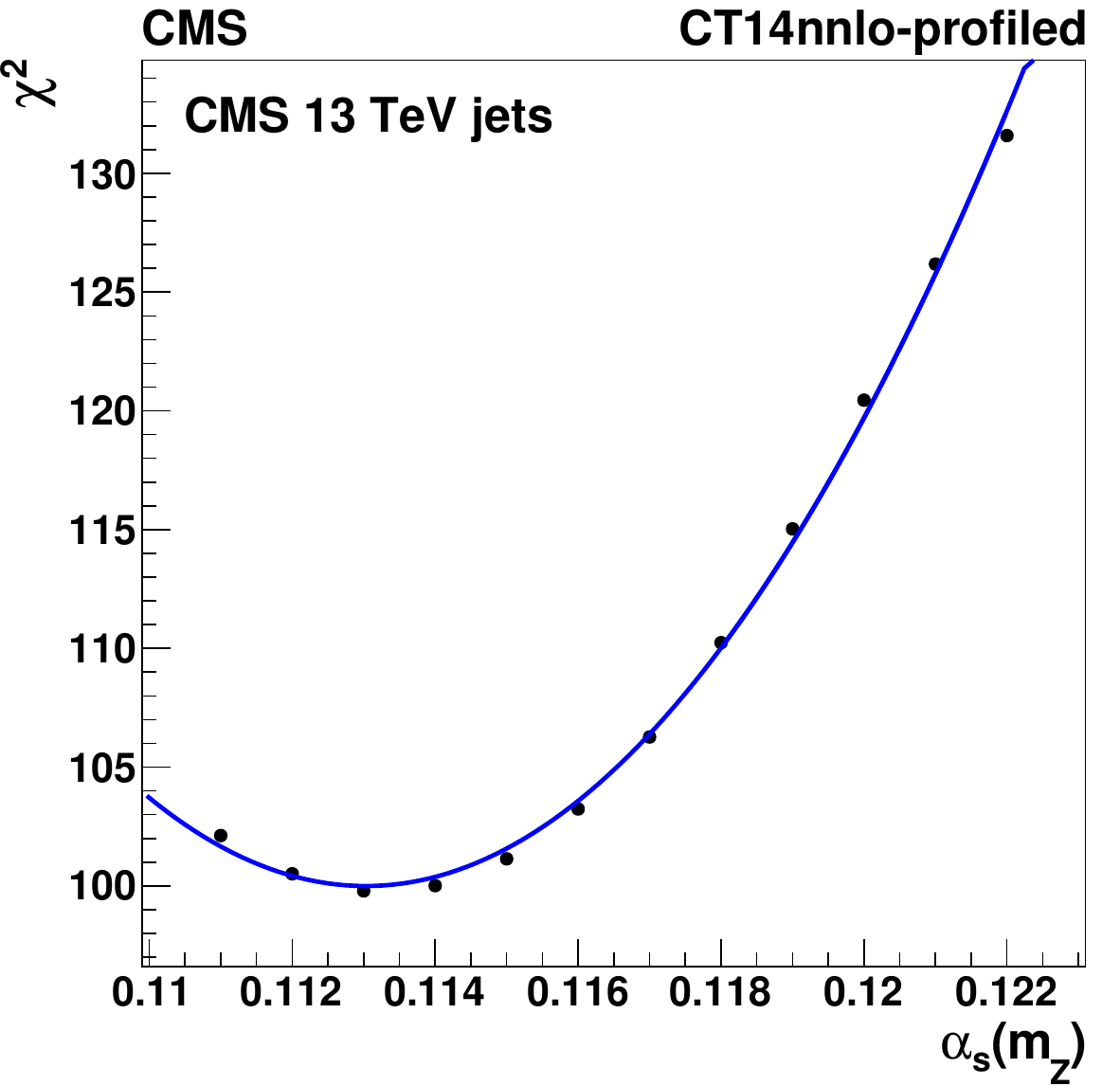}
\caption{The $\chi^2$ obtained in profiling of CT14 PDF \alpSZ series using the CMS inclusive jet cross section at $\sqrt{s}=13\TeV$ at NLO (left) and NNLO (right).}
\label{profiling_as_jets}
\end{figure}

In addition to profiling the PDFs, the impact of the inclusive jet measurements on the extraction of the strong coupling constant is investigated. For this purpose, the \alpS-series of the CT14 PDFs at NLO and NNLO is used, where the value of \alpSZ was varied from 0.1110 to 0.1220. Note that any possible \alpSZ dependence of the $k$-factors could not be accounted for. The individual profiling is performed for each of the PDF members in the \alpS series and the resulting $\chi^2$ is shown in Fig.~\ref{profiling_as_jets} for both NLO and NNLO. The optimal value of \alpSZ and its uncertainty is obtained by a parabolic fit as presented in Fig.~\ref{profiling_as_jets}. The impact of the scale uncertainty on the result is investigated by varying \mur and \muf in the theoretical predictions for the jet cross section. The $\chi^2$ scan is performed for each scale choice individually. The values obtained for the strong coupling are 
$\alpSZ = 0.1170 \pm 0.0018\,\text{(PDF)} \pm 0.0035\,\text{(scale)}$
at NLO and 
$\alpSZ = 0.1130 \pm 0.0016\,\text{(PDF)} \pm 0.0014\,\text{(scale)}$ 
at NNLO. The NLO result is in good agreement with the world average~\cite{PDG2020}. 

\begin{figure}[htb!]
\includegraphics[width=0.459\textwidth]{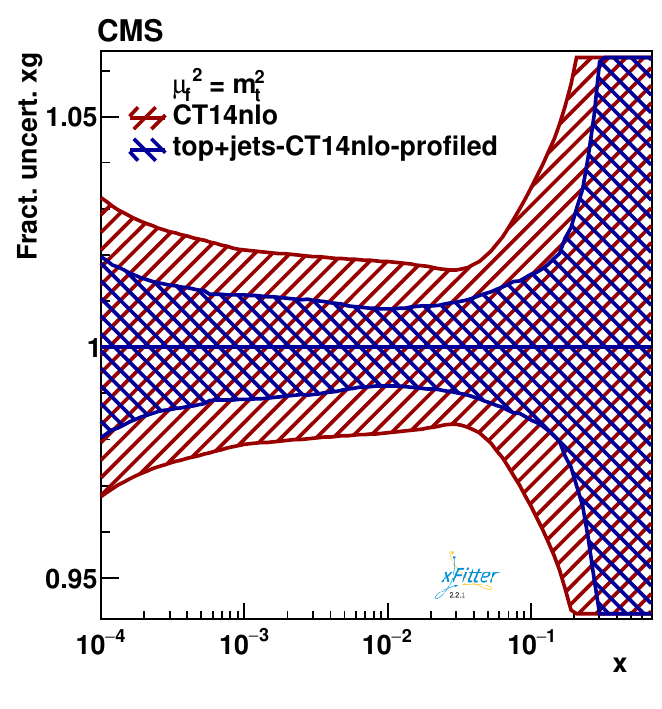}
\raisebox{1.6mm}{\includegraphics[width=0.487\textwidth]{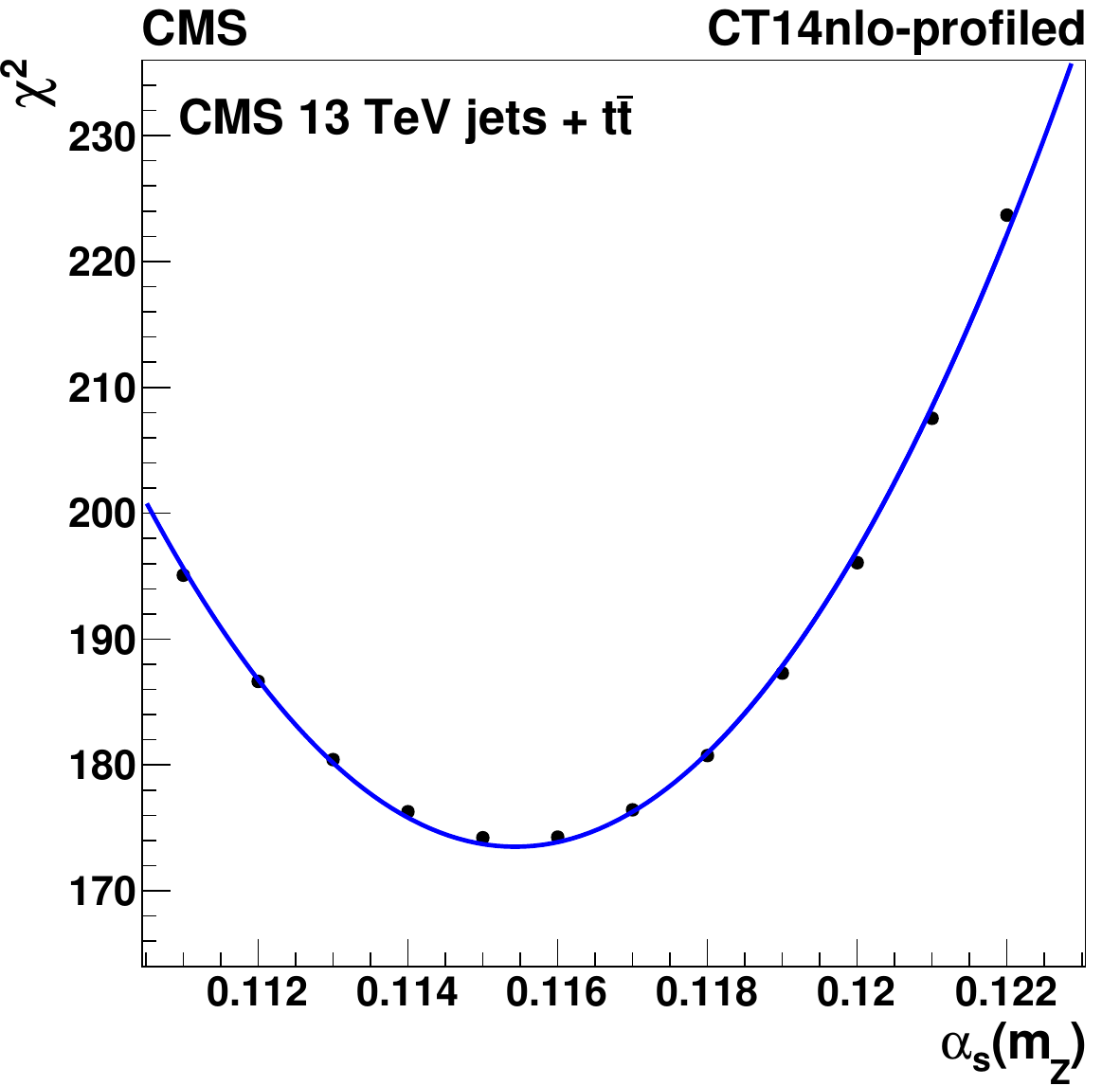}}
\caption{Fractional uncertainty in the gluon distribution (left), shown as a function of $x$ for the scale $\muf=\mt$. The profiling is performed using CT14nlo PDF at NLO, by using the CMS inclusive jet and the triple-differential \ttbar cross sections at $\sqrt{s}=13\TeV$.  The original (profiled) uncertainty is shown in red (blue). The $\chi^2$ (right) obtained in profiling of CT14 PDF \alpSZ series using the same data as in (left).}
\label{profiling_jet_top}
\end{figure}

The profiling analysis is repeated by using the triple-differential CMS \ttbar cross section of Ref.~\cite{Sirunyan:2019zvx} together with the inclusive jet cross section. Consistent with the available theoretical prediction for the \ttbar measurements, this analysis is performed at NLO. The results are shown in Fig.~\ref{profiling_jet_top}, where the uncertainty in the profiled gluon distribution is presented in comparison to that of the original CT14 PDF. The reduction of the uncertainty in the gluon distribution at high $x$ is stronger than in the case when only the CMS inclusive jet cross section is used. This is expected from the additional sensitivity of the \ttbar production to the gluon distribution at high $x$. Also the \alpSZ scan is shown in Fig.~\ref{profiling_jet_top}, now using both CMS data sets. The resulting NLO value of the strong coupling is $\alpSZ = 0.1154 \pm 0.0009\,\text{(PDF)} \pm 0.0015\,\text{(scale)}$, consistent with the result of Ref.~\cite{Sirunyan:2019zvx}. The additional sensitivity of the \ttbar production to the strong coupling becomes visible in the reduced PDF uncertainties. The profiled pole mass of the top quark results in $\mtpole = 170.3 \pm 0.5\,\text{(PDF)} + 0.2\,\text{(scale)}\GeV$, consistent with the value obtained in Ref.~\cite{Sirunyan:2019zvx}.

\begin{figure}[htb!]
\centering
\includegraphics[width=0.65\textwidth]{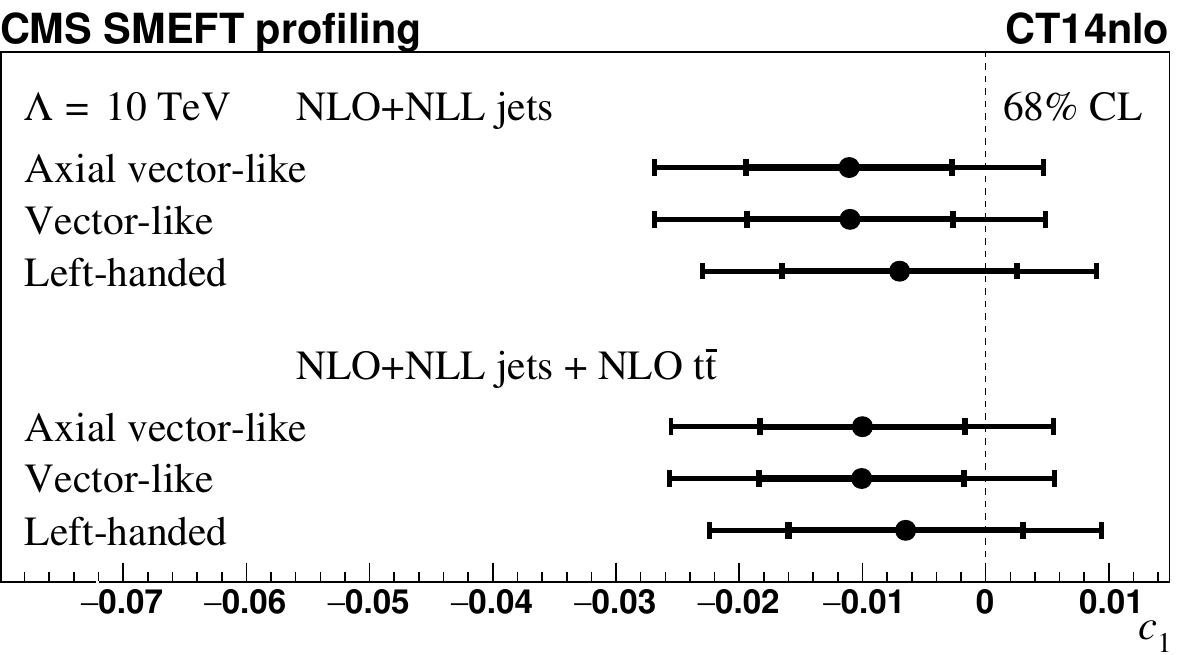}
\caption{The profiled Wilson coefficient $c_1$ for the contact interaction models, assuming the left-handed, vector-like, and axial vector-like scenarios, as obtained in the profiling analysis using NLO+NLL calculation for the jet production and the CT14nlo PDF set. The value of $\Lambda = 10\TeV$ is assumed. The results are obtained using the CMS measurements of inclusive jet cross section and of normalised triple-differential \ttbar cross section at $\sqrt{s}=13\TeV$. The inner error bar shows the PDF uncertainty at 68\% \CL, while the outer error bar represents the total uncertainty, obtained from the PDF and scale uncertainties, added in quadrature.}
\label{CI_profiling}
\end{figure}

Furthermore, the profiling analysis is repeated assuming the SMEFT prediction for the inclusive jet production cross section. While the results of the profiled PDFs remain unchanged with respect to the SM results described above, the Wilson coefficient $c_1$ is profiled, assuming the value of the scale of the new interaction $\Lambda=10\TeV$, and the results are summarised in the Fig.~\ref{CI_profiling}.

The resulting values of $c_1$ are consistent with zero within uncertainties for all investigated CI models, demonstrating a good description of the data by the SM. Since the SMEFT computation is applied only to the inclusive jet production cross section, the $c_1$ results are independent of the inclusion of \ttbar data. Once the relevant calculations for these data become available, a more global SMEFT interpretation would become possible.

\clearpage

\subsection{Results of the full QCD fit in SM at NNLO}

\begin{figure}[htbp!]
\centering
\includegraphics[width=0.47\textwidth]{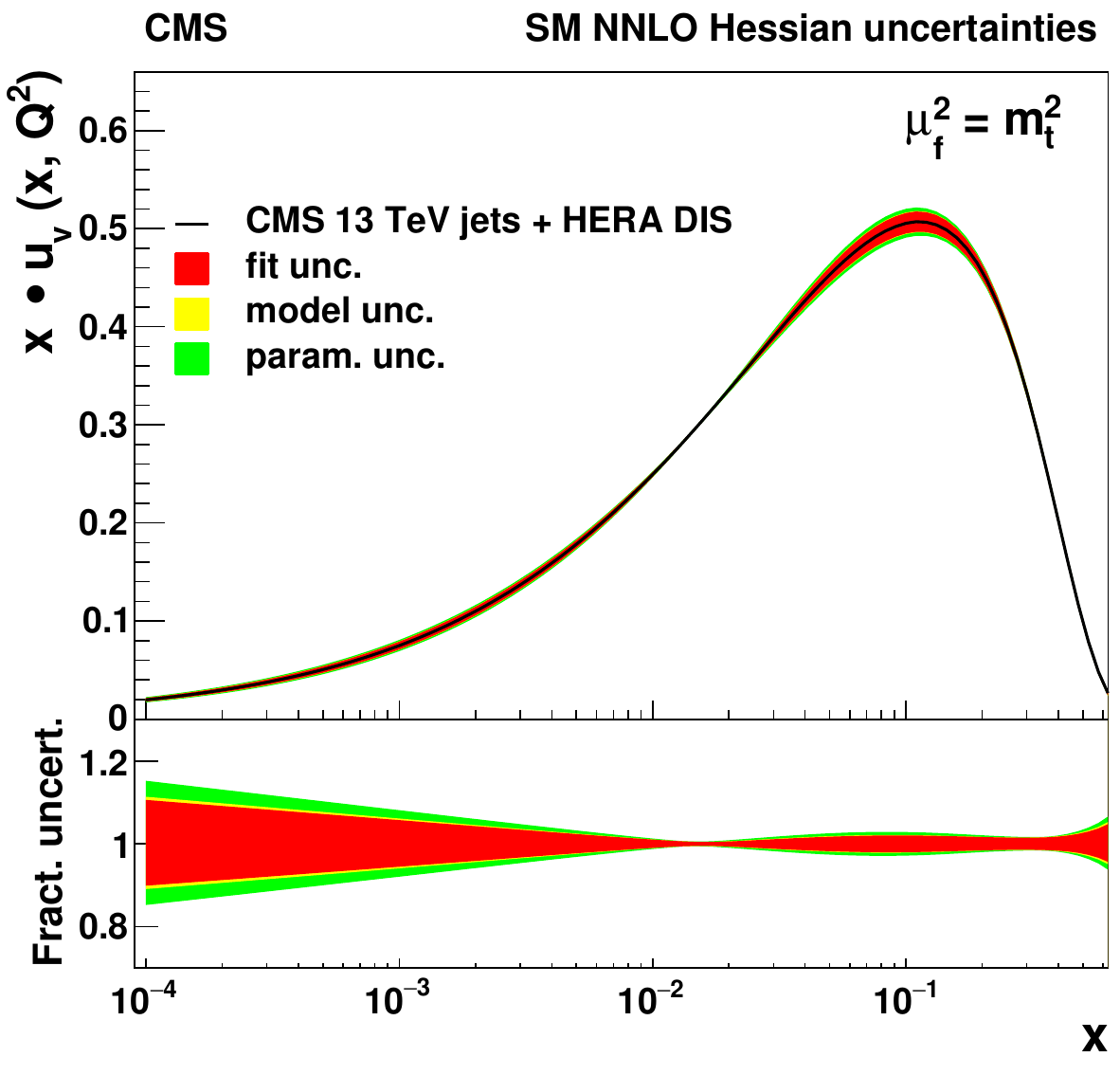}
\includegraphics[width=0.47\textwidth]{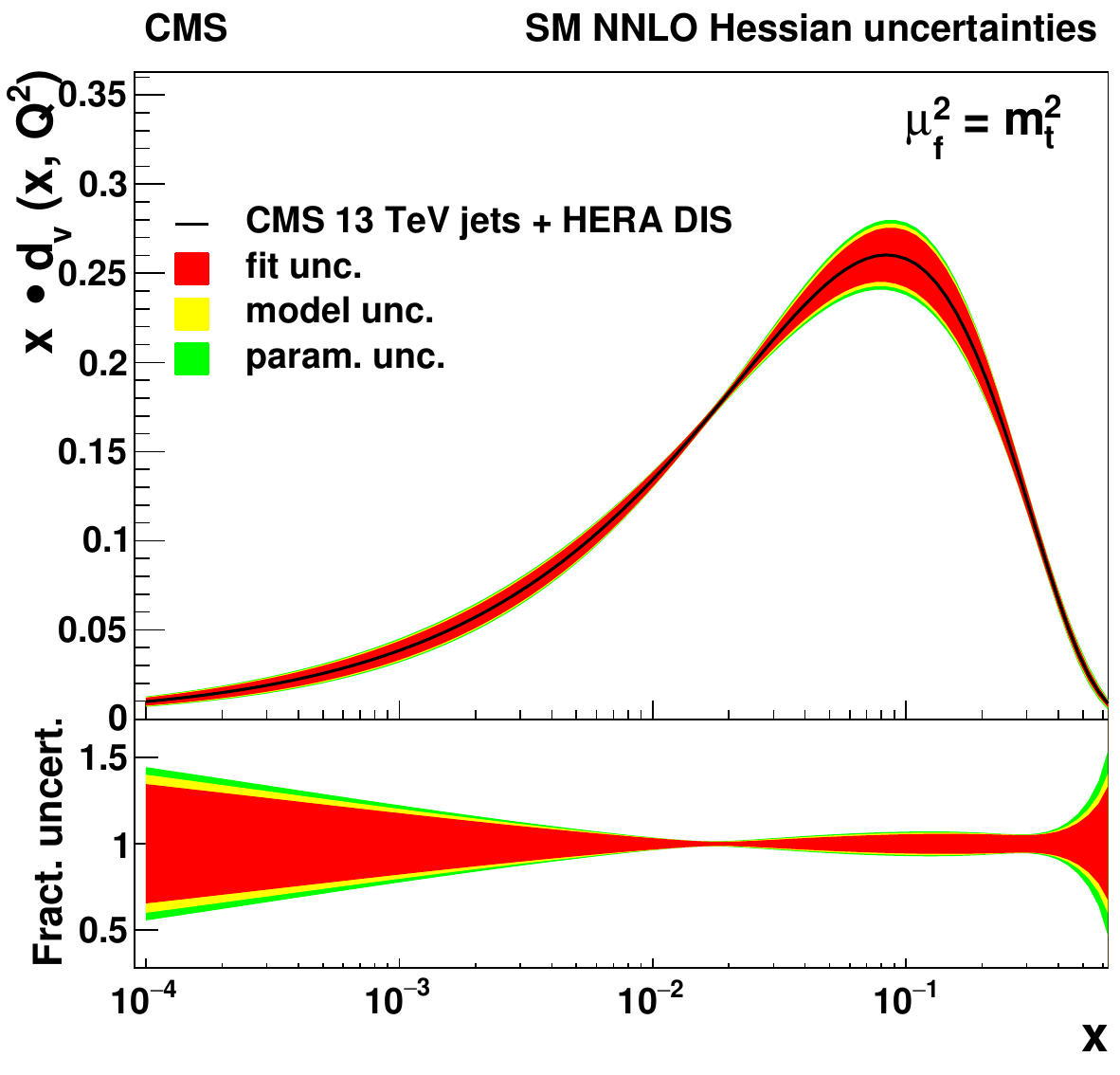}\\
\includegraphics[width=0.47\textwidth]{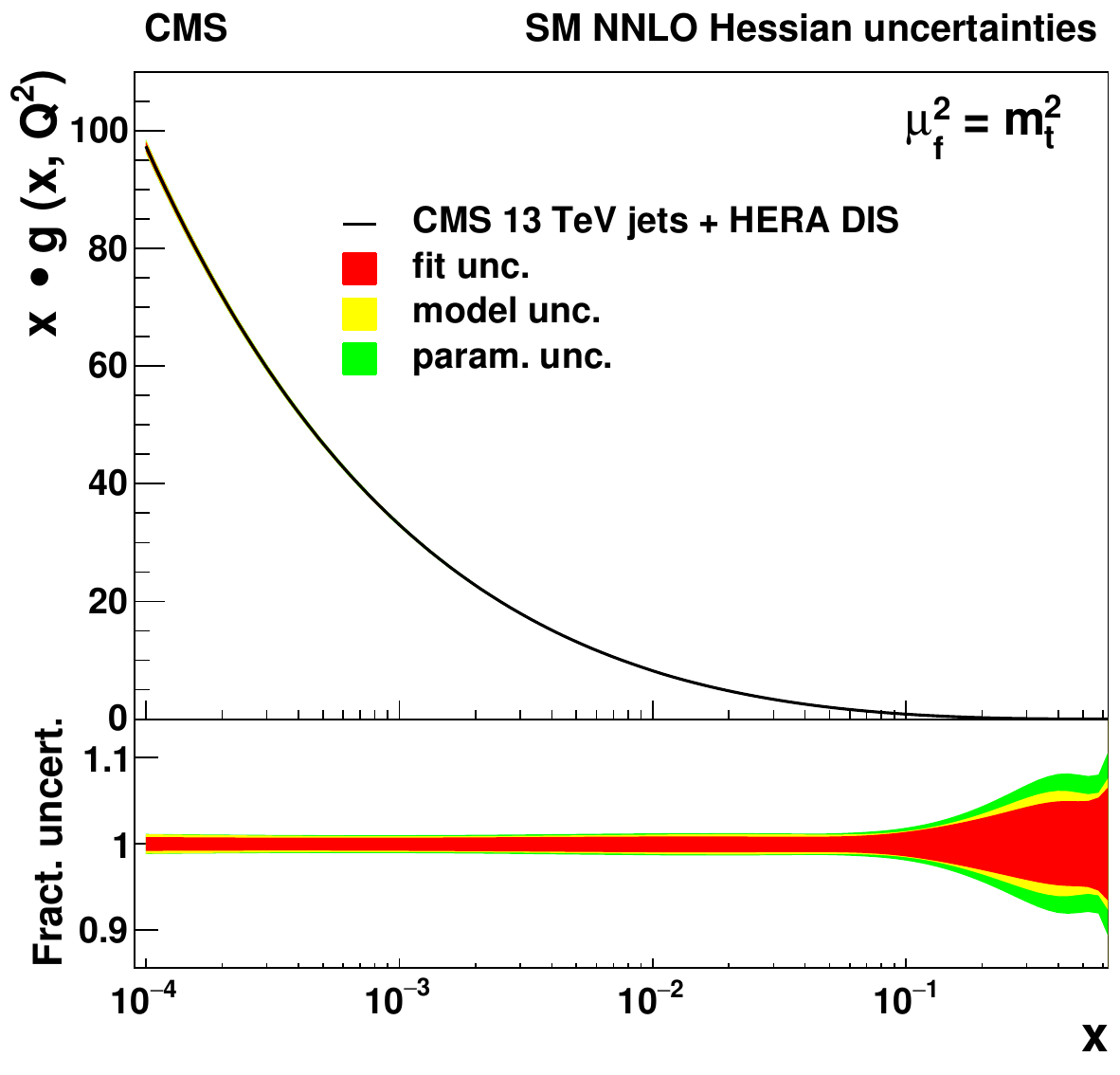}
\includegraphics[width=0.47\textwidth]{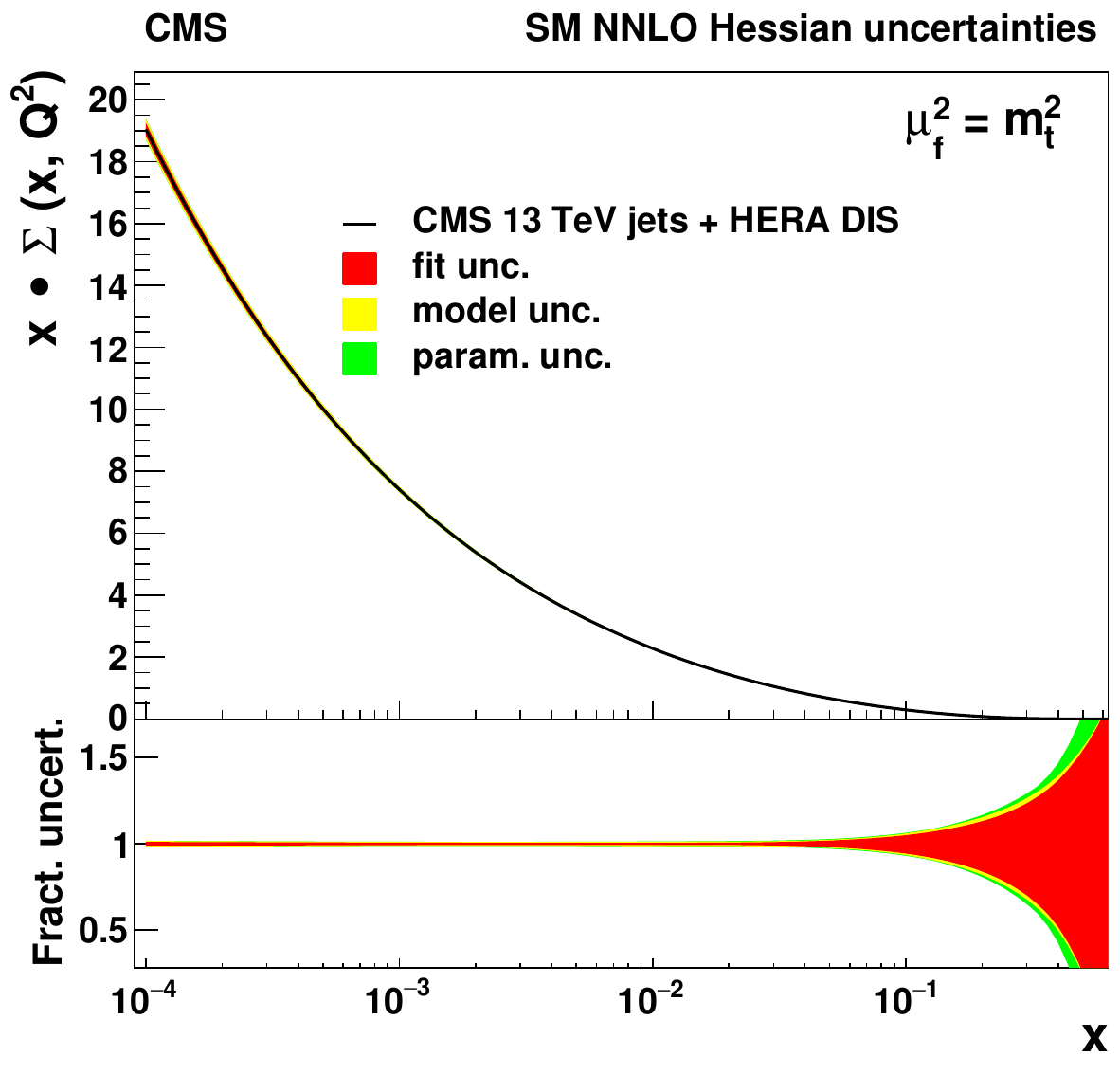}
\caption{The \PQu-valence (upper left), \PQd-valence (upper right), gluon (lower left), and sea quark (lower right) distributions, shown as a function of $x$ at the scale $\muf=m_t^2$, resulting from the NNLO fit using HERA DIS together with the CMS inclusive jet cross section at $\sqrt{s}=13\TeV$. Contributions of fit, model, and parameterisation uncertainties for each PDF are shown. In the lower panels, the relative uncertainty contributions are presented.}
\label{NNLO_HERA+CMS_breakdown}
\end{figure}

The present measurement of the inclusive jet production cross section is used together with the inclusive DIS cross section of HERA in a full QCD analysis at NNLO. The PDF parameterisation at the starting scale, resulting from the scan as described in Section~\ref{QCD_analysis_strategy} reads:
\begin{align}
x \Pg(x) &= A_\Pg
            x^{B_\Pg}
            (1-x)^{C_\Pg}
            (1 + D_\Pg x + E_\Pg x^2),
\label{NNLO_ResultingPDFparameterisation_g}\\
x \PQu_v(x) &= A_{\PQu_v}
               x^{B_{\PQu_v}}
               (1-x)^{C_{\PQu_v}}
               (1 + E_{\PQu_v} x^2),
\label{NNLO_ResultingPDFparameterisation_uv}\\
x \PQd_v(x) &= A_{\PQd_v}
               x^{B_{\PQd_v}}
               (1-x)^{C_{\PQd_v}},
\label{NNLO_ResultingPDFparameterisation_dv}\\
x \Ubar(x) &= A_{\Ubar}
              x^{B_{\Ubar}}
              (1-x)^{C_{\Ubar}}
              (1 + D_{\Ubar} x),
\label{NNLO_ResultingPDFparameterisation_U}\\
x \Dbar(x) &= A_{\Dbar}
              x^{B_{\Dbar}}
              (1-x)^{C_{\Dbar}}
              (1 + E_{\Dbar} x^2).
\label{NNLO_ResultingPDFparameterisation_D}
\end{align}
The resulting PDFs are shown in Fig.~\ref{NNLO_HERA+CMS_breakdown}, illustrating the contributions from the fit, model and the parameterisation uncertainties. The value of \alpSZ is obtained simultaneously with the PDFs and corresponds to 
\begin{equation}
\alpSZ = 0.1170
               \pm 0.0014\,\text{(fit)}
               \pm 0.0007\,\text{(model)}
               \pm 0.0008\,\text{(scale)}
               \pm 0.0001\,\text{(param.)},
\label{NNLO_alphas}
\end{equation}
which agrees with the previous extractions of the strong coupling constant at NNLO at hadron colliders~\cite{Andreev:2017vxu, Sirunyan:2018goh}, of which it has best precision, to date.

\begin{figure}[htbp!]
\centering
\includegraphics[width=0.47\textwidth]{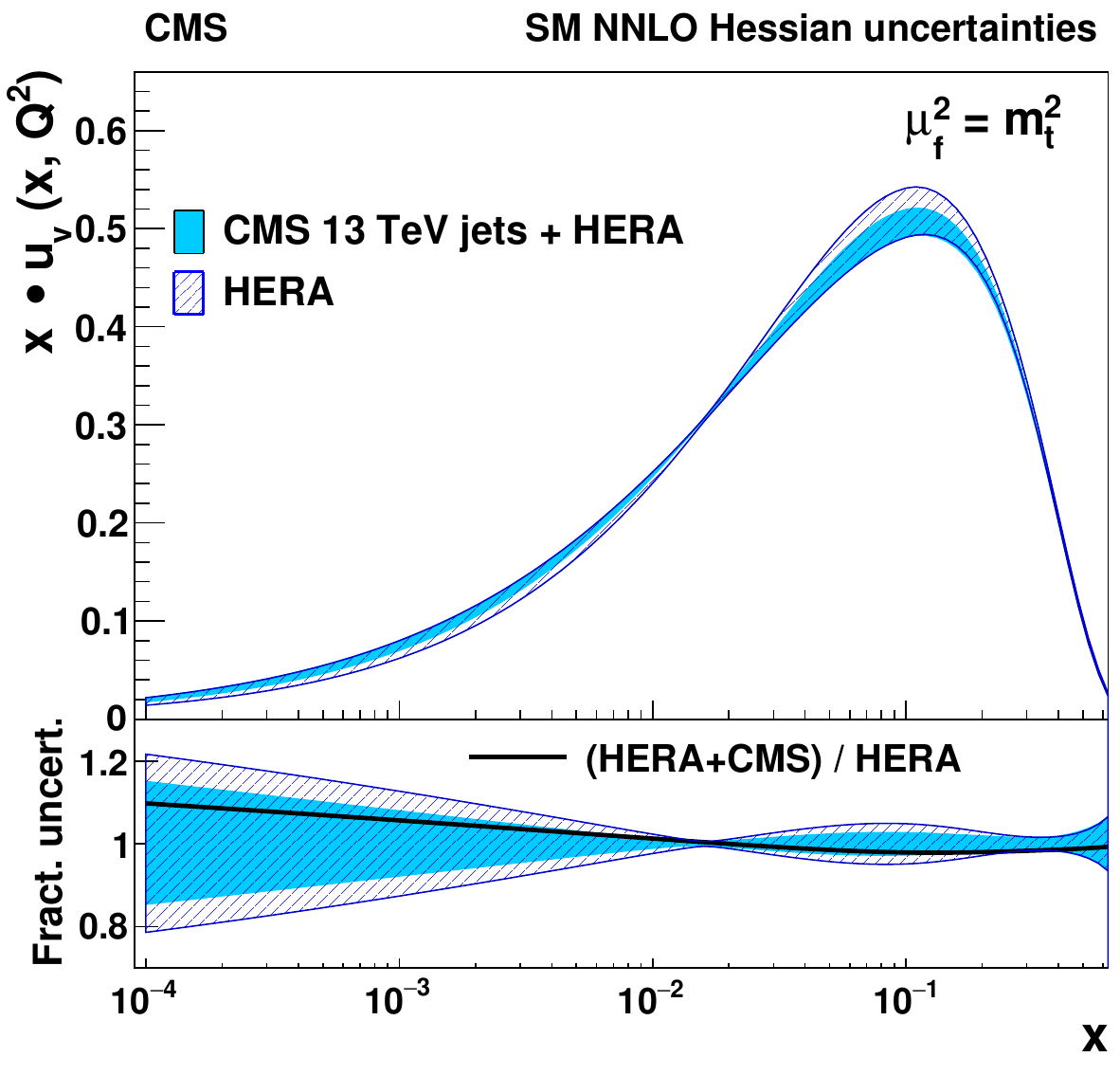}
\includegraphics[width=0.47\textwidth]{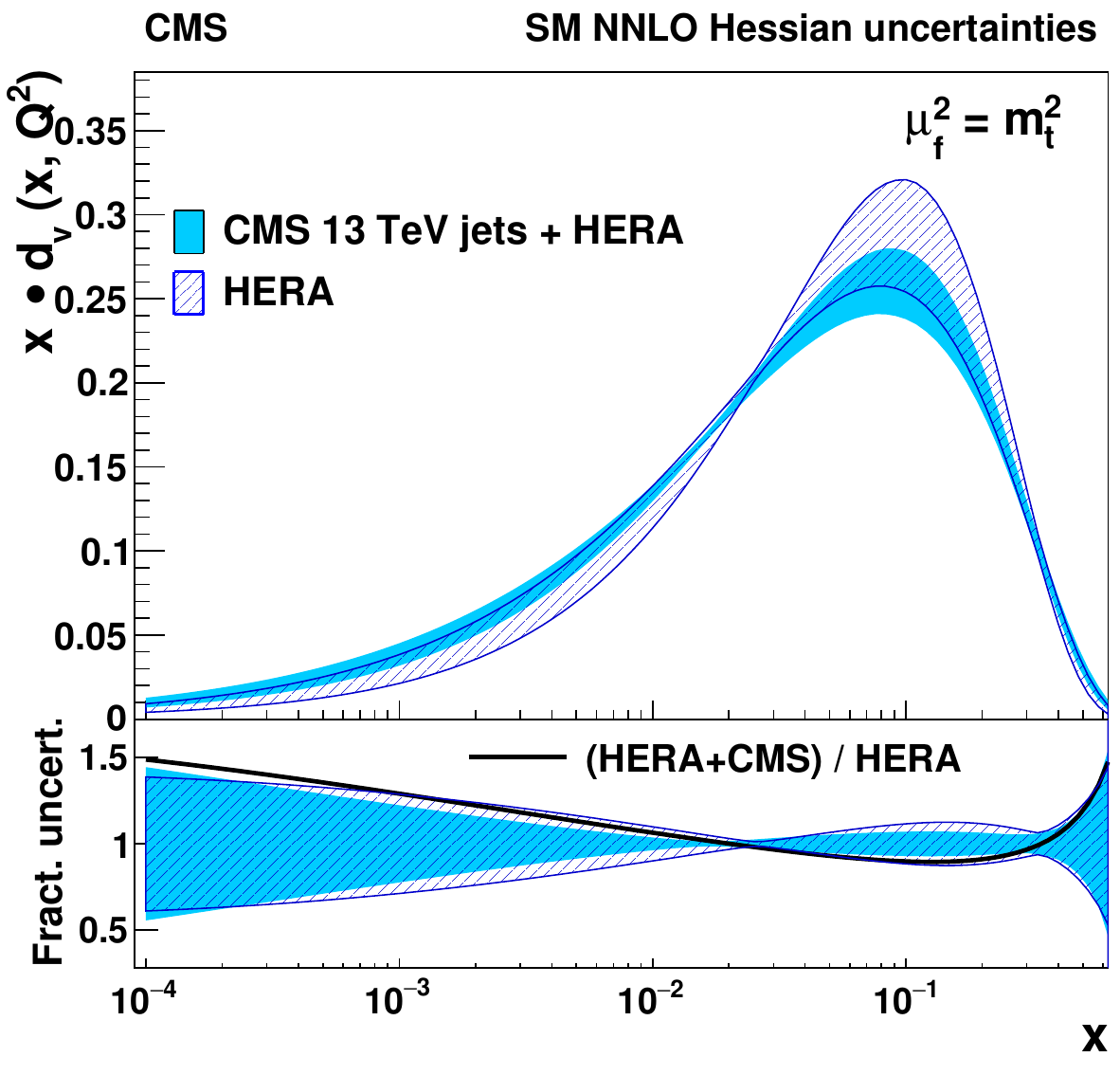}\\
\includegraphics[width=0.47\textwidth]{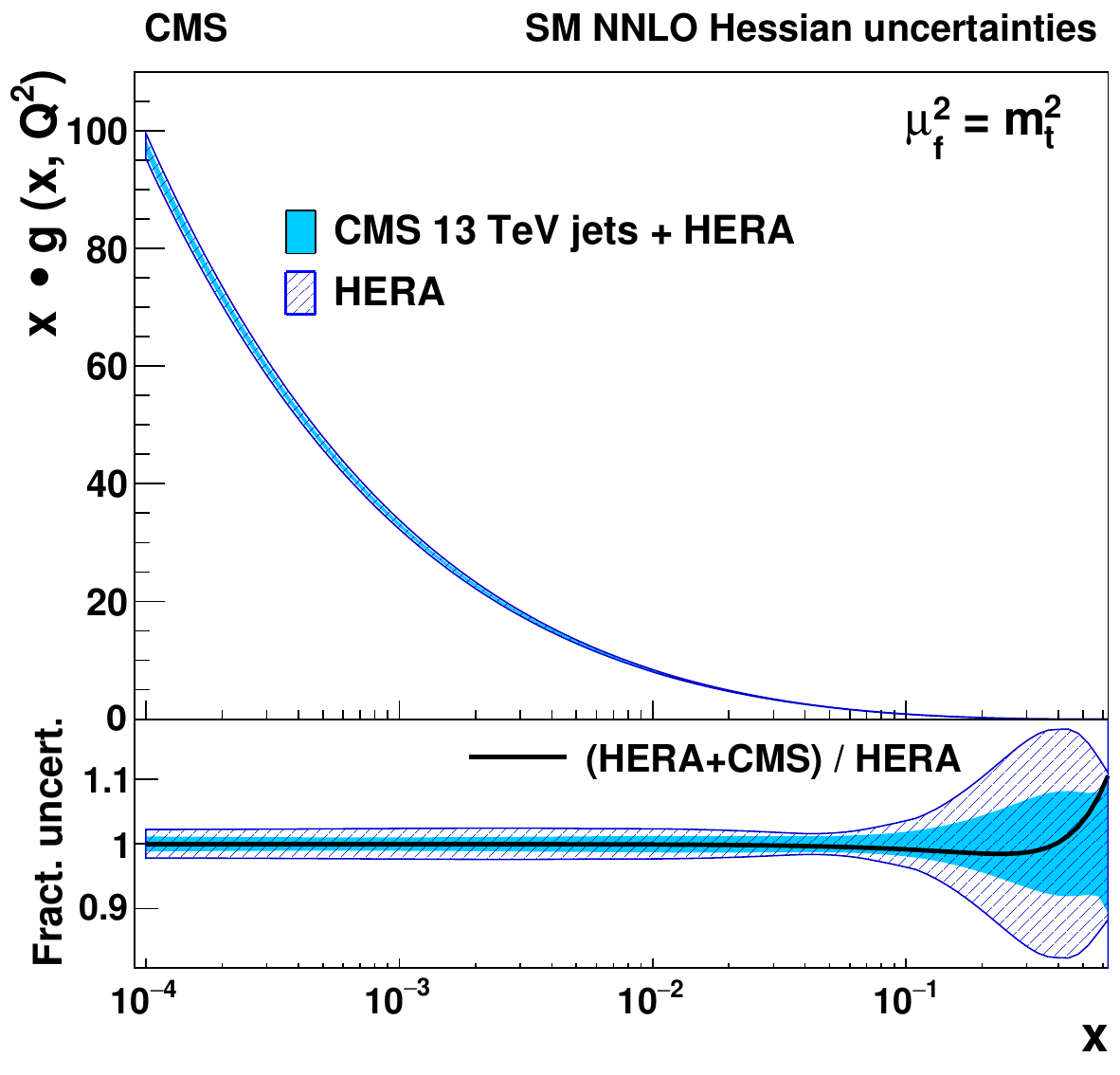}
\includegraphics[width=0.47\textwidth]{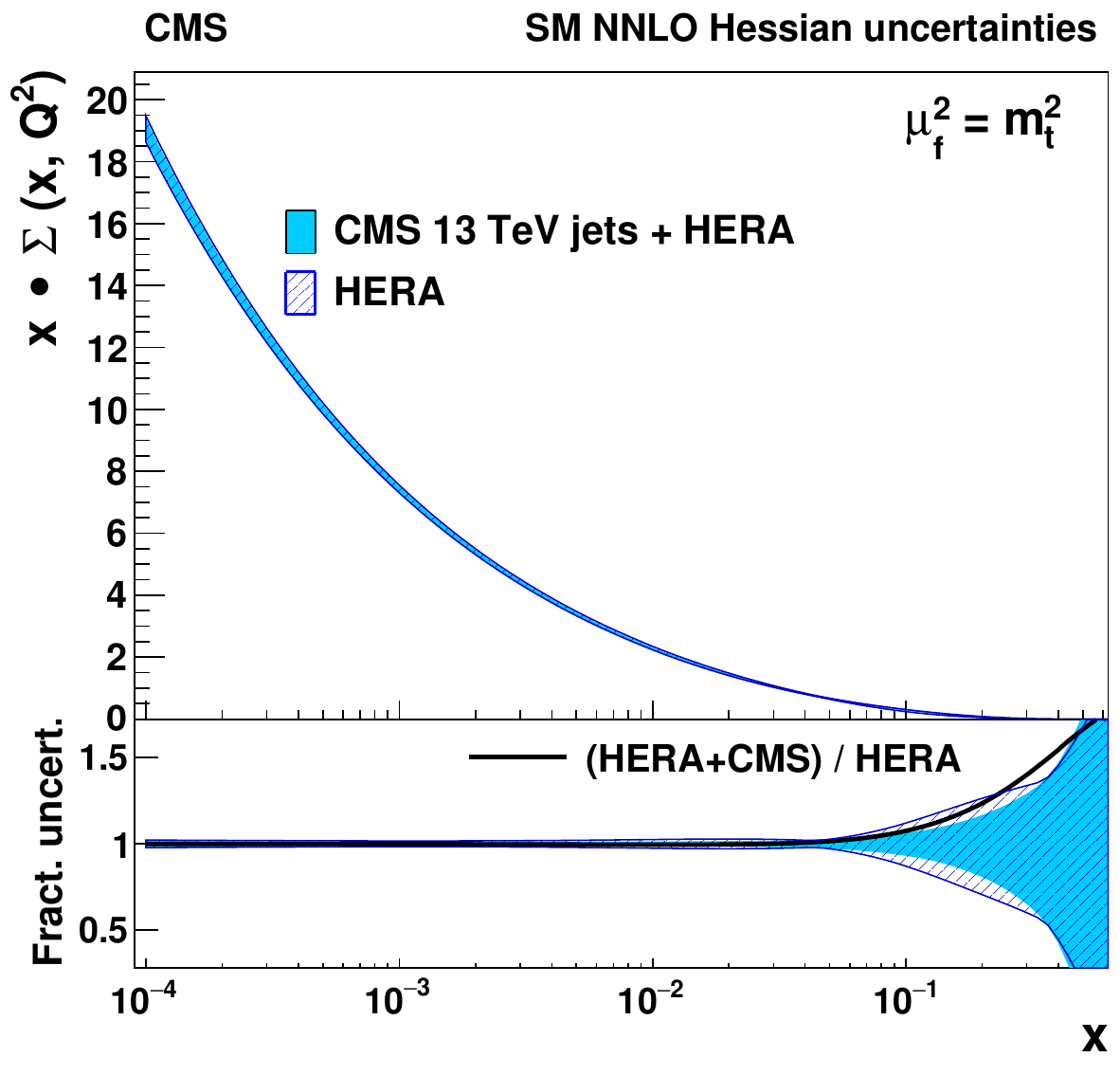}
\caption{The \PQu-valence (upper left), \PQd-valence (upper right), gluon (lower left), and sea quark (lower right) distributions, shown as a function of $x$ at the scale $\muf=m_t^2$. The filled (hatched) band represents the results of the NNLO fit using HERA DIS and the CMS inclusive jet cross section at $\sqrt{s}=13\TeV$ (using the HERA DIS data only). The PDFs are shown with their total uncertainty.  In the lower panels, the comparison of the relative PDF uncertainties is shown for each distribution. The dashed line corresponds to the ratio of the central PDF values of the two variants of the fit.}
\label{NNLO_HERA_vs_HERA+CMS}
\end{figure}

The impact of the present CMS jet data in a full QCD fit (HERA+CMS fit) is demonstrated by comparing of the resulting PDFs with an alternative fit, where only the HERA data is used (HERA-only fit). Since the inclusive DIS data have much lower sensitivity to the value of \alpSZ, it is fixed in the HERA-only fit to the result of the HERA+CMS fit. The comparison of the resulting PDFs is presented in Fig.~\ref{NNLO_HERA_vs_HERA+CMS}. The uncertainty is significantly reduced once the CMS jet measurements are included.

The global and partial $\chi^2$ values for each data set, for HERA-only and HERA+CMS fits, are listed in Table~\ref{QCD_analysis_partial_chi2_NNLO}, where the $\chi^2$ values illustrate a general agreement among all the data sets. The somewhat high $\chi^2/\Ndof$ values for the combined DIS data are very similar to those observed in Ref.~\cite{Abramowicz:2015mha}, where they are investigated in detail.

\begin{table}[htbp!]
    \topcaption{Partial $\chi^2$ per number of data points \Ndp and the global $\chi^2$ per degree of freedom, \Ndof, as obtained in the QCD analysis at NNLO of HERA+CMS jet data and HERA-only data. In the DIS data, the proton beam energy is given as $E_\Pp$ and the electron energy is 27.5\GeV.}
    \label{QCD_analysis_partial_chi2_NNLO}
    \centering
    \begin{tabular}{l l c c}
                  & ~ & HERA-only                        & HERA+CMS \\
        Data sets & ~ & Partial $\chi^2/\Ndp$ & Partial $\chi^2/\Ndp$ \\
        \hline
        HERA I+II neutral current& $\Pep\Pp$, $E_\Pp=920\GeV$     & 378/332      & 375/332   \\
        HERA I+II neutral current& $\Pep\Pp$, $E_\Pp=820\GeV$     & 60/63        & 60/63     \\
        HERA I+II neutral current& $\Pep\Pp$, $E_\Pp=575\GeV$     & 201/234      & 201/234   \\
        HERA I+II neutral current& $\Pep\Pp$, $E_\Pp=460\GeV$     & 208/187      & 209/187   \\
        HERA I+II neutral current& $\Pem\Pp$, $E_\Pp=920\GeV$     & 223/159      & 227/159   \\
        HERA I+II charged current& $\Pep\Pp$, $E_\Pp=920\GeV$     & 46/39        & 46/39     \\
        HERA I+II charged current& $\Pem\Pp$, $E_\Pp=920\GeV$     & 55/42        & 56/42     \\[\cmsTabSkip]
        CMS inclusive jets 13\TeV& $0.0 < \abs{y} < 0.5$          & \NA          & 13/22     \\ 
                                 & $0.5 < \abs{y} < 1.0$          & \NA          & 31/21     \\
                                 & $1.0 < \abs{y} < 1.5$          & \NA          & 18/19     \\
                                 & $1.5 < \abs{y} < 2.0$          & \NA          & 14/16     \\
        \hline
        Correlated $\chi^2$      &                                & 66           & 83        \\
        Global $\chi^2/\Ndof$    &                                & 1231/1043    & 1321/1118 \\
    \end{tabular}
\end{table}

\clearpage

\subsection{Results of the SMEFT fit at NLO}

To illustrate the possibility of simultaneous extraction of the SM parameters as PDFs, \alpSZ, and \mtpole, together with the constraints on the physics beyond the SM, the present CMS measurements of the inclusive jet cross section, the triple-differential normalised CMS \ttbar cross section at 13\TeV, and the HERA DIS cross sections are used in a SMEFT fit. Here, the SM prediction for the inclusive jet cross section is modified to account for CI as described in Section~\ref{QCD_analysis_theory}. 
The parameterisation is reinvestigated, as explained in Section~\ref{QCD_analysis_strategy}, and results in
\begin{align}
x \Pg(x) &= A_\Pg
            x^{B_\Pg}
            (1-x)^{C_\Pg}
            (1 + E_\Pg x^2),
\label{ResultingPDFparameterisation_g}\\
x \PQu_v(x) &= A_{\PQu_v}
               x^{B_{\PQu_v}}
               (1-x)^{C_{\PQu_v}}
               (1 + D_{\PQu_v}x + E_{\PQu_v}x^2),
\label{ResultingPDFparameterisation_uv}\\
x \PQd_v(x) &= A_{\PQd_v}
               x^{B_{\PQd_v}}
               (1-x)^{C_{\PQd_v}}
               (1 + D_{\PQd_v}x),
\label{ResultingPDFparameterisation_dv}\\
x \Ubar(x) &= A_{\Ubar}
              x^{B_{\Ubar}}
              (1-x)^{C_{\Ubar}},
\label{ResultingPDFparameterisation_U}\\
x \Dbar(x) &= A_{\Dbar}
              x^{B_{\Dbar}}
              (1-x)^{C_{\Dbar}}.
\label{ResultingPDFparameterisation_D}
\end{align}

First, the analysis is performed in the standard model. Then, alternatively, the SMEFT fit is done. Both SM and SMEFT fits are performed at NLO to be consistent with the order of the theoretical prediction for the \ttbar data and for the CI corrections to the SM Lagrangian, although the SM prediction for the inclusive jet cross section is available at NNLO. The partial and global $\chi^2$ values for the SM and SMEFT fits are listed in Table~\ref{QCD_analysis_partial_chi2_scan3}. The fits with all CI models and various $\Lambda$ values resulted in very similar $\chi^2$ values.

\begin{table}[htbp!]
    \topcaption[Partial $\chi^2$s full]{Partial $\chi^2$ per number of data points \Ndp and the global $\chi^2$ per degree of freedom, \Ndof, as obtained in the QCD analysis of HERA DIS data and the CMS measurements of inclusive jet production and the normalised triple-differential \ttbar production at $\sqrt{s}=13\TeV$, obtained in SM and SMEFT analyses.}
    \label{QCD_analysis_partial_chi2_scan3}
    \centering
    \begin{tabular}{l l  c c}
                                 &      & SM fit & SMEFT fit            \\
        Data sets                &      & Partial $\chi^2/\Ndp$ & Partial $\chi^2/\Ndp$ \\
        \hline
        HERA I+II neutral current& $\Pep\Pp$, $E_\Pp=920\GeV$  & 402/332    & 404/332   \\
        HERA I+II neutral current& $\Pep\Pp$, $E_\Pp=820\GeV$  & 60/63      & 60/63     \\
        HERA I+II neutral current& $\Pep\Pp$, $E_\Pp=575\GeV$  & 198/234    & 198/234   \\
        HERA I+II neutral current& $\Pep\Pp$, $E_\Pp=460\GeV$  & 208/187    & 208/187   \\
        HERA I+II neutral current& $\Pem\Pp$, $E_\Pp=920\GeV$  & 223/159    & 223/159   \\
        HERA I+II charged current& $\Pep\Pp$, $E_\Pp=920\GeV$  & 46/39      & 46/39     \\
        HERA I+II charged current& $\Pem\Pp$, $E_\Pp=920\GeV$  & 55/42      & 54/42     \\[\cmsTabSkip]
        CMS 13\TeV \ttbar 3D     &                             & 23/23      & 23/23     \\[\cmsTabSkip]
        CMS inclusive jets 13\TeV& $0.0 < \abs{y} < 0.5$       & 13/22      & 20/22     \\ 
                                 & $0.5 < \abs{y} < 1.0$       & 28/21      & 27/21     \\
                                 & $1.0 < \abs{y} < 1.5$       & 13/19      & 11/19     \\
                                 & $1.5 < \abs{y} < 2.0$       & 33/16      & 28/16     \\
        \hline
        Correlated $\chi^2$      &                             & 121        & 115       \\
        Global $\chi^2/\Ndof$    &                             & 1411/1141  & 1401/1140 \\
    \end{tabular}
\end{table}

\begin{figure}[htbp!]
\centering
\includegraphics[width=0.47\textwidth]{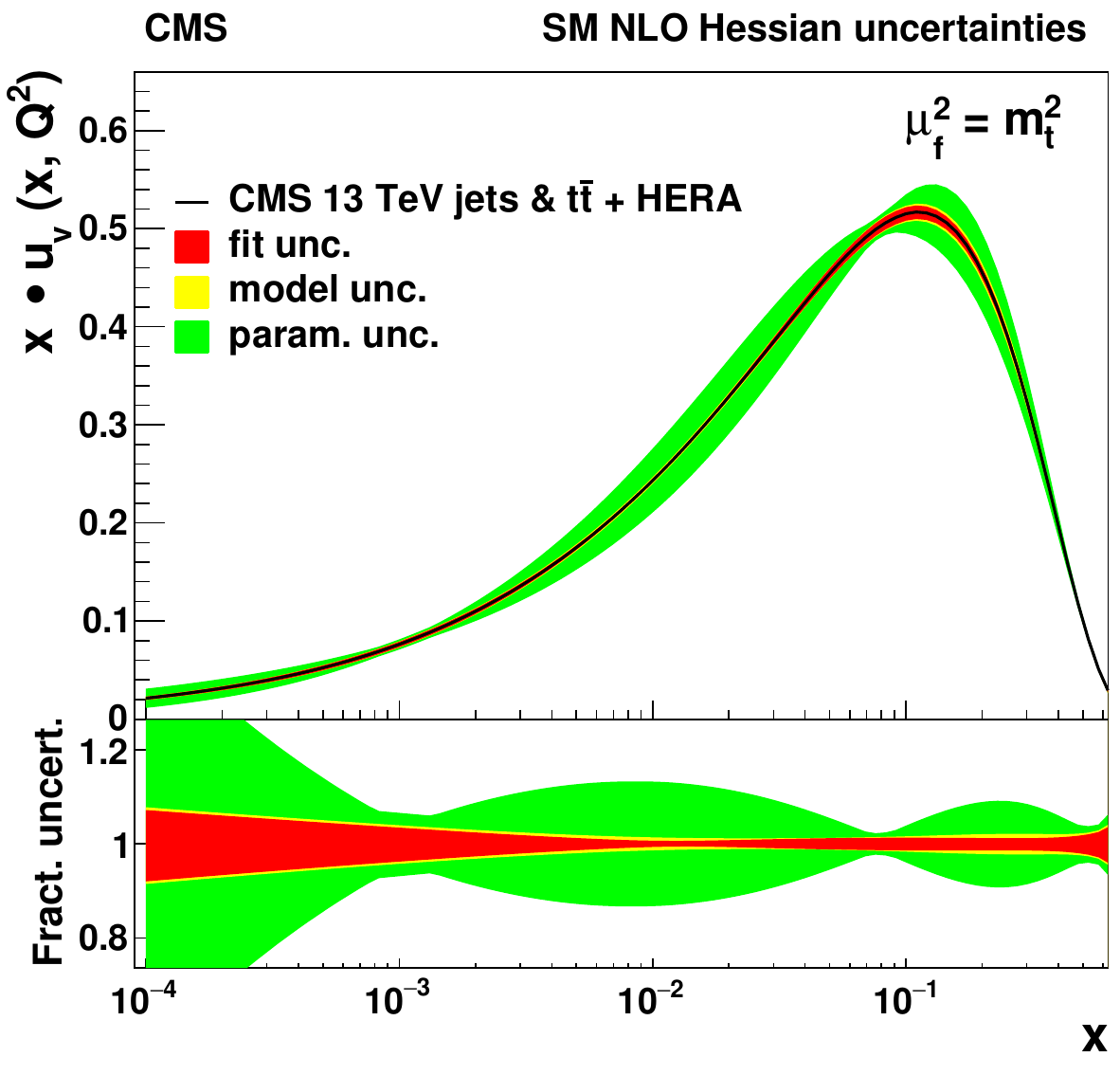}
\includegraphics[width=0.47\textwidth]{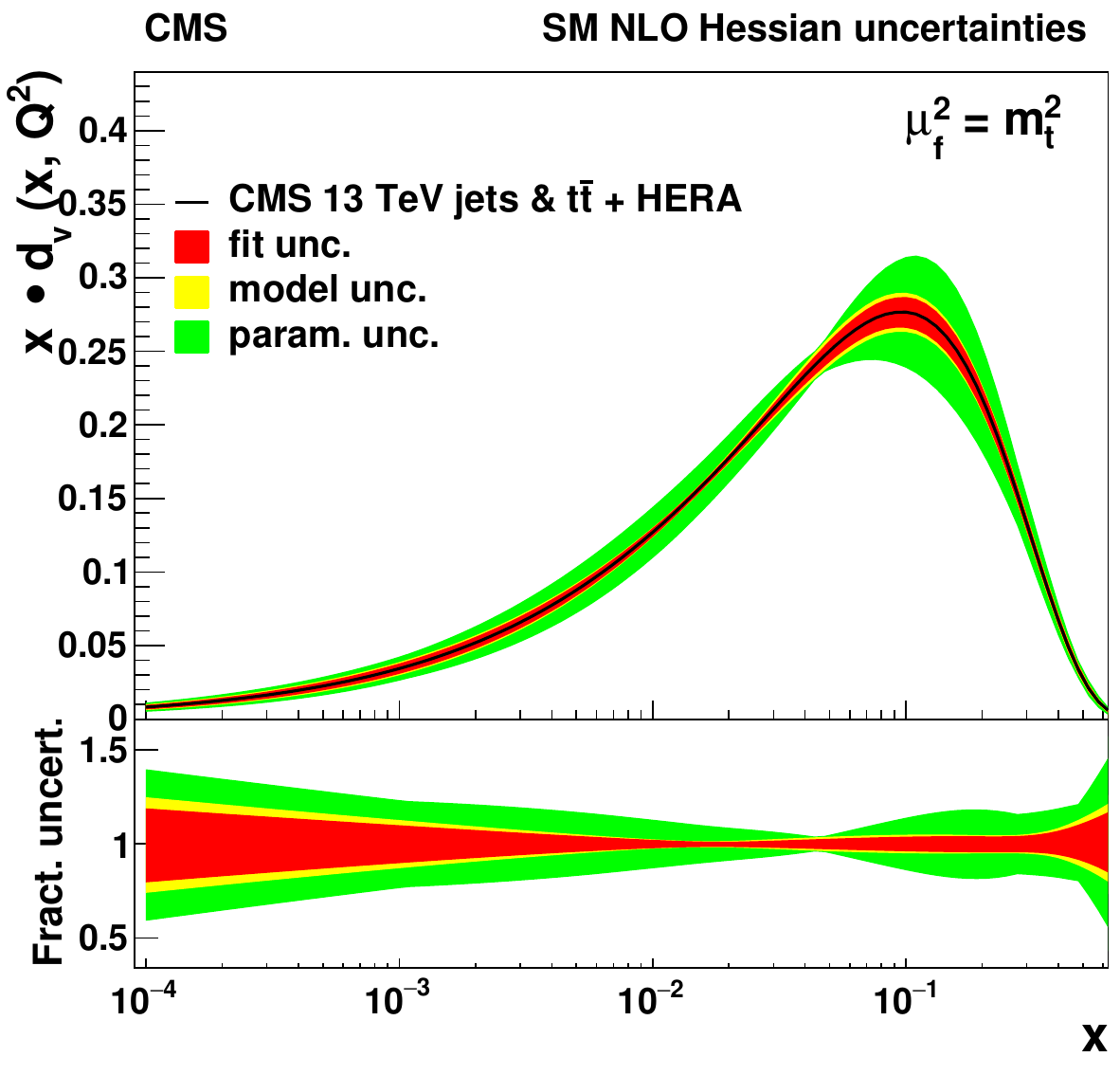}\\
\includegraphics[width=0.47\textwidth]{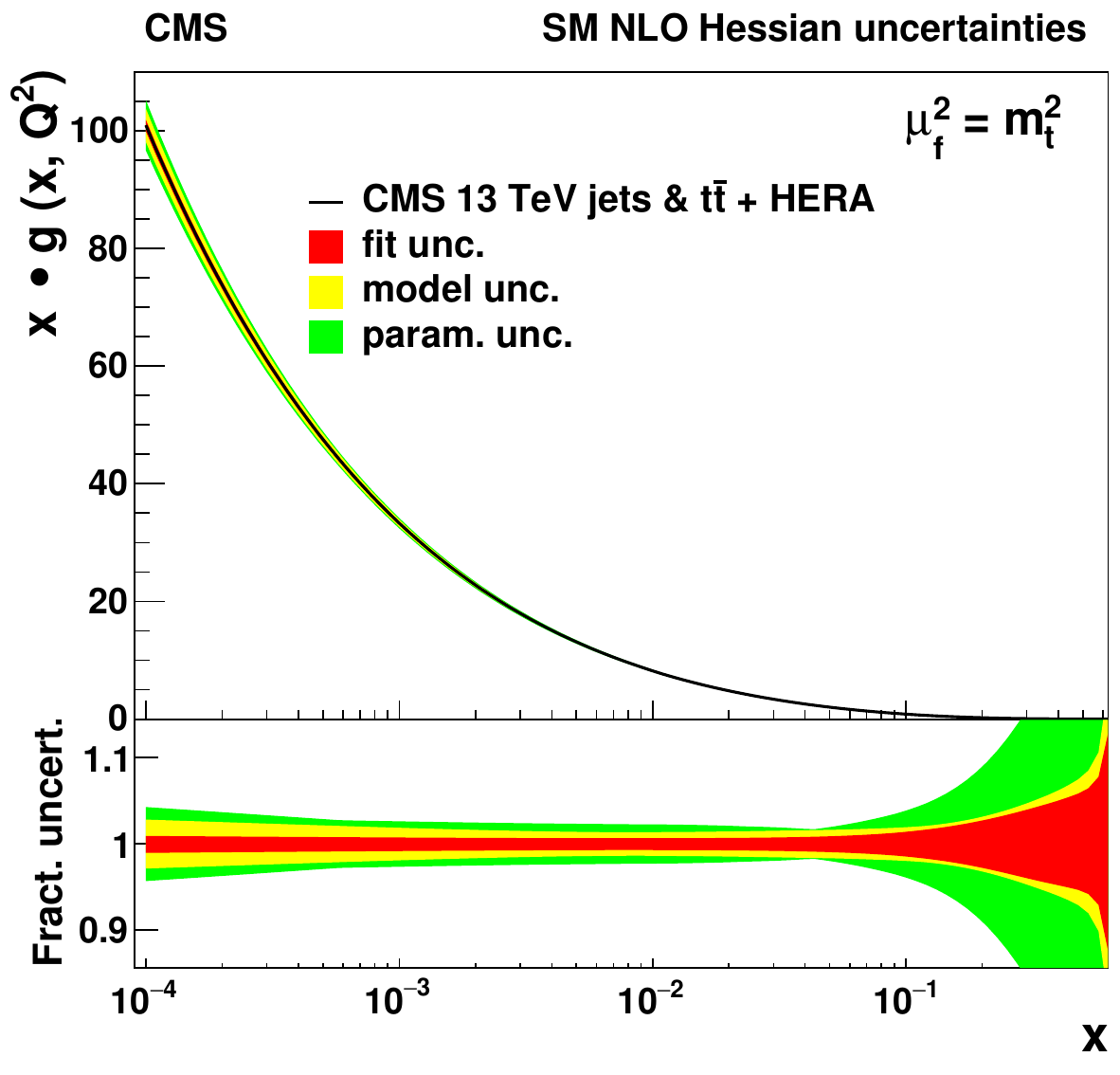}
\includegraphics[width=0.47\textwidth]{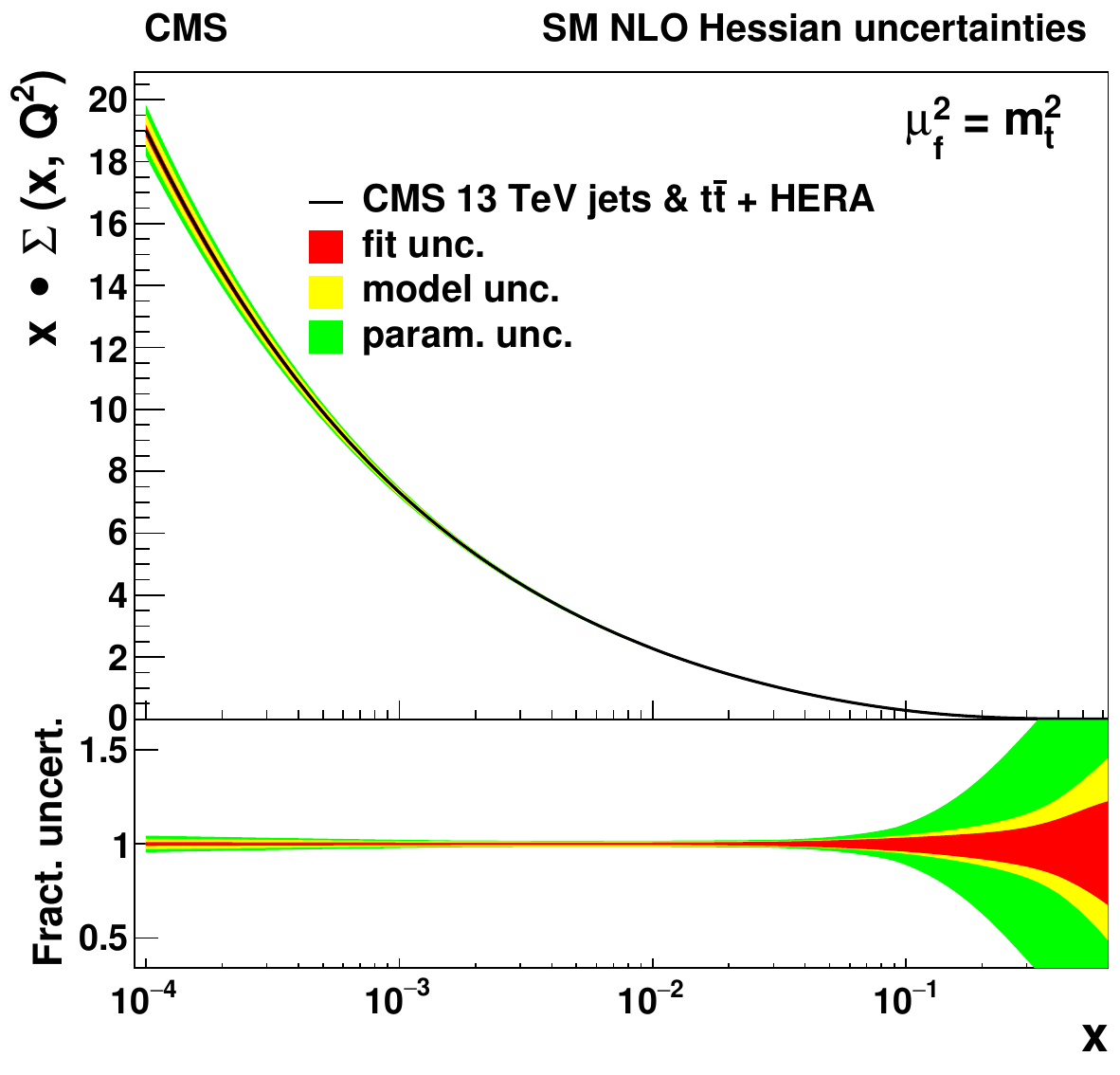}
\caption{The \PQu-valence (upper left), \PQd-valence (upper right), gluon (lower left), and sea quark (lower right) distributions, shown as functions of $x$ at the scale $\muf=\mt^2$, resulting 
from the SM fit using HERA DIS together with the CMS inclusive jet cross section and the normalised triple-differential cross section of \ttbar production at $\sqrt{s}=13\TeV$. Contributions of fit, model, and parameterisation uncertainties for each PDF are shown. In the lower panels, the relative uncertainty contributions are presented.}
\label{scan3_HERA+CMS_breakdown}
\end{figure}

The PDFs resulting from SM NLO fit are presented in Fig.~\ref{scan3_HERA+CMS_breakdown} demonstrating the contributions of the fit, model, and parameterisation uncertainties. The NLO values of \alpSZ and of \mtpole are determined simultaneously with the PDFs as 
$\alpSZ= 0.1188 \pm 0.0017\,\text{(fit)} \pm 0.0004\,\text{(model)} \pm 0.0025\,\text{(scale)} \pm 0.0001\,\text{(param.)}$, 
and 
$\mtpole = 170.4 \pm 0.6\,\text{(fit)} \pm 0.1\,\text{(model)} \pm 0.1\,\text{(scale)} \pm 0.1\,\text{(param.)}\GeV$. 
These are consistent with earlier CMS results~\cite{Khachatryan:2016mlc} and Ref.~\cite{Sirunyan:2019zvx}, respectively. The uncertainty in the value of \alpSZ is dominated by the scale variation, which is significantly larger than in the NNLO result of Eq.~\eqref{NNLO_alphas}. 

\begin{figure}[htbp!]
\centering
\includegraphics[width=0.47\textwidth]{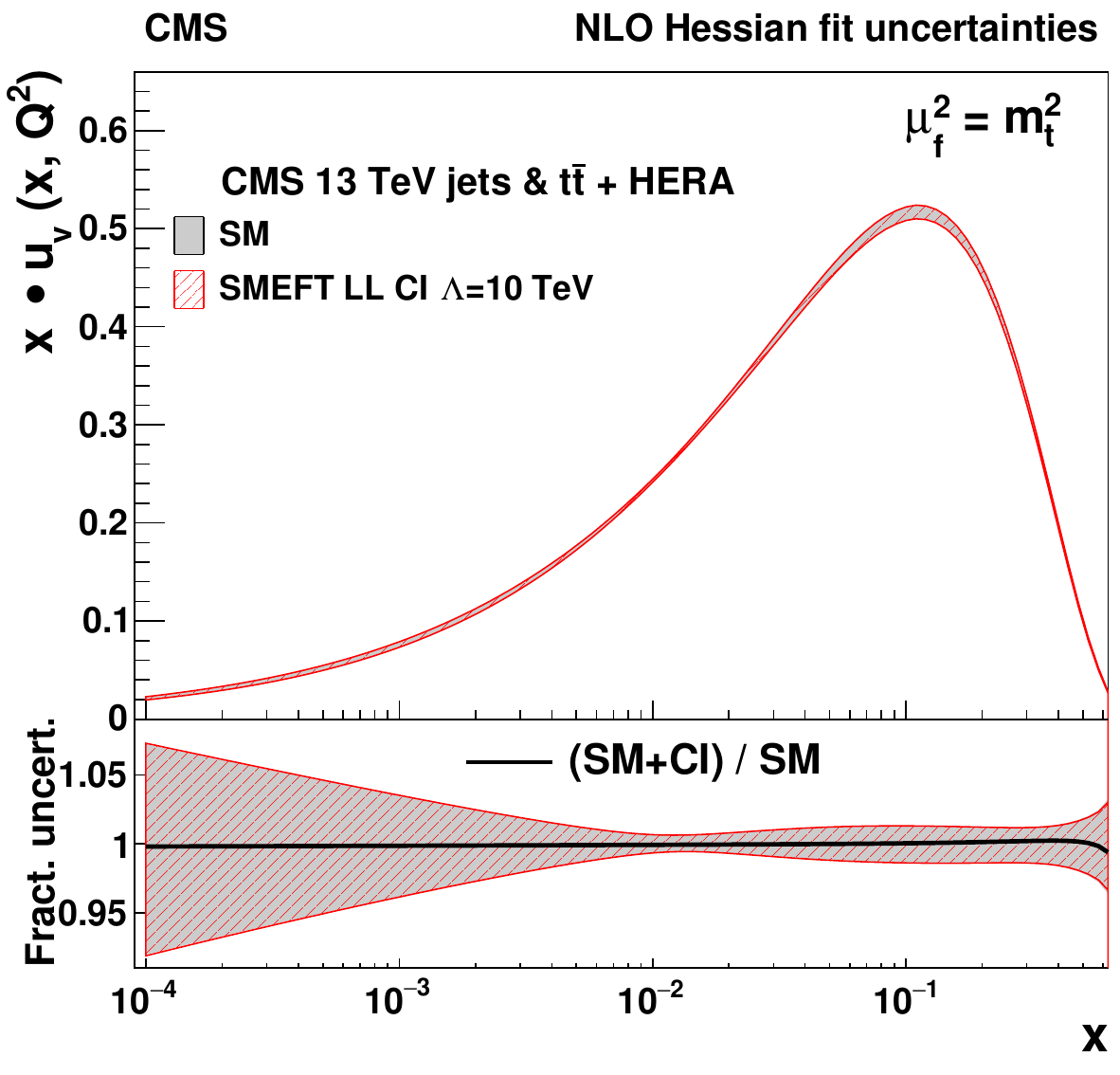}
\includegraphics[width=0.47\textwidth]{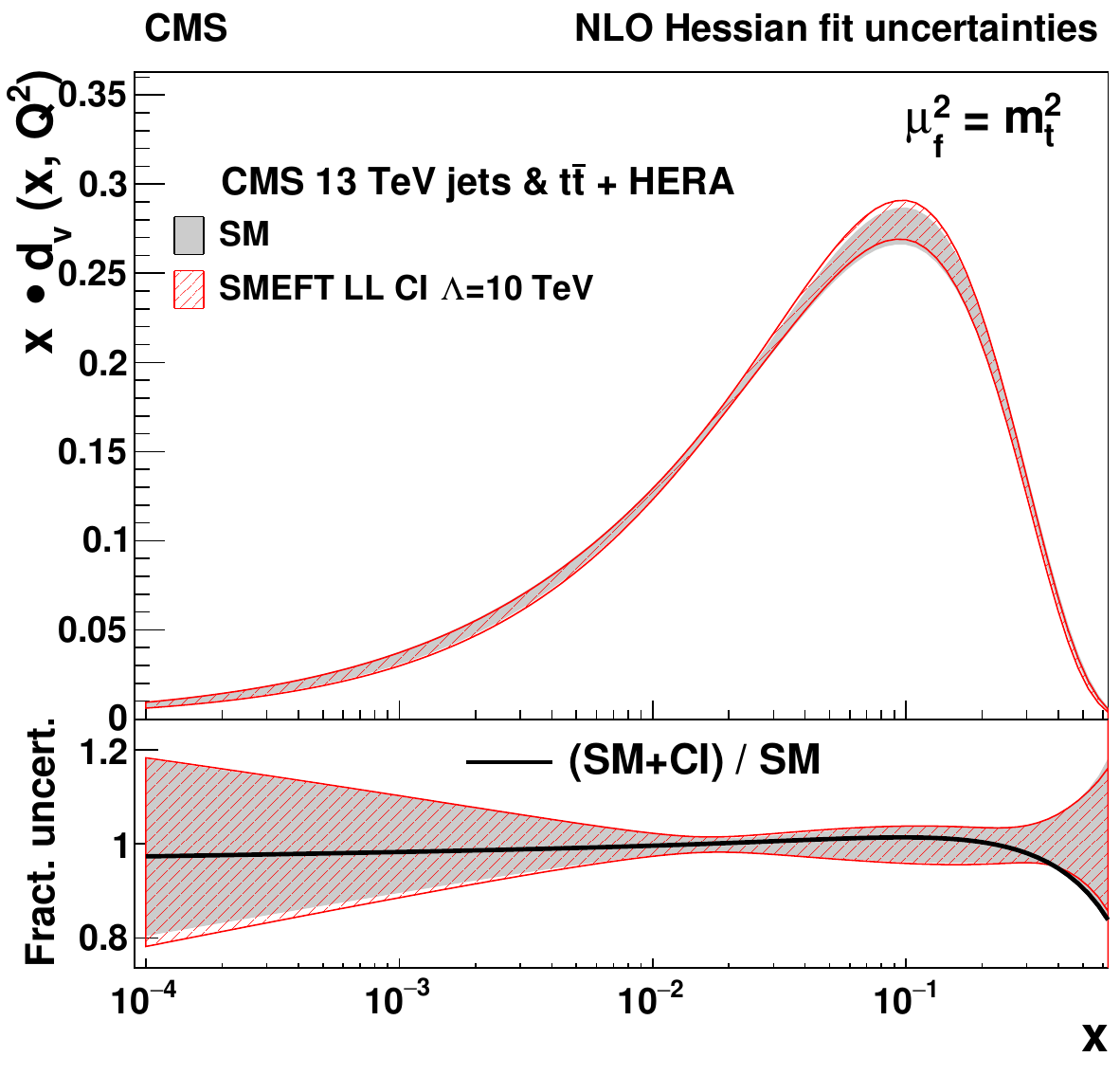}\\
\includegraphics[width=0.47\textwidth]{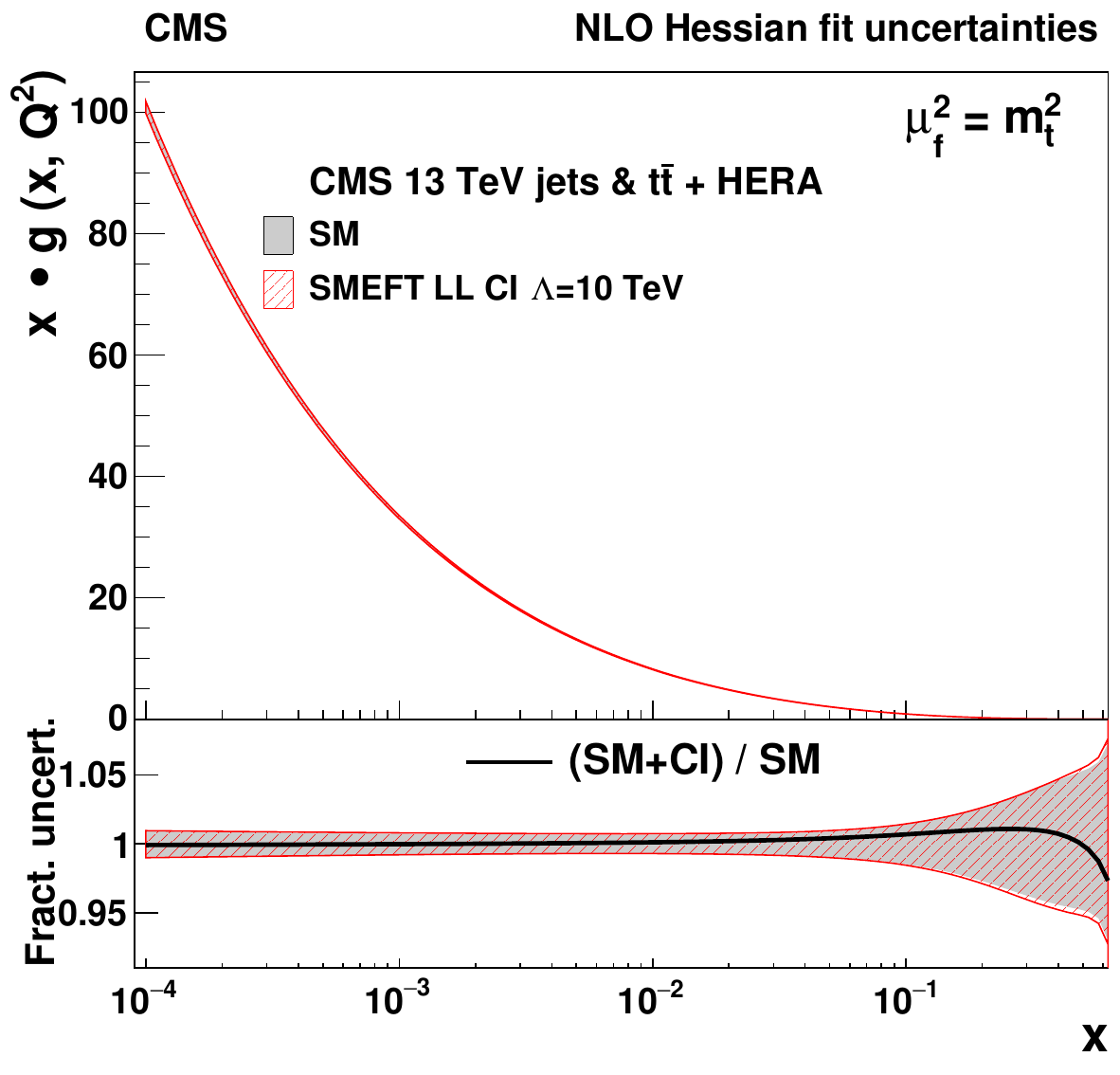}
\includegraphics[width=0.47\textwidth]{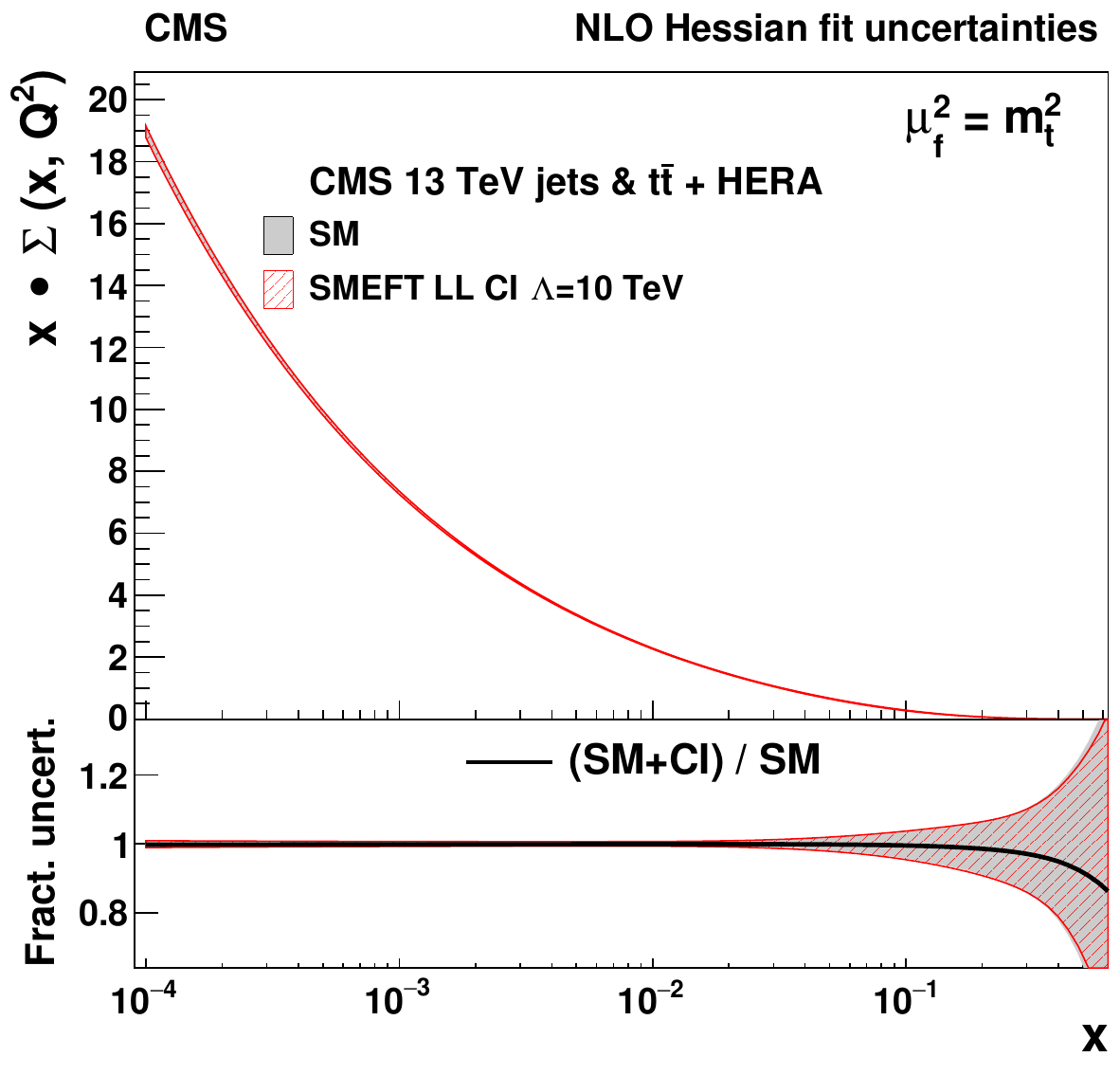}
\caption{The \PQu-valence (upper left), \PQd-valence (upper right), gluon (lower left), and sea quark (lower right) distributions, shown as functions of $x$ at the scale $\muf^2=\mt^2$, resulting from the fits with and without the CI terms. The SMEFT fit is performed with the left-handed CI model with $\Lambda=10\TeV$. }
\label{CI_QCD_analysis_PDFs}
\end{figure}

In the SMEFT fit, the Wilson coefficient $c_1$ is introduced as a new free parameter, assuming different values for the scale of the new interaction $\Lambda = 5$, 10, 13, 20, and 50\TeV. In all SMEFT fits, the $\chi^2$ is reduced by about 10, with just the addition of $c_1$ as an additional free parameter. Independent of the value of $\Lambda$, the strong coupling constant and the top quark mass in these SMEFT fits result to 
$\alpSZ = 0.1187 \pm 0.0016\,\text{(fit)} \pm 0.0005\,\text{(model)} \pm 0.0023\,\text{(scale)} \pm 0.0018\,\text{(param.)}$
, and 
$\mtpole = 170.4 \pm 0.6\,\text{(fit)} \pm 0.1\,\text{(model)} \pm 0.1\,\text{(scale)} \pm 0.2\,\text{(param.)}\GeV$. These values agree well with those obtained in the SM fit, and have larger parameterisation uncertainties because of increased flexibility in the SMEFT fit. 
The PDFs resulting from the SMEFT fits at different values of $\Lambda$ and for different CI models agree with each other and with the PDFs in the SM fit. 
In Fig.~\ref{CI_QCD_analysis_PDFs} the PDFs, which are obtained in the SM fit or in the SMEFT fit using the left-handed CI model, are compared. The PDFs are shown only with their fit uncertainty obtained by using the Hessian method.

\begin{figure}[htbp!]
\centering
\includegraphics[width=0.47\textwidth]{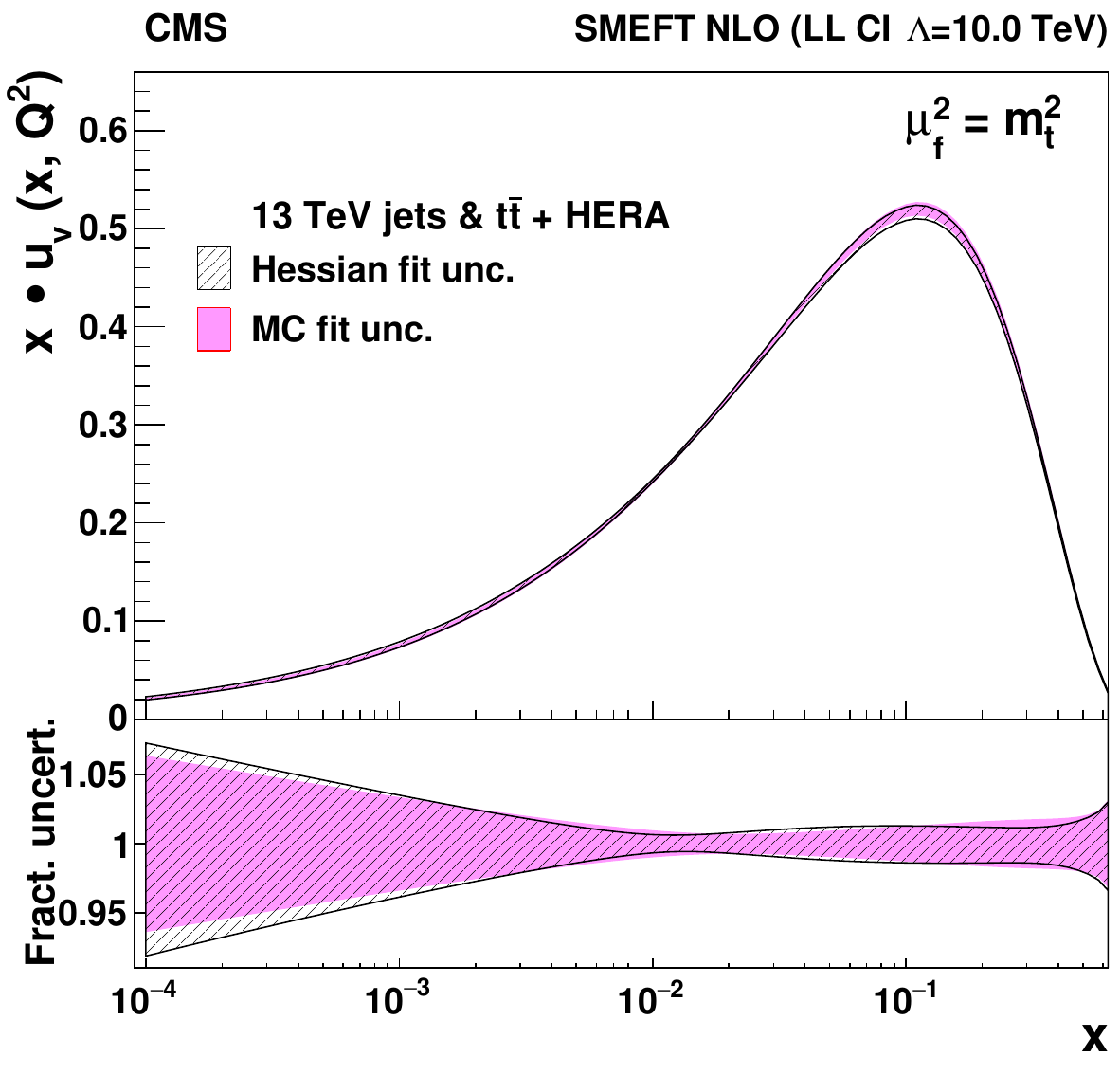}
\includegraphics[width=0.47\textwidth]{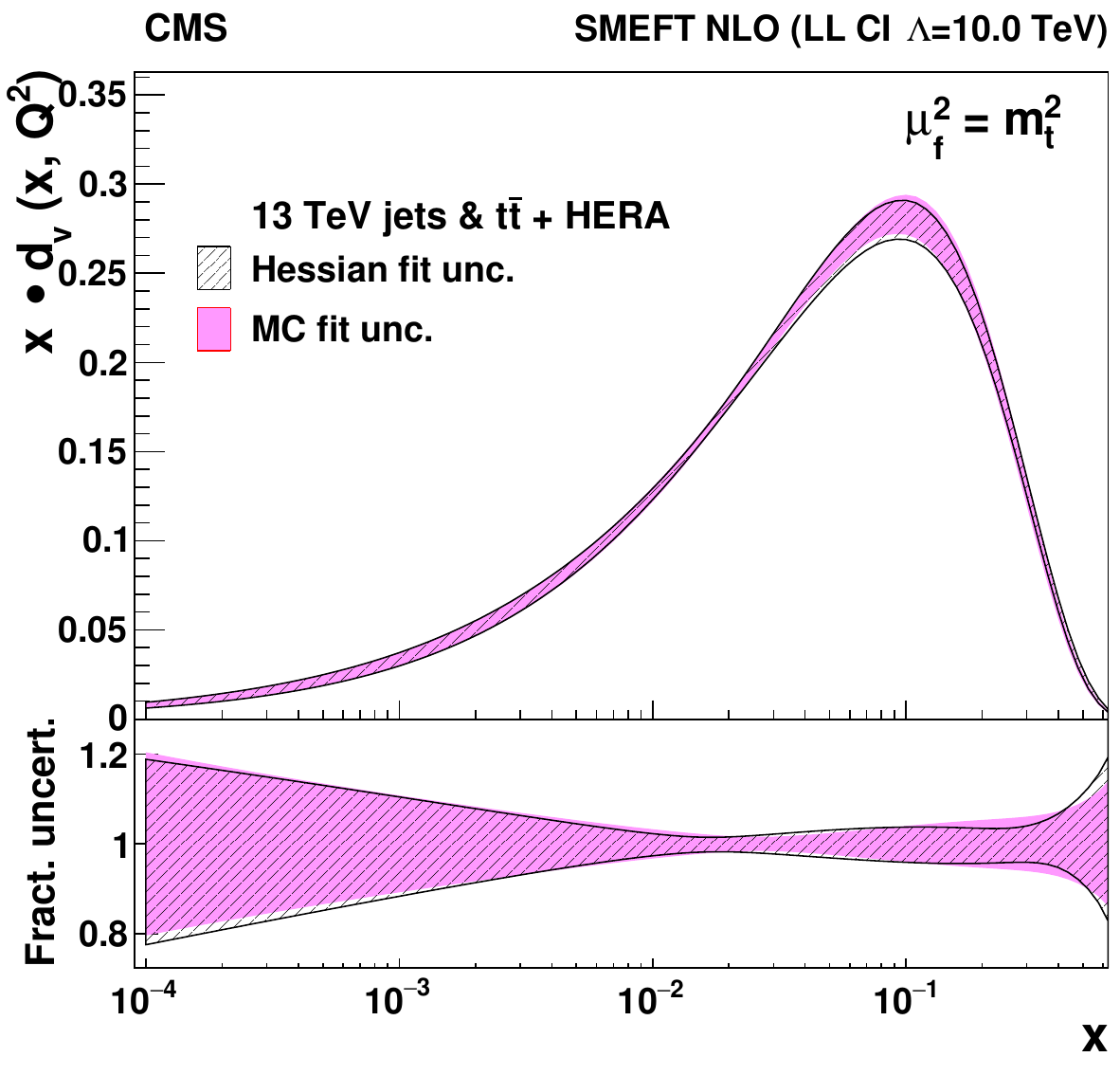}\\
\includegraphics[width=0.47\textwidth]{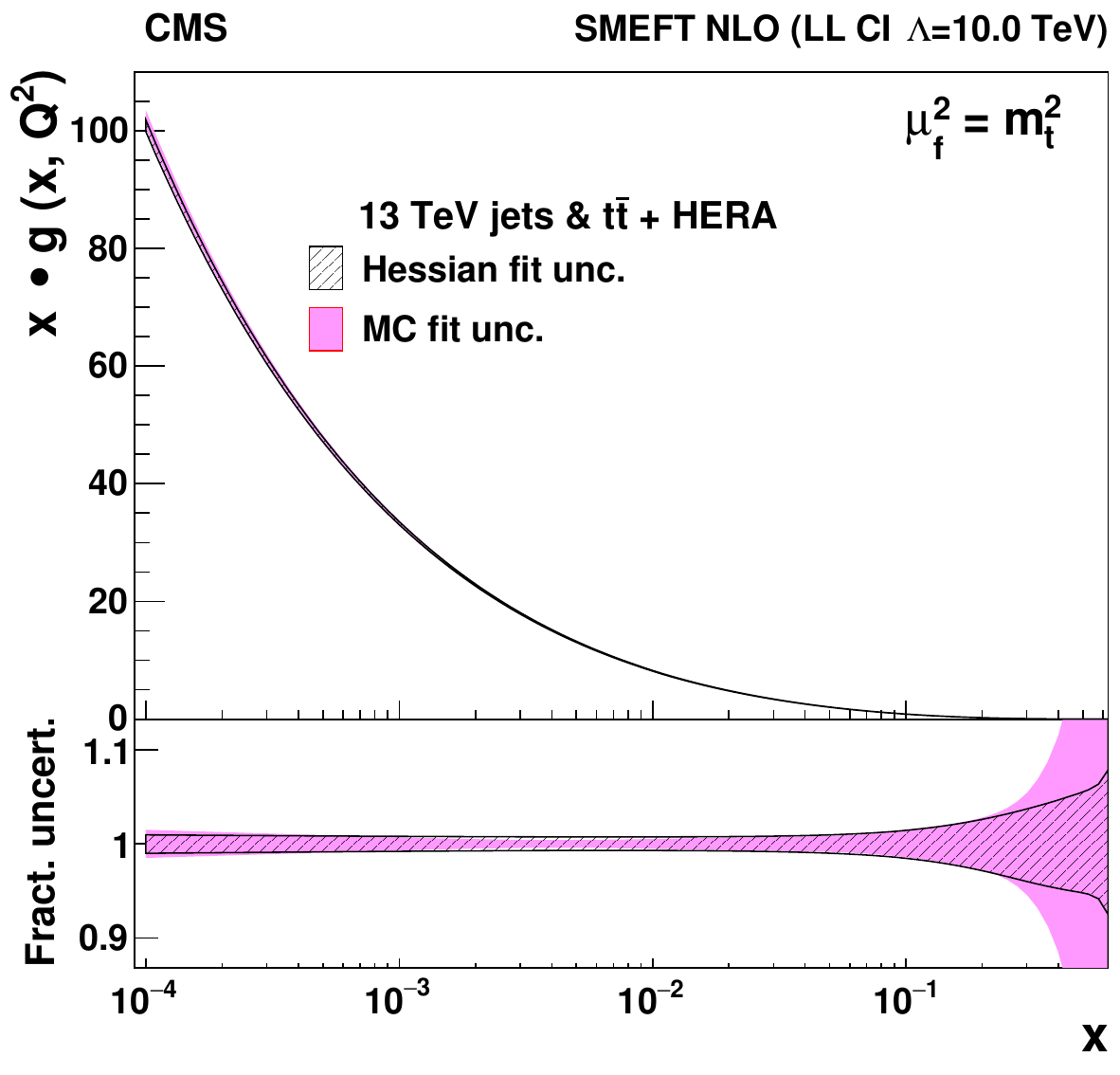}
\includegraphics[width=0.47\textwidth]{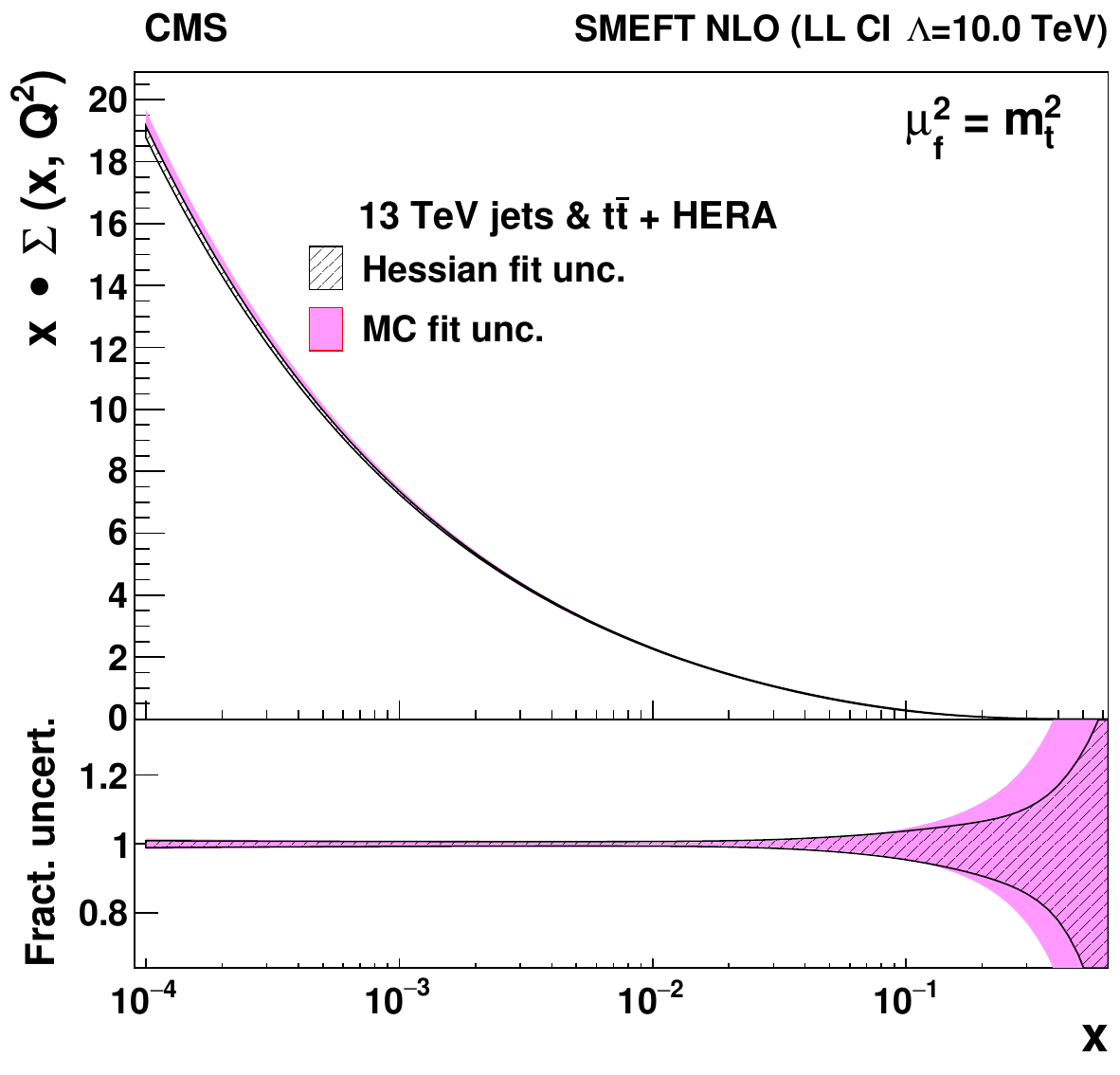}
\caption{The \PQu-valence (upper left), \PQd-valence (upper right), gluon (lower left), and sea quark (lower right) distributions, shown as a function of $x$ at the scale $\muf^2=\mt^2$, resulting from the SMEFT fit with the left-handed CI model with $\Lambda=10\TeV$. The PDFs are shown with the fit uncertainties obtained by the Hessian (solid blue) and Monte Carlo (solid red) methods.}
\label{CI_QCD_analysis_MC_PDFs}
\end{figure}

To account for possible non-Gaussian tails, the PDF uncertainties are alternatively obtained by using the MC method, based on 800 replicas. The Hessian and the MC uncertainties in the SMEFT fit are shown in Fig.~\ref{CI_QCD_analysis_MC_PDFs}. The uncertainties obtained by using the MC method are larger at high $x$, which might suggest non-Gaussian tails in the PDF uncertainties. However this is not reflected in the uncertainty in $c_1$ coefficients; the respective uncertainties obtained by Hessian or MC methods agree well.   

\begin{figure}[htp!]
\centering
\includegraphics[width=0.75\textwidth]{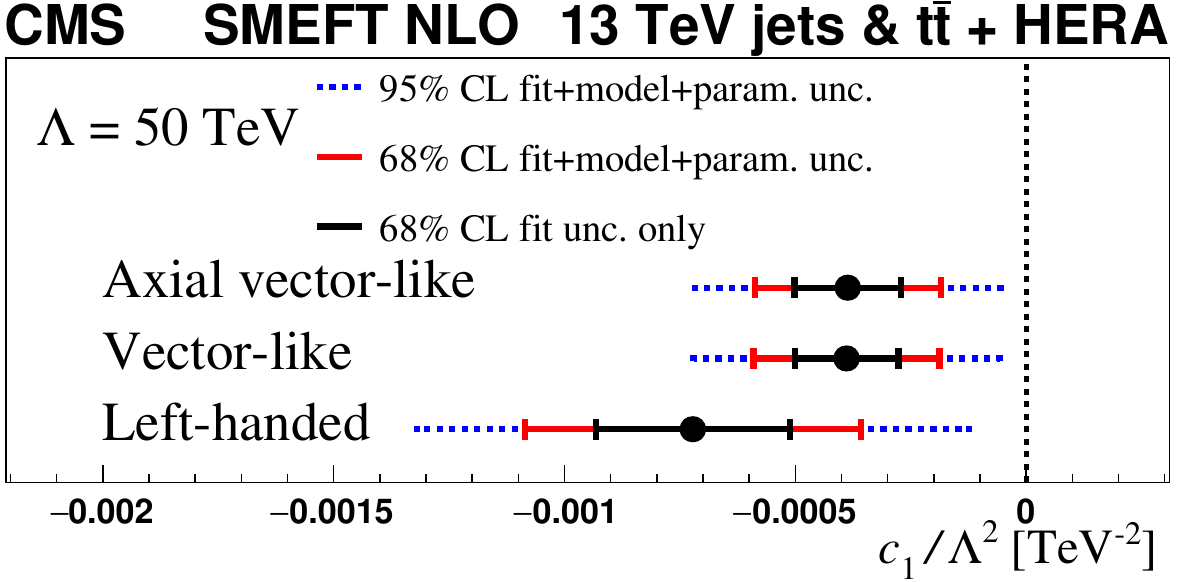}
\caption{The Wilson coefficients $c_1$ obtained in the SMEFT analysis at NLO, divided by $\Lambda^2$, for $\Lambda=50\TeV$. The solid (dashed) lines represent the total uncertainty at 68 (95)\% \CL. The inner (outer) error bars show the fit (total) uncertainty at 68\% \CL.}
\label{CI_fit}
\end{figure}

The Wilson coefficients $c_1$ are obtained for different assumptions on the value of $\Lambda$, as listed in Table~\ref{WilsonCoeffTab}. All SMEFT fits lead to negative $c_1$, which would translate into a positive interference with the SM gluon exchange. However, the differences from the SM ($c_1=0$) are not statistically significant. The ratio $c_1/\Lambda^2$ is illustrated for $\Lambda=50\TeV$ in Fig.~\ref{CI_fit} and is observed to remain constant for various values of $\Lambda$.

\begin{table}[htbp!]
\topcaption{The values and uncertainties of the fitted Wilson coefficients $c_1$ for various scales $\Lambda$. The fit uncertainties are obtained by using the Hessian method.}
\label{WilsonCoeffTab}
\centering
\begin{tabular}{c c c c c c c}
Scale            & CI model         & $c_1$    & Fit     & Model    & Scale   & Param.  \\
\hline
~                &Left-handed       &$ -0.017$ &$0.0047$ & $0.0001$ & $0.004$ & $0.002$ \\
$\Lambda=5\TeV$  &Vector-like       &$ -0.009$ &$0.0026$ & $0.0001$ & $0.002$ & $0.001$ \\
~                &Axial vector-like &$ -0.009$ &$0.0025$ & $0.0001$ & $0.002$ & $0.001$ \\[\cmsTabSkip]
~                &Left-handed       &$-0.068$  &$0.019$  & $0.003$  & $0.016$ & $0.009$ \\
$\Lambda=10\TeV$ &Vector-like       &$-0.037$  &$0.011$  & $0.002$  & $0.008$ & $0.006$ \\
~                &Axial vector-like &$-0.036$  &$0.011$  & $0.003$  & $0.008$ & $0.005$ \\[\cmsTabSkip]
~                &Left-handed       &$-0.116$  &$0.033$  & $0.006$  & $0.026$ & $0.015$ \\
$\Lambda=13\TeV$ &Vector-like       &$-0.063$  &$0.018$  & $0.004$  & $0.015$ & $0.008$ \\
~                &Axial vector-like &$-0.062$  &$0.018$  & $0.003$  & $0.014$ & $0.008$ \\[\cmsTabSkip]
~                &Left-handed       &$-0.28$   &$0.08$   & $0.01$   & $0.06$  & $0.04$  \\
$\Lambda=20\TeV$ &Vector-like       &$-0.15$   &$0.04$   & $0.01$   & $0.04$  & $0.02$  \\
~                &Axial vector-like &$-0.15$   &$0.04$   & $0.01$   & $0.04$  & $0.02$  \\[\cmsTabSkip]
~                &Left-handed       &$-1.8$    &$0.53$   & $0.08$   & $0.42$  & $0.23$  \\
$\Lambda=50\TeV$ &Vector-like       &$-1.0$    &$0.28$   & $0.05$   & $0.23$  & $0.13$  \\
~                &Axial vector-like &$-1.0$    &$0.29$   & $0.04$   & $0.23$  & $0.13$  \\
\end{tabular}
\end{table}

Conventional searches for CI fix the values of Wilson coefficient to +1 (-1) for a destructive (constructive) interference with the SM gluon exchange, and impose exclusion limits on the scale $\Lambda$~\cite{PDG2020}. The results obtained in the present analysis that indicate negative Wilson coefficients, with $\abs{c_1}$ close to 1 for $\Lambda=50\TeV$, can be translated into a 95\% \CL exclusion limit for the left-handed model with constructive interference, corresponding to $\Lambda>24\TeV$. The most stringent comparable result is obtained in the analysis of dijet cross section at $\sqrt{s}=13\TeV$ by the ATLAS Collaboration~\cite{Aaboud:2017yvp}, in which the 95\% \CL exclusion limits for purely left-handed CI of 22\TeV for constructive interference, and of 30\TeV for destructive interference are obtained.

As already observed in the results of profiling, the results of the full fit show agreement between the measurements and the SM prediction. 
Since the parameterisation and the PDF uncertainties differ in the profiling analysis and in the full fit, a direct quantitative comparison of these results is not possible. The advantage of the full fit with respect to the profiling is in the properly considered, and therefore mitigated, correlations between the QCD parameters and the PDFs. The full SMEFT fit assures that the possible BSM effects are not absorbed in the PDFs and in turn, into the SM prediction, which is the basis for the search for new physics. 

\section{Summary}

\label{sec:summary}

In this paper, the measurement of the double-differential inclusive jet cross sections in proton-proton collisions at $\sqrt{s}=13\TeV$ is presented as a function of the jet transverse momentum $\pt$ and the jet rapidity $\abs{y}$ for jets reconstructed using the anti-$\kt$ clustering algorithm with a distance parameter $R$ of $0.4$ and $0.7$. The phase space covers jet \pt from 97\GeV up to 3.1\TeV and jet rapidity up to $\abs{y}=2.0$.
The measured jet cross sections are compared with predictions of perturbative quantum chromodynamics (pQCD) at next-to-next-to-leading order (NNLO) and next-to-leading order (NLO) with the next-to-leading-logarithmic (NLL) resummation correction, using various sets of parton distribution functions (PDFs).
A strong impact of the measurement on determination of the parton distributions is observed, expressed by significant differences among the theoretical predictions using different PDF sets, and by large corresponding uncertainties. 

To investigate the impact of the measurements on the PDFs and the strong coupling constant \alpS, a QCD analysis is performed, where the jet production cross section with $R=0.7$ is used together with the HERA measurements of deep inelastic scattering. Significant improvement of the accuracy of the PDFs by using the present measurement in the QCD analysis is demonstrated in a profiling analysis using the CT14 PDF set and in the full PDF fit. 

The value of the strong coupling constant at the \PZ boson mass is extracted in a QCD analysis at NNLO using the inclusive jet cross sections in proton-proton collisions for the first time, and results in 
$\alpSZ= 0.1170 \pm 0.0014\,\text{(fit)} \pm 0.0007\,\text{(model)} \pm 0.0008\,\text{(scale)} \pm 0.0001\,\text{(parametrisation)}$.

The QCD analysis is also performed at NLO, where the CMS measurement of the normalised triple-differential top quark-antiquark production cross section at $\sqrt{s}=13\TeV$ is used in addition. In this analysis, the PDFs, the values of the strong coupling constant, and of the top quark pole mass \mtpole are extracted simultaneously with 
$\alpSZ = 0.1188 \pm 0.0017\,\text{(fit)} \pm 0.0004\,\text{(model)} \pm 0.0025\,\text{(scale)} \pm 0.0001\,\text{(parameterisation)}$, 
dominated by the scale uncertainty, and 
$\mtpole = 170.4 \pm 0.6\,\text{(fit)} \pm 0.1\,\text{(model)} \pm 0.1\,\text{(scale)} \pm 0.1\,\text{(parameterisation)}\GeV$.
The resulting values of \alpSZ agree with the world average and the previous CMS results using the jet measurements. The value of \mtpole agrees well with the result of the previous CMS analysis using the triple-differential cross section of the top quark-antiquark pair production. Although the inclusive jet production is not directly sensitive to \mtpole, the resulting value is improved by the additional constraint on the gluon distribution and on \alpSZ provided by the jet measurements.

Furthermore, an alternative QCD analysis is performed with the same data, where the standard model Lagrangian is modified by the 
introduction of effective terms related to 4-quark contact interactions. In the analysis, the Wilson coefficients for the contact interactions are extracted for different values assumed for the scale $\Lambda$ of the new interaction. The results are translated into a 95\% confidence level exclusion limit for the left-handed model with constructive interference, corresponding to $\Lambda>24\TeV$. These results are compatible with the standard model and the previous limits obtained at the LHC using jet production. The advantage of the present approach is the simultaneous extraction of PDFs, thereby mitigating possible bias in the interpretation of the measurements in terms of physics beyond the standard model. 

Tabulated results are provided in the HEPData record for this analysis~\cite{HEPdata}.

\begin{acknowledgments}

    We congratulate our colleagues in the CERN accelerator departments for the excellent performance of the LHC and thank the technical and administrative staffs at CERN and at other CMS institutes for their contributions to the success of the CMS effort. In addition, we gratefully acknowledge the computing centres and personnel of the Worldwide LHC Computing Grid and other centres for delivering so effectively the computing infrastructure essential to our analyses. Finally, we acknowledge the enduring support for the construction and operation of the LHC, the CMS detector, and the supporting computing infrastructure provided by the following funding agencies: BMBWF and FWF (Austria); FNRS and FWO (Belgium); CNPq, CAPES, FAPERJ, FAPERGS, and FAPESP (Brazil); MES and BNSF (Bulgaria); CERN; CAS, MoST, and NSFC (China); MINCIENCIAS (Colombia); MSES and CSF (Croatia); RIF (Cyprus); SENESCYT (Ecuador); MoER, ERC PUT and ERDF (Estonia); Academy of Finland, MEC, and HIP (Finland); CEA and CNRS/IN2P3 (France); BMBF, DFG, and HGF (Germany); GSRI (Greece); NKFIA (Hungary); DAE and DST (India); IPM (Iran); SFI (Ireland); INFN (Italy); MSIP and NRF (Republic of Korea); MES (Latvia); LAS (Lithuania); MOE and UM (Malaysia); BUAP, CINVESTAV, CONACYT, LNS, SEP, and UASLP-FAI (Mexico); MOS (Montenegro); MBIE (New Zealand); PAEC (Pakistan); MSHE and NSC (Poland); FCT (Portugal); JINR (Dubna); MON, RosAtom, RAS, RFBR, and NRC KI (Russia); MESTD (Serbia); SEIDI, CPAN, PCTI, and FEDER (Spain); MOSTR (Sri Lanka); Swiss Funding Agencies (Switzerland); MST (Taipei); ThEPCenter, IPST, STAR, and NSTDA (Thailand); TUBITAK and TAEK (Turkey); NASU (Ukraine); STFC (United Kingdom); DOE and NSF (USA).

    \hyphenation{Rachada-pisek} Individuals have received support from the Marie-Curie programme and the European Research Council and Horizon 2020 Grant, contract Nos.\ 675440, 724704, 752730, 758316, 765710, 824093, 884104, and COST Action CA16108 (European Union); the Leventis Foundation; the Alfred P.\ Sloan Foundation; the Alexander von Humboldt Foundation; the Belgian Federal Science Policy Office; the Fonds pour la Formation \`a la Recherche dans l'Industrie et dans l'Agriculture (FRIA-Belgium); the Agentschap voor Innovatie door Wetenschap en Technologie (IWT-Belgium); the F.R.S.-FNRS and FWO (Belgium) under the ``Excellence of Science -- EOS" -- be.h project n.\ 30820817; the Beijing Municipal Science \& Technology Commission, No. Z191100007219010; the Ministry of Education, Youth and Sports (MEYS) of the Czech Republic; the Deutsche Forschungsgemeinschaft (DFG), under Germany's Excellence Strategy -- EXC 2121 ``Quantum Universe" -- 390833306, and under project number 400140256 - GRK2497; the Lend\"ulet (``Momentum") Programme and the J\'anos Bolyai Research Scholarship of the Hungarian Academy of Sciences, the New National Excellence Program \'UNKP, the NKFIA research grants 123842, 123959, 124845, 124850, 125105, 128713, 128786, and 129058 (Hungary); the Council of Science and Industrial Research, India; the Latvian Council of Science; the Ministry of Science and Higher Education and the National Science Center, contracts Opus 2014/15/B/ST2/03998 and 2015/19/B/ST2/02861 (Poland); the Funda\c{c}\~ao para a Ci\^encia e a Tecnologia, grant CEECIND/01334/2018 (Portugal); the National Priorities Research Program by Qatar National Research Fund; the Ministry of Science and Higher Education, projects no. 14.W03.31.0026 and no. FSWW-2020-0008, and the Russian Foundation for Basic Research, project No.19-42-703014 (Russia); the Programa Estatal de Fomento de la Investigaci{\'o}n Cient{\'i}fica y T{\'e}cnica de Excelencia Mar\'{\i}a de Maeztu, grant MDM-2015-0509 and the Programa Severo Ochoa del Principado de Asturias; the Stavros Niarchos Foundation (Greece); the Rachadapisek Sompot Fund for Postdoctoral Fellowship, Chulalongkorn University and the Chulalongkorn Academic into Its 2nd Century Project Advancement Project (Thailand); the Kavli Foundation; the Nvidia Corporation; the SuperMicro Corporation; the Welch Foundation, contract C-1845; and the Weston Havens Foundation (USA).
\end{acknowledgments}

\bibliography{auto_generated}
\clearpage
\appendix
\numberwithin{table}{section}
\numberwithin{figure}{section}
\section{Supplemental material: comparison to NLO}\label{app:suppMat}

\begin{figure}[ht]
    \includegraphics[width=0.5\textwidth,page=1]{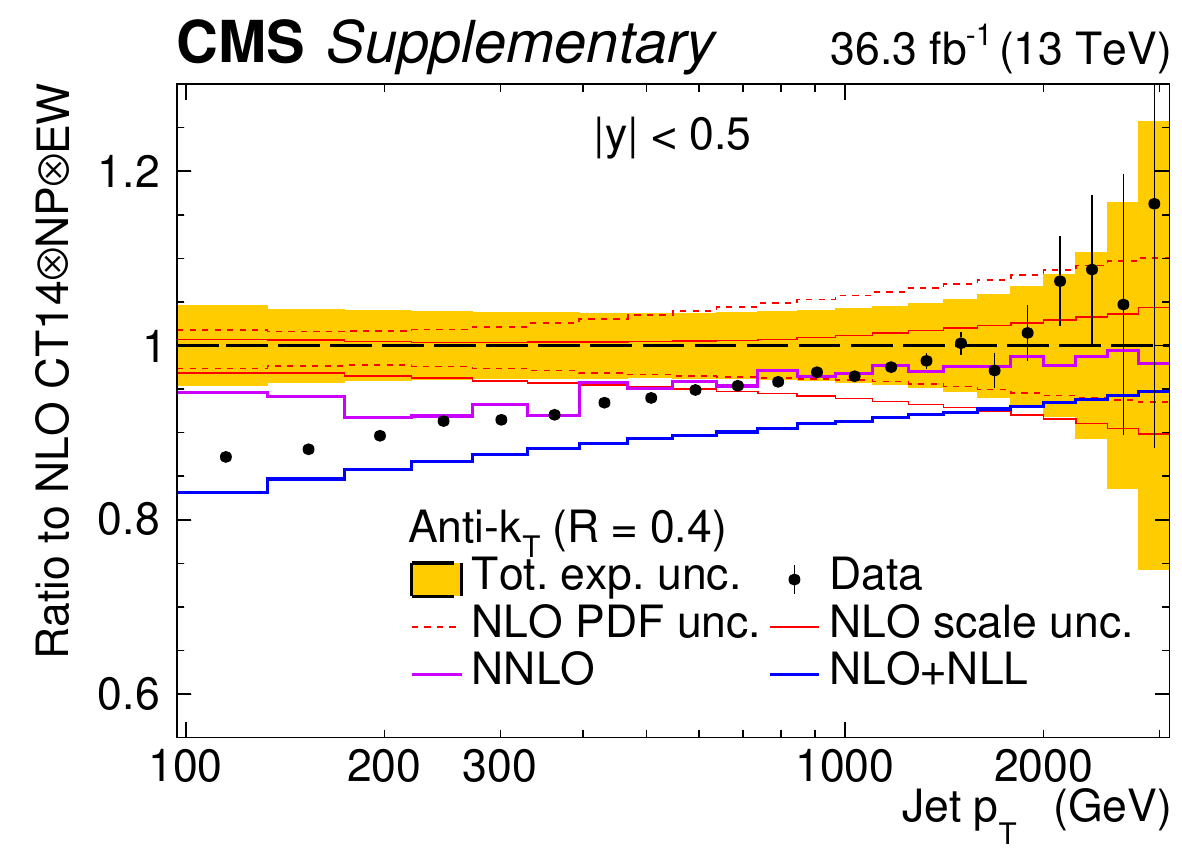}
    \includegraphics[width=0.5\textwidth,page=2]{Figure_A_001.pdf}
    
    \includegraphics[width=0.5\textwidth,page=3]{Figure_A_001.pdf}
    \includegraphics[width=0.5\textwidth,page=4]{Figure_A_001.pdf}

    \caption[Ratio to predictions at different orders ($R=0.4$)]{Cross sections of inclusive jet production for distance parameter $R=0.4$ as a function of transverse momentum of the individual jet in bins
of absolute rapidity $\abs{y}$, compared to the theoretical predictions at NLO, NLO+NLL, and NNLO. All results are normalised to the prediction at NLO. The measurement (solid symbols) is presented with the statistical uncertainties (vertical error bars), while the systematic uncertainty is represented by a yellow filled band, centered at 1. The NLO (black dashed line) and NLO+NLL (blue solid line) predictions are obtained using CT14nlo PDF. The PDF (dotted red line) and scale (solid red line) uncertainties are shown for the NLO prediction. The NNLO calculation (purple solid line) is obtained using CT14nnlo PDF.}
    \label{suppfig:16ak4}
\end{figure}

\begin{figure}
    \includegraphics[width=0.5\textwidth,page=1]{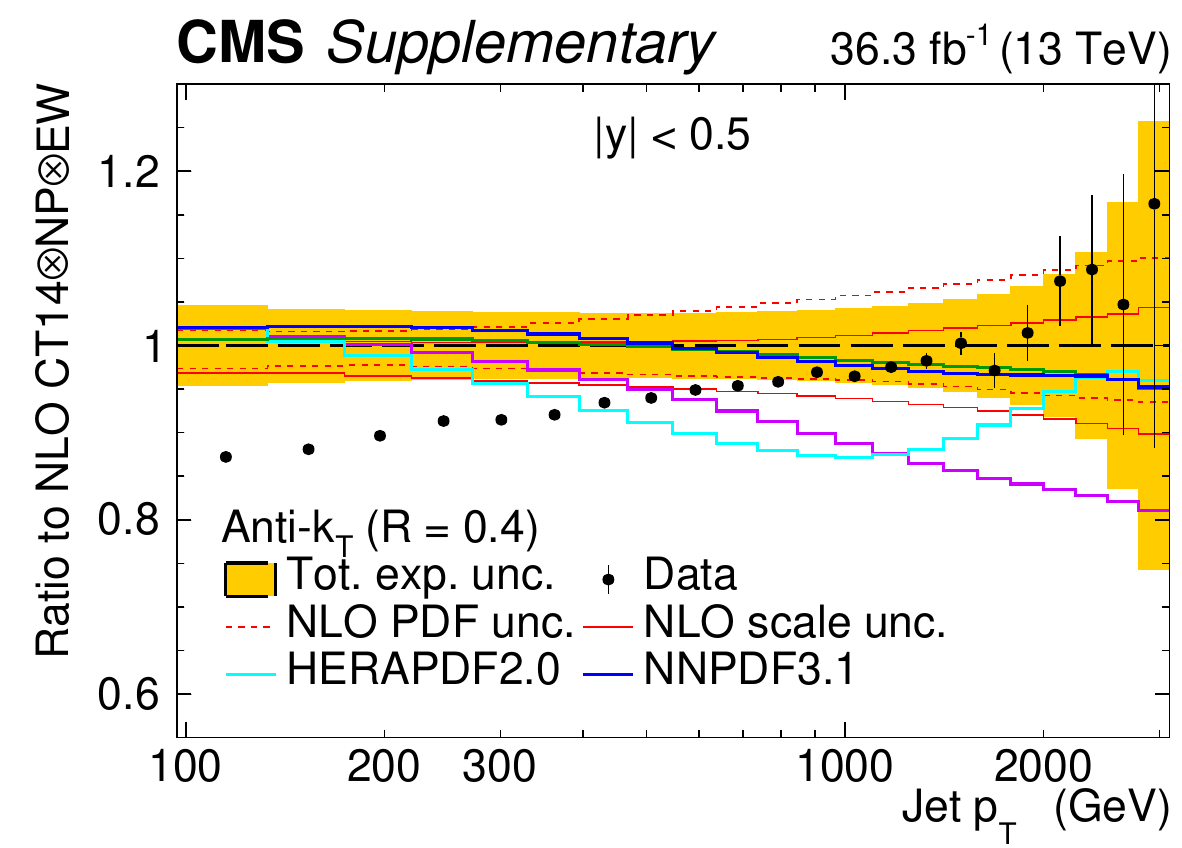}
    \includegraphics[width=0.5\textwidth,page=2]{Figure_A_002.pdf}
                                                                 
    \includegraphics[width=0.5\textwidth,page=3]{Figure_A_002.pdf}
    \includegraphics[width=0.5\textwidth,page=4]{Figure_A_002.pdf}

    \caption[Ratio to NLO predictions with different PDFs ($R=0.4$)]{Cross sections of inclusive jet production for distance parameter $R=0.4$ as a function of transverse momentum of the individual jet in bins
of absolute rapidity $\abs{y}$, compared to the theoretical predictions at NLO using different PDFs (lines of different colors). All results are normalised to the prediction at NLO obtained using CT14nlo PDF (black dashed line). The measurement (solid symbols) is presented with the statistical uncertainties (vertical error bars), while the systematic uncertainty is represented by a yellow filled band, centered at 1. 
    }
    \label{suppfig:16ak4pdf}
\end{figure}

\begin{figure}
    \includegraphics[width=0.5\textwidth,page=1]{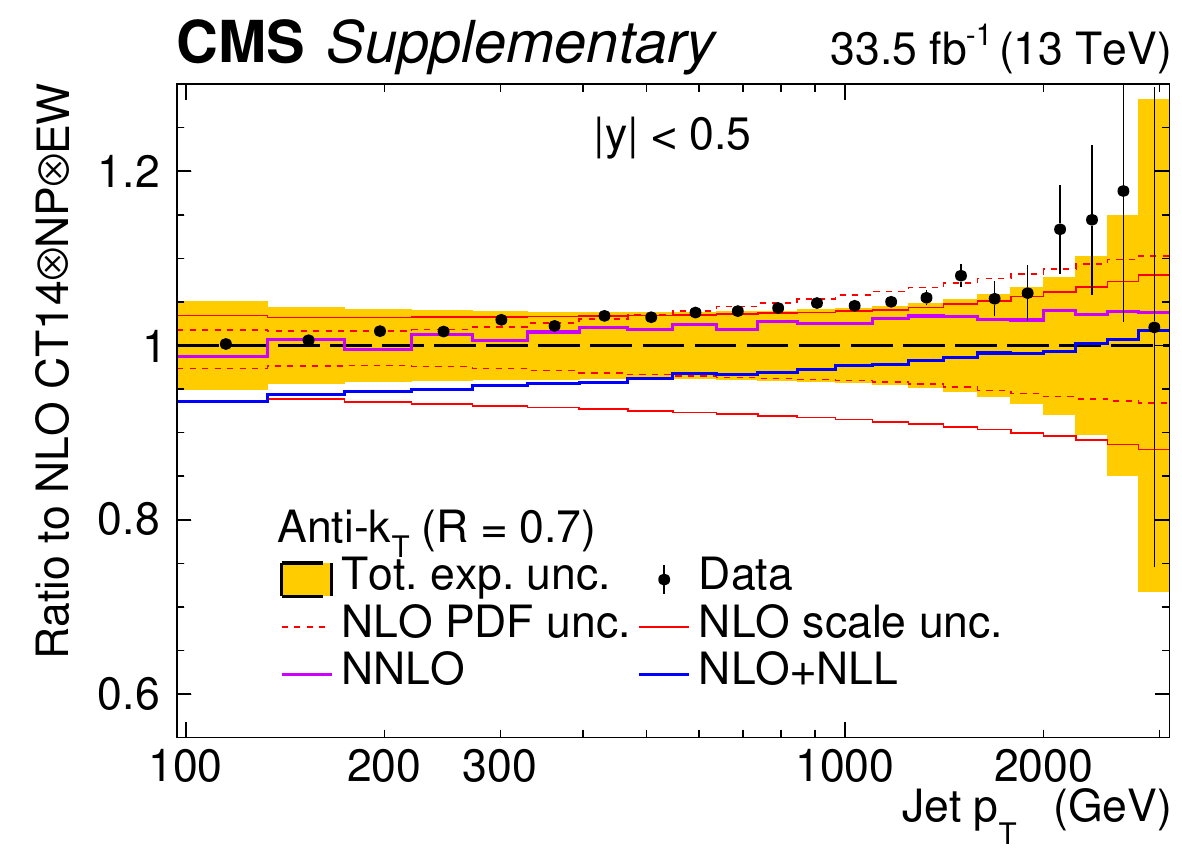}
    \includegraphics[width=0.5\textwidth,page=2]{Figure_A_003.pdf}

    \includegraphics[width=0.5\textwidth,page=3]{Figure_A_003.pdf}
    \includegraphics[width=0.5\textwidth,page=4]{Figure_A_003.pdf}

    \caption[Ratio to predictions at different orders ($R=0.7$)]{Same as Fig.~\ref{suppfig:16ak4} for the distance parameter $R=0.7$.
    }
    \label{suppfig:16ak7}
\end{figure}

\begin{figure}
    \includegraphics[width=0.5\textwidth,page=1]{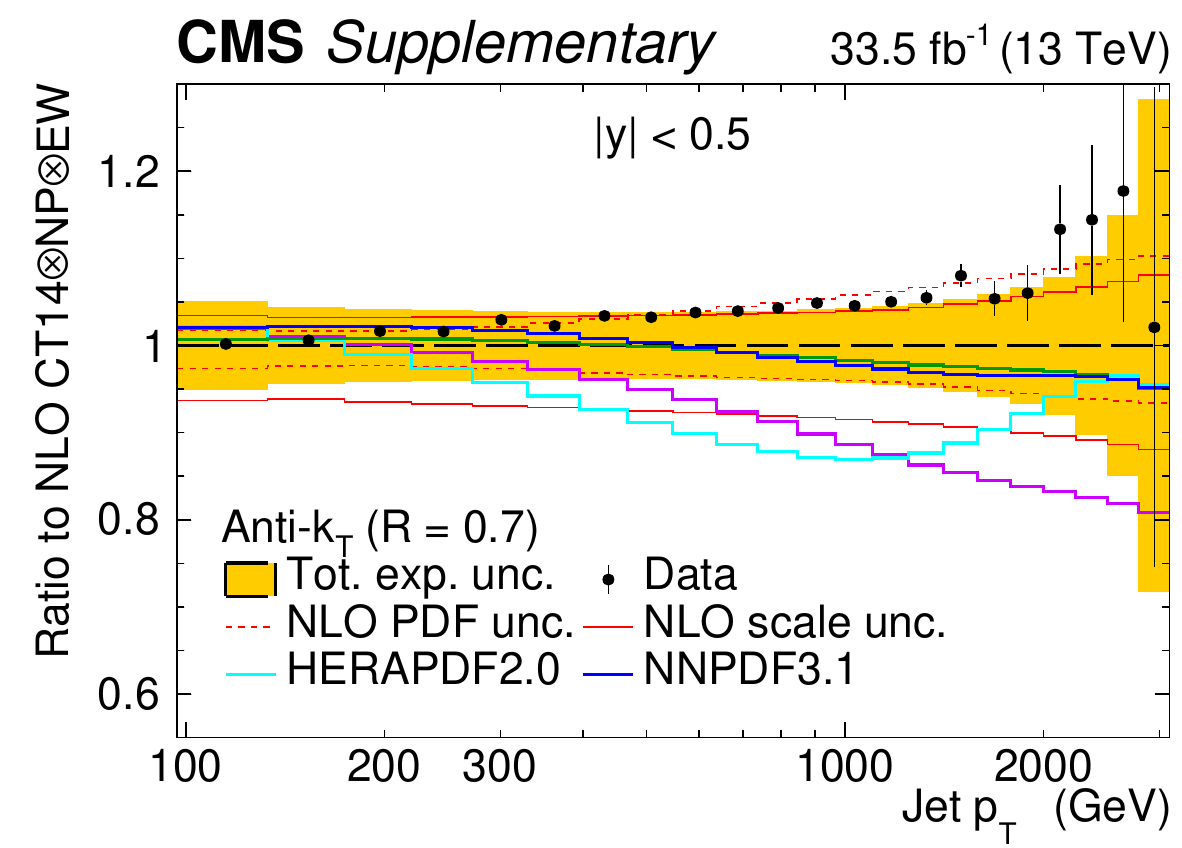}
    \includegraphics[width=0.5\textwidth,page=2]{Figure_A_004.pdf}
                                                                 
    \includegraphics[width=0.5\textwidth,page=3]{Figure_A_004.pdf}
    \includegraphics[width=0.5\textwidth,page=4]{Figure_A_004.pdf}

    \caption[Ratio to NLO predictions with different PDFs ($R=0.7$)]{Same as Fig.~\ref{suppfig:16ak4pdf} for the distance parameter $R=0.7$.}
    \label{suppfig:16ak7pdf}
\end{figure}

\clearpage
\section{Addendum: QCD analysis at NNLO using NNLO interpolation grids}

\begin{figure}[ht]
    \centering
    \includegraphics[width=\textwidth]{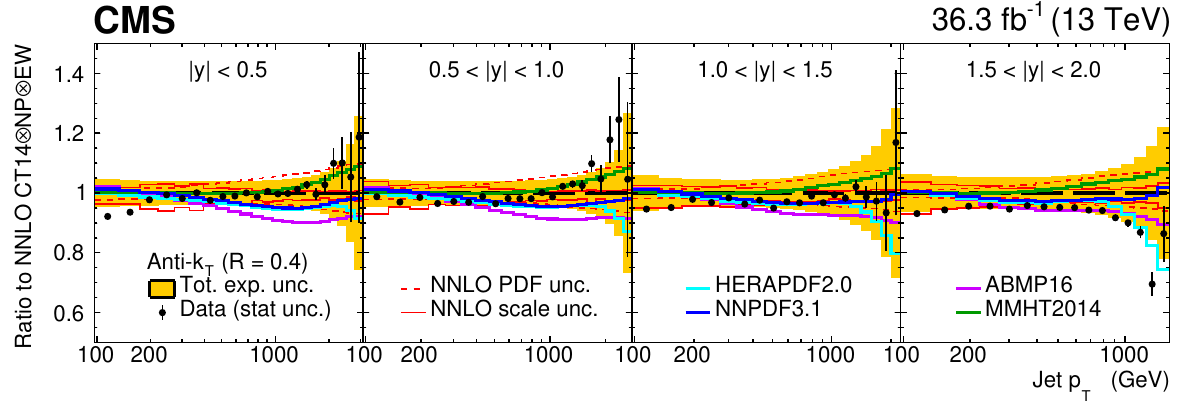}

    \includegraphics[width=\textwidth]{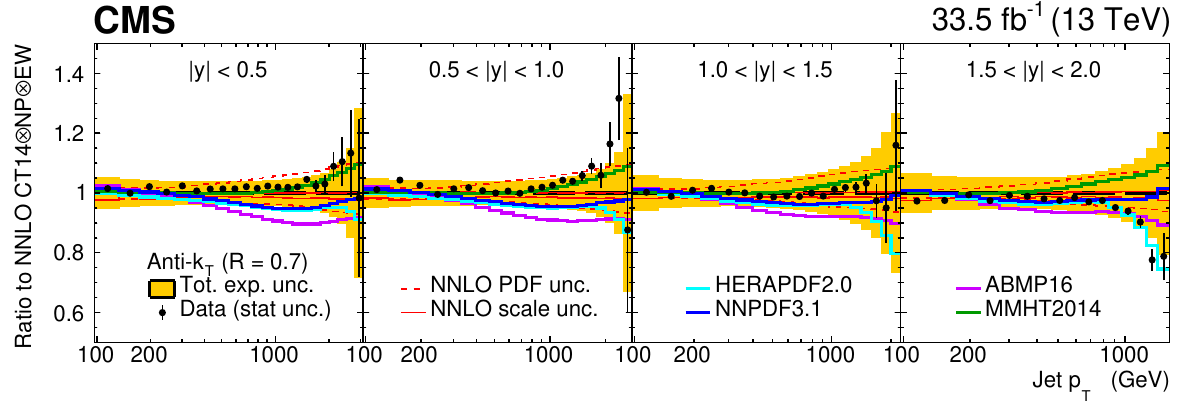}

    \caption[Comparison of result to NNLO with interpolation tables]{The double-differential cross section of inclusive jet production, as a function of \pt and \absy, for jets clustered using the anti-\kt algorithm with $R=0.4$ (upper panel) and $R=0.7$ (lower panel), presented as ratios to the QCD predictions. The data points are shown by filled circles, with statistical uncertainties shown by vertical error bars, while the total experimental uncertainty is centred at one and is presented by the orange band.
The data are divided by the NNLO prediction corrected for NP and EW effects, using CT14nnlo PDF and choosing jet \pt as renormalisation and factorisation scale.
NNLO predictions obtained with alternative PDF sets are displayed in different colours as a ratio to the central prediction using CT14nnlo.
    }
    \label{fig:comparisonToNNLO}
\end{figure}

The QCD analysis at NNLO is repeated by using the NNLO interpolation grids for the double-differential inclusive jet cross section~\cite{Britzger:2022lbf}, which were released after the journal publication of the original analysis. The NNLOJET calculation used to derive these grids is based on the leading-colour and leading-flavour-number approximation and does not include the most recent subleading colour contributions. However, these contributions were reported in Ref.~\cite{Chen:2022tpk} to be very small in inclusive jet production, in particular for a jet size of $R=0.7$. The grids also contain an estimate of the numerical integration uncertainty of around 1\% or less. To account for point-to-point fluctuations, this uncertainty, after consultation with the authors of NNLOJET, has been increased by a factor of two; however, its impact in the fit is negligible.
A comparison of the measurement with predictions using various PDFs is shown in Fig.~\ref{fig:comparisonToNNLO}.
Although the PDF parametrisation remains identical, higher precision in PDF and QCD parameters is expected by using NNLO grids consistently in the QCD analysis. These new results supersede those obtained by using the $k$-factor technique.  

\begin{figure}[ht]
\centering
\includegraphics[width=0.47\textwidth]{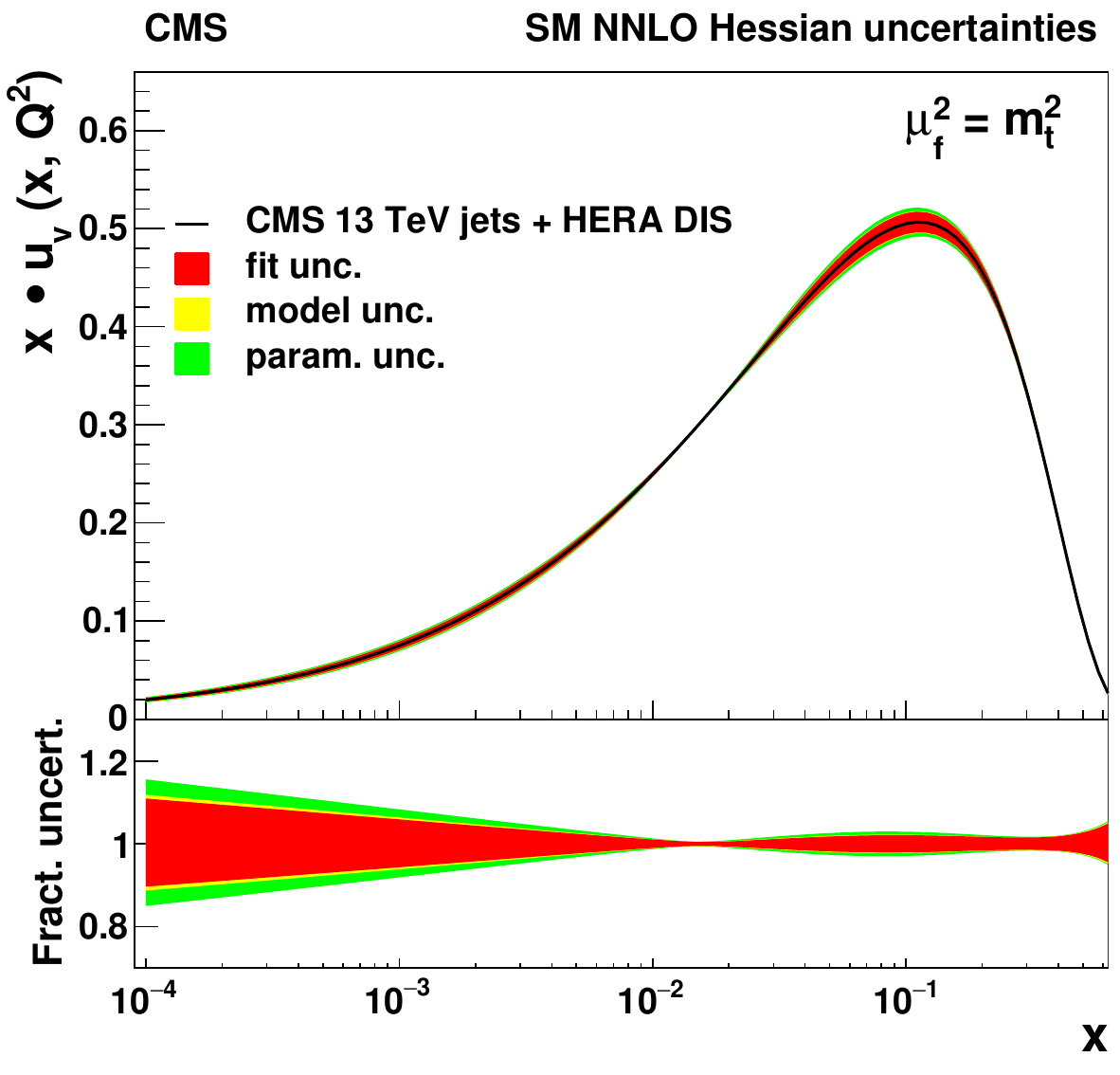}
\includegraphics[width=0.47\textwidth]{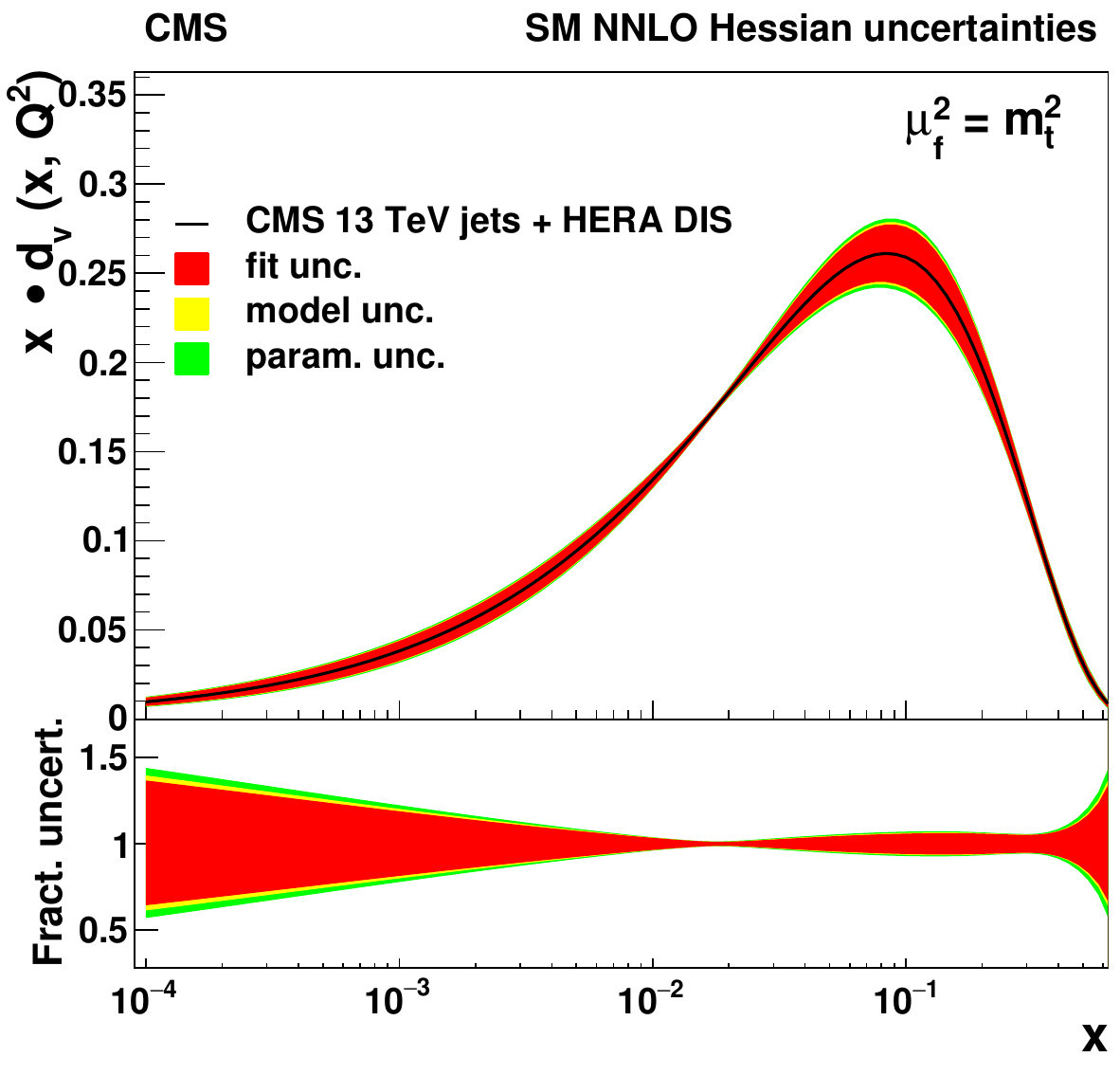}\\
\includegraphics[width=0.47\textwidth]{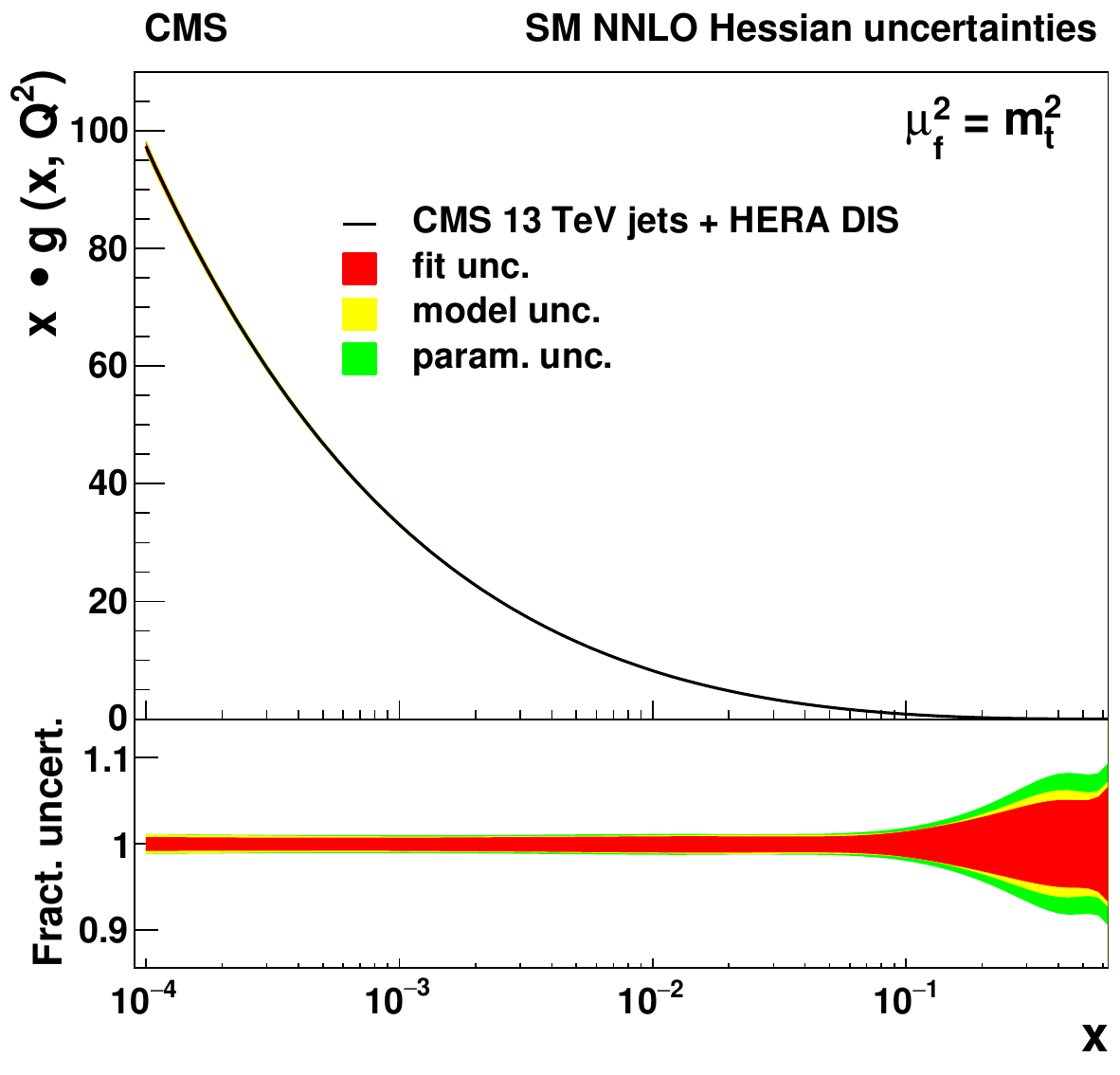}
\includegraphics[width=0.47\textwidth]{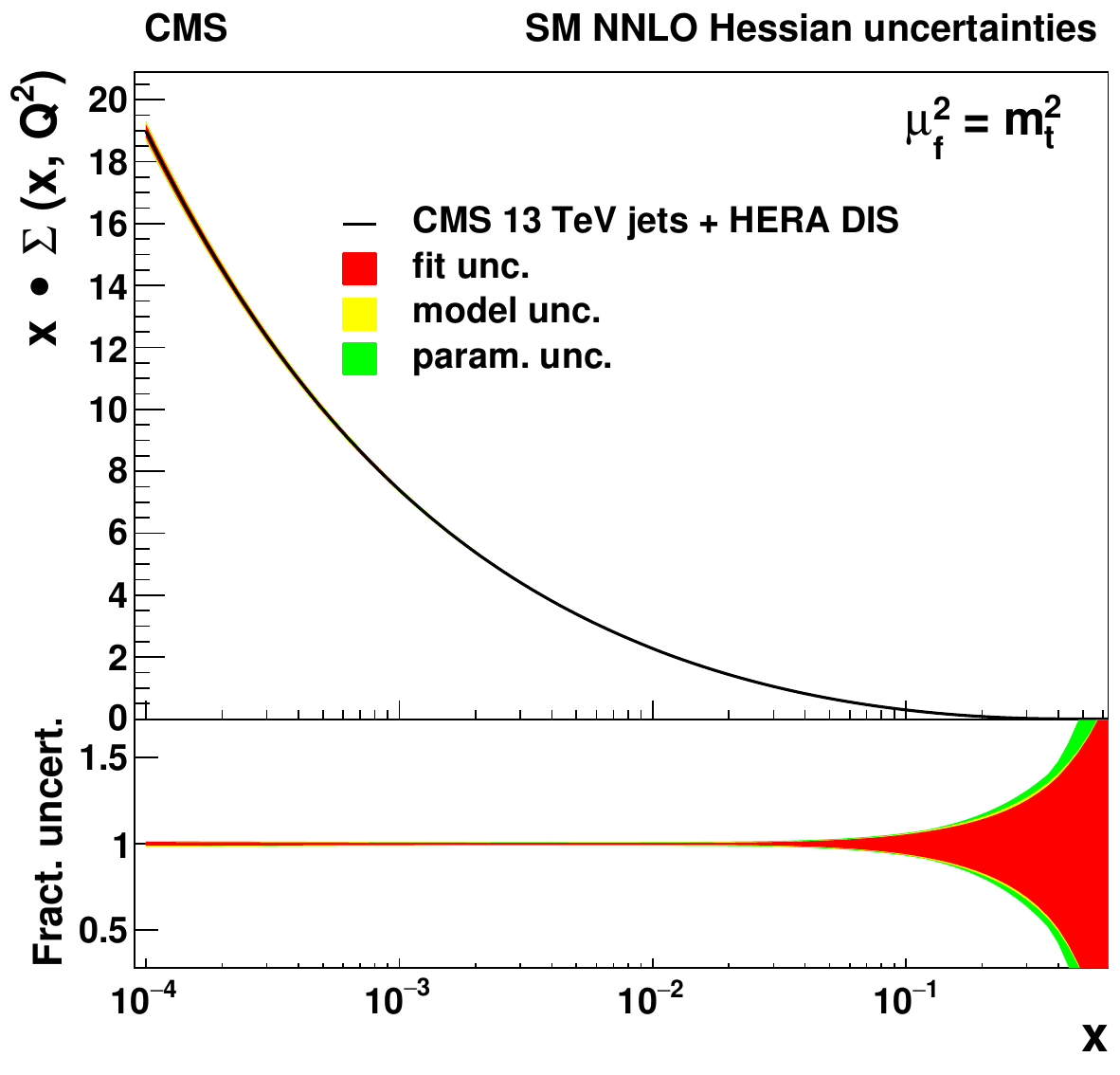}
\caption{The \PQu-valence (upper left), \PQd-valence (upper right), gluon (lower left), and sea quark (lower right) distributions shown as a function of $x$ at the scale $\muf=\mt$, resulting from the NNLO fit using HERA DIS data together with the CMS inclusive jet cross section at $\sqrts=13\TeV$. The prediction for the inclusive jet cross section is obtained using NNLO interpolation grids. Contributions of the fit, model, and parametrisation uncertainties for each PDF are shown. In the lower panels, the relative uncertainty contributions are presented.}
\label{NNLO_grid_HERA+CMS_breakdown}
\end{figure}

The PDFs from the QCD analysis at NNLO of the CMS inclusive jet production and HERA DIS cross sections are shown in Fig.~\ref{NNLO_grid_HERA+CMS_breakdown}, illustrating the contributions of the fit, model, and parametrisation uncertainties. 

{\tolerance=800
The value of the strong coupling constant \alpSZ is extracted simultaneously with the PDFs and corresponds to $\alpSZ = 0.1166 \pm 0.0014\,\text{(fit)} \pm 0.0007\,\text{(model)} \pm 0.0004\,\text{(scale)}\pm 0.0001\,\text{(param.)}$,
showing improved precision with respect to the NNLO result obtained using the $k$-factor technique. 
The global and partial \chisq values for each data set in the NNLO fits using the interpolation grids are listed in Table~\ref{QCD_analysis_partial_chi2_NNLO_grid}, where the \chisq values illustrate a general agreement among all the data sets.
\par}

\begin{table}[ht]
    \topcaption{Partial \chisq per number of data points, \Ndp, and the global \chisq per degree of freedom, \Ndof, as obtained in the QCD analysis at NNLO of HERA+CMS jet data, using NNLO interpolation grids for the 13\TeV inclusive jet cross section. In the DIS data, the proton beam energy is given as \Ep and the electron energy is 27.5\GeV.}
    \label{QCD_analysis_partial_chi2_NNLO_grid}
    \centering
    \begin{tabular}{l l | c}
                                  &  & HERA+CMS \\
        Data sets & ~ & Partial $\chisq/\Ndp$ \\
        \hline                                                                                                     
        HERA I+II neutral current                       & $\Pep\Pp$, $\Ep=920\GeV$     & 376/332   \\
        HERA I+II neutral current                       & $\Pep\Pp$, $\Ep=820\GeV$     & 60/63     \\
        HERA I+II neutral current                       & $\Pep\Pp$, $\Ep=575\GeV$     & 202/234   \\
        HERA I+II neutral current                       & $\Pep\Pp$, $\Ep=460\GeV$     & 209/187   \\
        HERA I+II neutral current                       & $\Pem\Pp$, $\Ep=920\GeV$     & 227/159   \\
        HERA I+II charged current                       & $\Pep\Pp$, $\Ep=920\GeV$     & 46/39     \\
        HERA I+II charged current                       & $\Pem\Pp$, $\Ep=920\GeV$     & 56/42     \\ [\cmsTabSkip]
        CMS inclusive jets 13\TeV                       & $0.0 < \absy < 0.5$                & 8.6/22    \\ 
                                                        & $0.5 < \absy < 1.0$                & 23/21     \\
                                                        & $1.0 < \absy < 1.5$                & 13/19     \\
                                                        & $1.5 < \absy < 2.0$                & 14/16     \\
        \hline
        Correlated \chisq                             &                               & 81        \\                                                                                                       
        Global $\chisq/\Ndof$                &                               & 1302/1118 \\
    \end{tabular}
\end{table}

\begin{figure}[ht]
\centering
\includegraphics[width=0.47\textwidth]{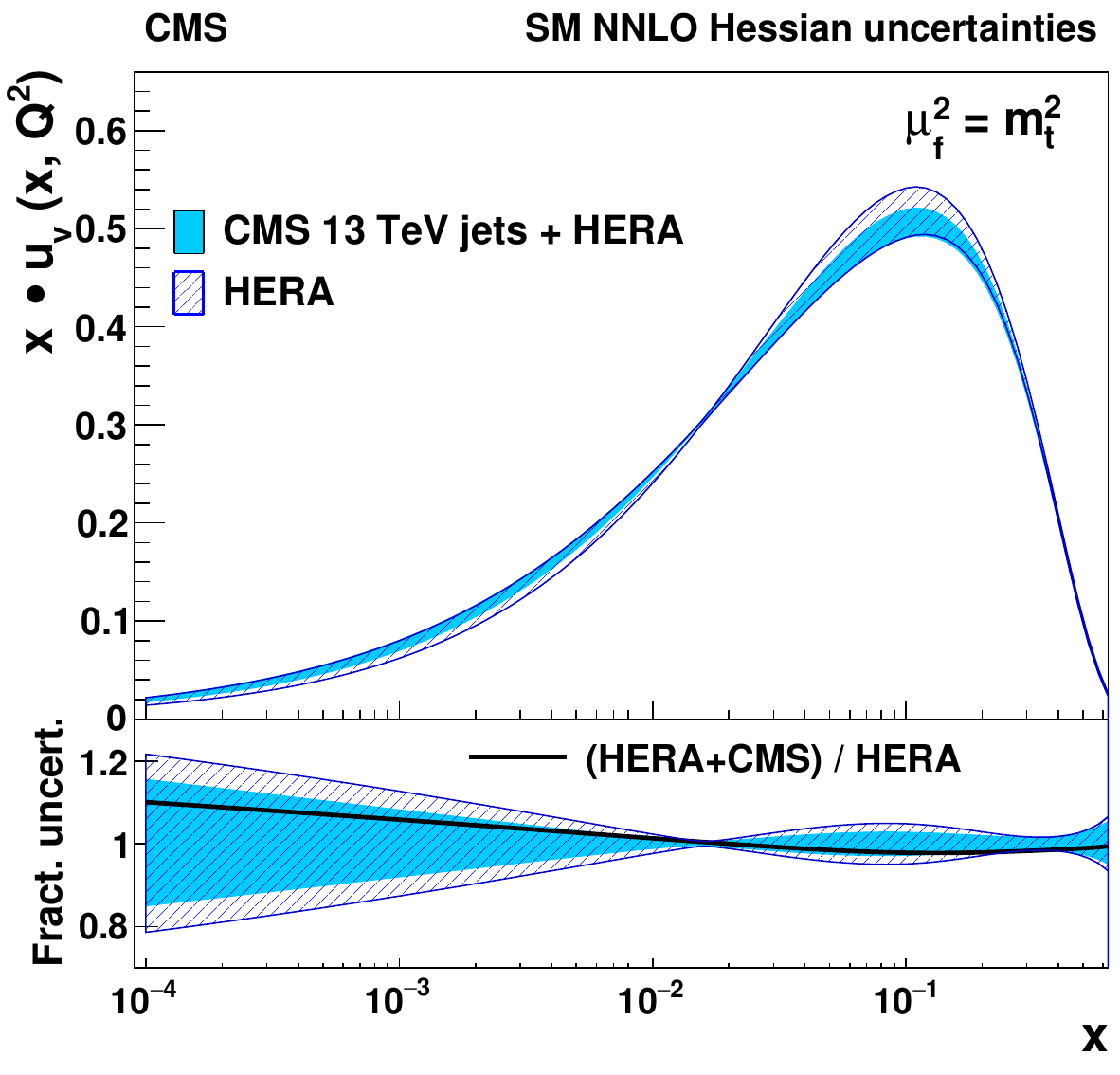}
\includegraphics[width=0.47\textwidth]{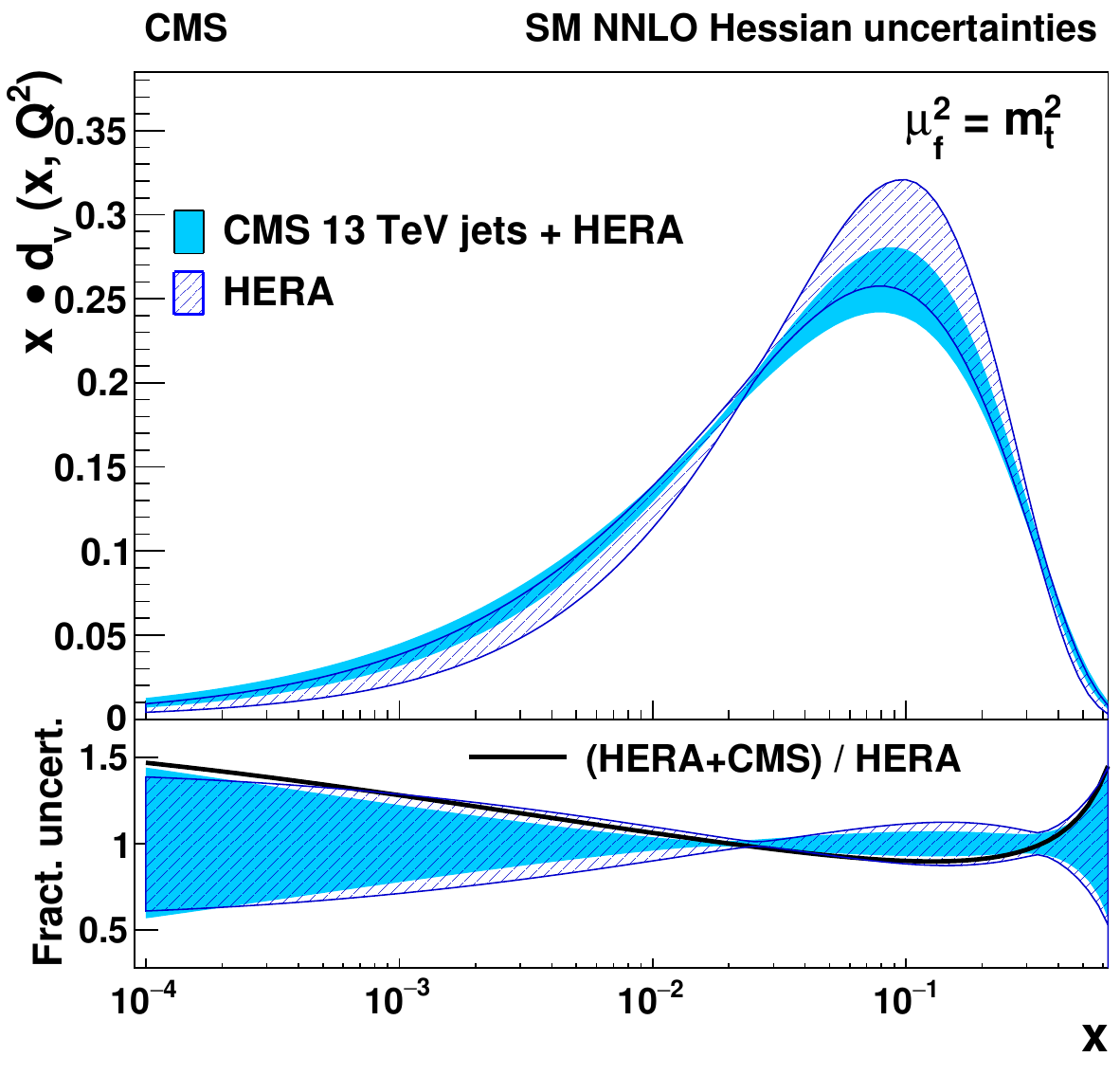}\\
\includegraphics[width=0.47\textwidth]{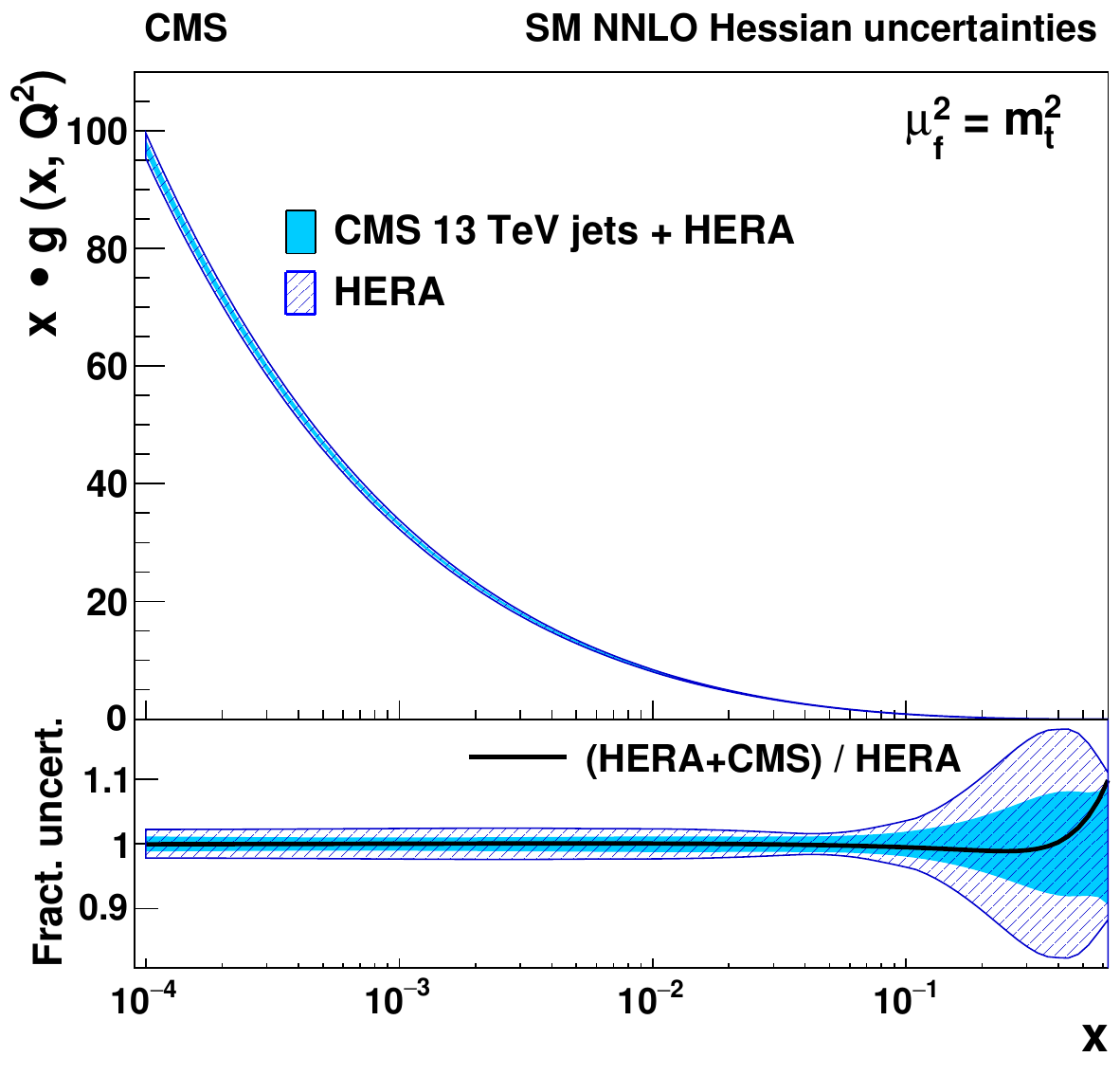}
\includegraphics[width=0.47\textwidth]{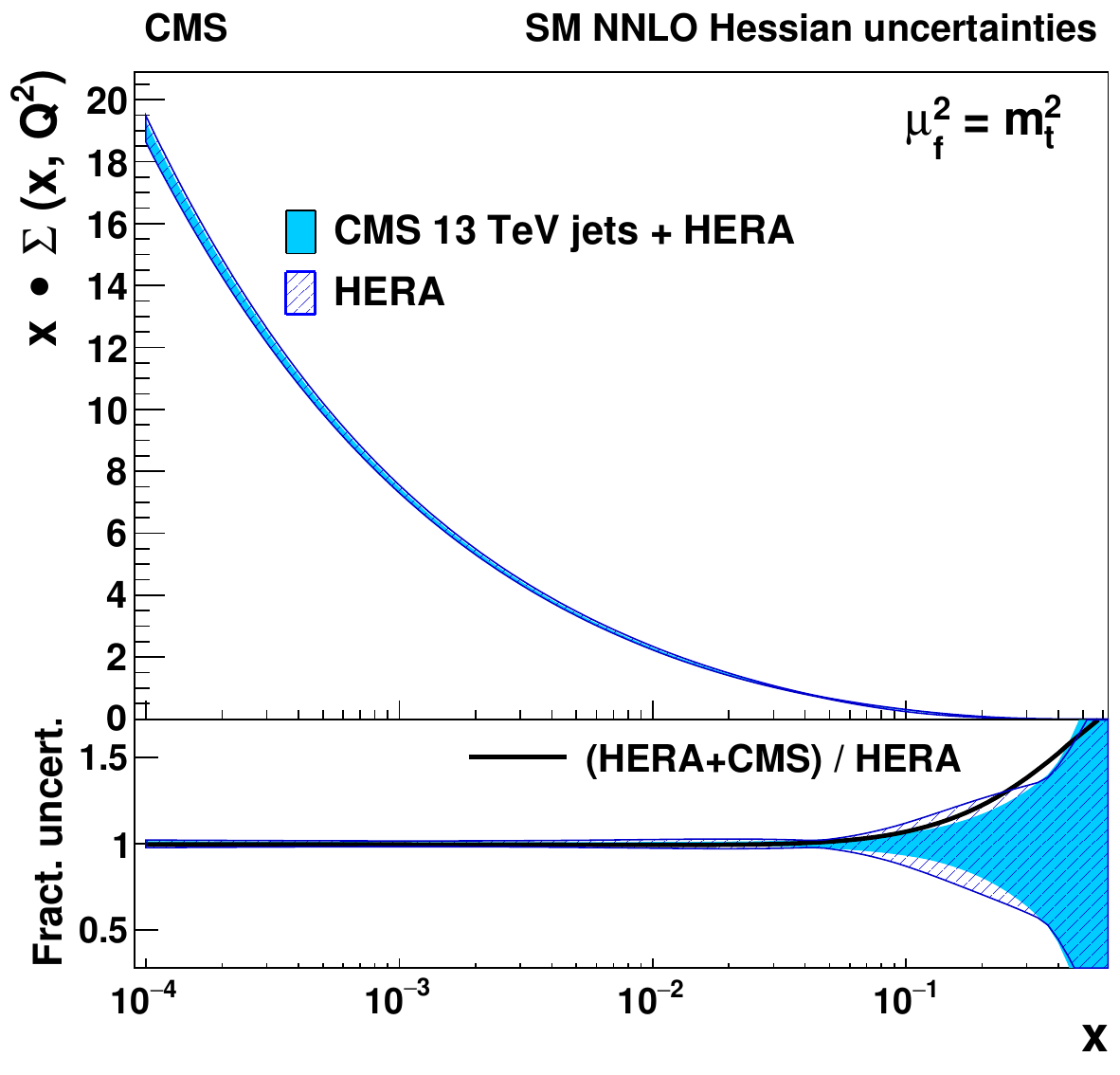}
\caption{The \PQu-valence (upper left), \PQd-valence (upper right), gluon (lower left), and sea quark (lower right) distributions shown as a function of $x$ at the scale $\muf=\mt$. The filled (hatched) band represents the results of the NNLO fit using HERA DIS and the CMS inclusive jet cross section at $\sqrts=13\TeV$  (using the HERA DIS data only). The PDFs are shown with their total uncertainty. The prediction for the inclusive jet cross section is obtained using NNLO interpolation grids. In the lower panels, the comparison of the relative PDF uncertainties is shown for each distribution. The line corresponds to the ratio of the central PDF values of the two variants of the fit.}
\label{NNLO_grid_HERA_vs_HERA+CMS}
\end{figure}

The impact of the CMS jet data in the QCD analysis (HERA+CMS fit) at NNLO is illustrated in Fig.~\ref{NNLO_grid_HERA_vs_HERA+CMS}, where the result is compared with the alternative fit using only the HERA DIS data (HERA-only fit). 

% \bibliography{auto_generated}

%%% END EDITABLE REGION %%%
% skeleton_end
\end{document}